\documentclass[10pt]{article}
\usepackage[french,english]{babel}
\usepackage{amsmath}
\usepackage{amsfonts}
\usepackage{amssymb}
\usepackage{graphicx}  
\usepackage{times}
\usepackage[makeroom]{cancel}
\begin{document}

%
%
September 6th 2016 \hfill 
\vskip 5cm
{\baselineskip 18pt
\begin{center}
{\bf THE 1-LOOP VACUUM POLARIZATION FOR A GRAPHENE-LIKE MEDIUM
\break
IN AN EXTERNAL MAGNETIC FIELD ;
\break
CORRECTIONS TO THE COULOMB POTENTIAL
}
\end{center}
}
\baselineskip 16pt
\arraycolsep 3pt  
%
\vskip .2cm
\centerline{
B.~Machet
     \footnote{Sorbonne Universit\'es, UPMC Univ Paris 06, UMR 7589,
LPTHE, F-75005, Paris, France}
     \footnote{CNRS, UMR 7589, LPTHE, F-75005, Paris, France.}
     \footnote{Postal address:
LPTHE tour 13-14, 4\raise 3pt \hbox{\tiny \`eme} \'etage,
          UPMC Univ Paris 06, BP 126, 4 place Jussieu,
          F-75252 Paris Cedex 05 (France)}
    \footnote{machet@lpthe.jussieu.fr}
     }
\vskip 1cm

{\bf Abstract:} I calculate the 1-loop vacuum polarization
$\Pi_{\mu\nu}(k,B,a)$
for a photon of momentum $k=(\hat k,k_3)$ interacting with the electrons of
a thin medium of thickness $2a$ simulating graphene, in the
presence of a constant and uniform external magnetic field $B$ orthogonal
to it (parallel to $k_3$). Calculations  are done with the techniques of Schwinger,
adapted to the geometry and Hamiltonian under scrutiny.
The situation gets more involved than for the electron self-energy 
because the photon is now allowed to also propagate outside the medium.
This makes $\Pi_{\mu\nu}$  factorize into a quantum, ``reduced''
$T_{\mu\nu}(\hat k,B)$
 and a transmittance function $V(k,a)$, in which
the geometry of the sample and the resulting confinement of the
$\gamma\,e^+\,e^-$ vertices play major roles. This drags the results away
from reduced QED$_{3+1}$ on a 2-brane.
The finiteness of $V$ at $k^2=0$ is an essential ingredient
to fulfill suitable renormalization condition for $\Pi_{\mu\nu}$ and to fix
the corresponding counterterms. Their connection with the
transversality of $\Pi_{\mu\nu}$ is investigated.
The corrections to the Coulomb potential and their dependence on $B$ 
strongly differ from QED$_{3+1}$.

\bigskip


\newpage
%
\section{Generalities. Framework of the calculations}

This study concerns the propagation of a photon (with incoming momentum
$k$) interacting with electrons belonging to a graphene-like medium of
thickness $2a$, and, more specially,
the 1-loop quantum corrections to its propagator..
They originate from the creation, inside the medium,
of virtual $e^+ e^-$ pairs which  propagate before annihilating,
again inside graphene.
The two $\gamma\,e^+e^-$ vertices are therefore geometrically constrained to lie  in the
interval $[-a, +a]$ along the direction $z$ of the magnetic field,
perpendicular to the surface of graphene.  This is best expressed by
evaluating the photon propagator in position space, and by integrating
the ``$z$'' coordinates of the two vertices from $-a$ to $+a$ instead of
the infinite interval of usual Quantum Field Theory (QFT).

The second feature that is implemented to mimic graphene is to deprive
the Hamiltonian of the Dirac electrons of its ``$\gamma_3
p_3$'' term (see for example \cite{Goerbig2011}).
 I shall not consider a Fermi velocity different from the
speed of light, nor additional degeneracies that usually take place in
graphene, and will furthermore consider  electrons to have a mass $m$,
that I shall let  go to $0$ at the end of the calculations.

\begin{figure}[h]
\begin{center}
\includegraphics[width=10 cm, height=7 cm]{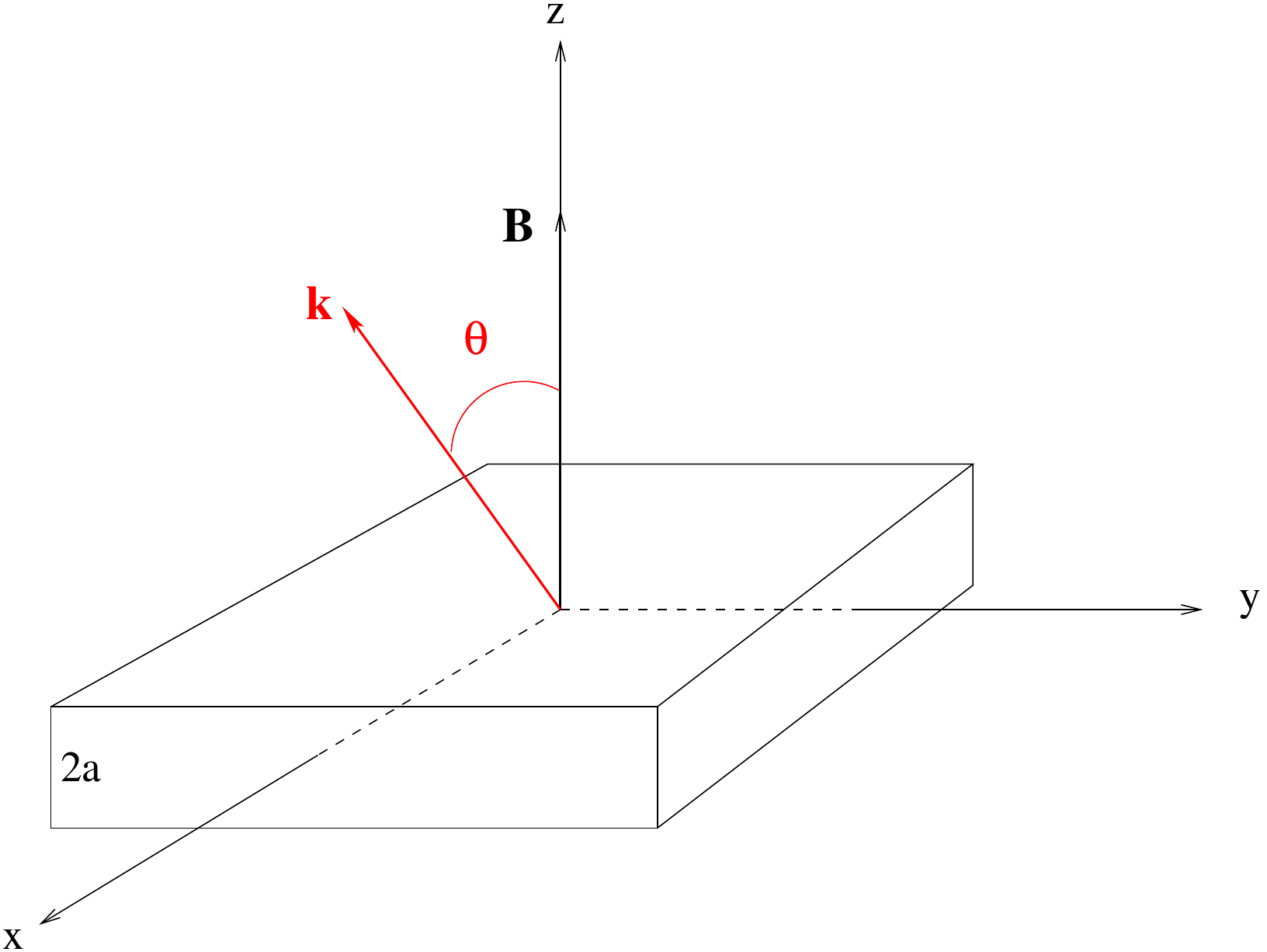}
\caption{$\vec B$ is perpendicular to the medium strip of width $2a$.
} \label{fig:setup}
\end{center}
\end{figure}

The setting is the following.
The constant and uniform
magnetic field  $\vec B$ is chosen to be parallel to the $z$ axis
and the wave vector $\vec k$ of the propagating photon
to lie in the $(x,z)$ plane (see Figure~\ref{fig:setup})
\footnote{When no ambiguity can occur, I shall often omit the arrow on
3-dimensional vectors, writing for example $B$ instead of $\vec B$.}.

The $(\vec B, \vec k)$ angle $\theta$ is the ``angle of incidence'';
the plane $(x,z)$ is the plane of incidence.

Calculations are performed with the techniques of Schwinger for Quantum
Field Theory in the presence of a constant and uniform external magnetic
field \cite{Schwinger1949} \cite{Schwinger1951}. They have  been
intensely used by Tsai in standard QED (see for example \cite{Tsai1974}).
Very careful and precise explanations of these techniques
have been given in the book by
Dittrich and Reuter \cite{DittrichReuter}, of invaluable help.

I shall in the following use ``hatted'' letters for vectors living in the
Lorentz subspace $(0,1,2)$ ($0$ being the time-like component, $1$, $2$ and
$3$ respectively the $x$, $y$ and $z$-like ones). For example
\begin{equation}
\hat k = (k^0,k^1,k^2,0),\quad k=(\hat k, k_3) = (k_0,k_1,k_2,k_3)=(k_0, \vec
k).
\end{equation}

Dirac $\gamma$ matrices and spinors are always 4-dimensional.
Throughout the paper I use the metric $(-1,+1,+1,+1)$ like in
\cite{Schwinger1949}, \cite{Schwinger1951}, \cite{Tsai1974} and \cite{DittrichReuter}.

I shall also use the following notations
\begin{equation}
\begin{split}
& k_\parallel=(k_0,0,0,k_3) \Rightarrow  k_\parallel^2 = -k_0^2 + k_3^2,\cr
& \hat k_\parallel = (k_0,0,0,0) \Rightarrow \hat k_\parallel ^2 = -k_0^2,\cr
& k_\perp = (0,k_1,k_2,0)=\hat k_\perp \Rightarrow k_\perp^2 = k_1^2 + k_2^2
= \hat k_\perp^2,\cr
& g^{\mu\nu}_\parallel = (-1,0,0,1),\quad g^{\mu\nu}_\perp=(0,1,1,0),\cr
& \hat g^{\mu\nu}=(-1,1,1,0),\quad
\hat g^{\mu\nu}_\parallel = (-1,0,0,0),\quad \hat g^{\mu\nu}_\perp =
(0,1,1,0)=g^{\mu\nu}_\perp,
\end{split}
\end{equation}
and
$\sigma^3 = \sigma^{12}= \frac{i}{2}[\gamma^1,\gamma^2]= diag(1,-1,1,-1)$
like in \cite{DittrichReuter} (it should not be confused with the $2 \times
2$ Pauli matrix).

When they are not needed, the factors $\hbar$ and $c$ will very often be
skipped.

The plan of this work is the following.

$\bullet$\ In section \ref{section:propagator}, I show, by working in position space,
how, due to the confinement of the $\gamma e^+ e^-$ vertices inside the
thin medium, the vacuum polarization $\Pi_{\mu\nu}(k,B)$ factorizes into  a
transmittance function $V(k, a)$ times  a ``reduced'' $T_{\mu\nu}(\hat
k, B)$; after giving an analytical expression for $V$, I show
its finiteness on mass-shell ($k^2=0$), which is, as shown later,
 essential for renormalization;
I also study its limit as $k_0 \to 0$, which is useful when calculating
the corrections to the Coulomb potential.

$\bullet$\ In section \ref{section:T}, I  get
the unrenormalized $T_{\mu\nu}^{bare}$ as a double integral; it is only
$(2+1)$-transverse.

$\bullet$\ In section \ref{section:renorm1}, I determine counterterms in
order that on mass-shell renormalization conditions for $\Pi_{\mu\nu}(k,B)$ are
satisfied. Only $(2+1)$-transversality is achieved.
The limits $B=0$ and $B\to \infty$ are studied in detail. Their massless
limit $m\to 0$ is smooth. $\Pi_{\mu\nu}$ is shown to vanish at $m=0, B\to
\infty$.

$\bullet$\ In section \ref{section:scalpot}, I calculate, at the limit $a
\to 0$, shown to be smooth, the corrections to
the Coulomb potential. I first show that, at $B=0$, it gets renormalized
by $1/(1+\alpha/2)$ while, at $B\to \infty$, the genuine Coulomb potential
is recovered. The interpolation between these two limits being smooth, 
sizable deviations from Coulomb are only  expected for strongly coupled
systems.

$\bullet$\ In section \ref{section:alter}, I investigate
whether, while still preserving on mass-shell renormalization conditions,
 counterterms can be adapted such that $(3+1)$-transversality is
achieved. A fist example introduces extra $B$-independent counterterms.
$(3+1)$-transversality is achieved at $B=0$ only; at $B\to \infty$,
$\Pi_{\mu\nu}$ does not vanish anymore at $m=0$. In the second example,
arguing that, {\em de facto}, by neglecting $B$-dependent boundary terms,
Schwinger introduces $B$-dependent counterterms,  I introduce 
counterterms that depend on the external $B$. At this price,
$(3+1)$-transversality can be achieved at any $B$ while $\Pi_{\mu\nu}$
vanishes at $B \to \infty$ independently of the limit $m\to 0$.

$\bullet$\ Section \ref{section:conclusion} concludes this works with
general remarks concerning the calculation, the fate of
dimensional reduction which is a well known phenomenon for QED$_{3+1}$ in
superstrong external $B$, and states numerous issues that have not been
tackled here and should be in future works.

$\bullet$\ The demonstration of the  master factorization formula
$\Pi_{\mu\nu} \sim V T_{\mu\nu}$ in position space, eq.~(\ref{eq:genform}), 
is detailed in Appendix \ref{section:genform}.  

Like in \cite{Machet2016-1} and \cite{Machet2016-2}, calculations are
exposed in details, with no ``gap'', such that it should not be a problem
for a dedicated reader to redo them.

\section{The photon propagator in $\boldsymbol x$-space and the
vacuum polarization $\boldsymbol{\Pi^{\mu\nu}}$ ; generalities}
\label{section:propagator}

\begin{figure}[h]
\begin{center}
\includegraphics[width=6 cm, height=3 cm]{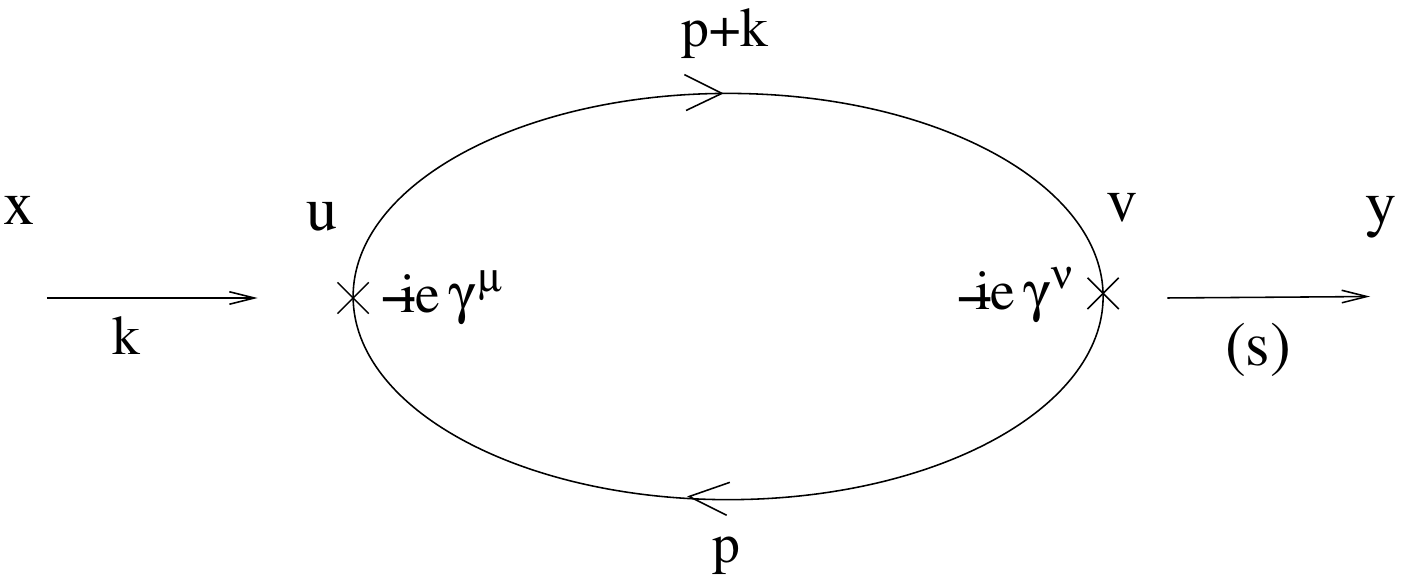}
\caption{The vacuum polarization $\Pi^{\mu\nu}(k)$.}
\label{fig:vacpol}
\end{center}
\end{figure}

The 1-loop vacuum polarization $\Pi_{\mu\nu}$  we determine by
calculating the photon propagator in position-space, while confining, at
the two vertices $\gamma\, e^+ e^-$, the corresponding $z$ coordinates
inside graphene, $z\in [-a,a]$.

It factorizes into $\Pi_{\mu\nu}(\hat k,k_3,\displaystyle\frac{y_3}{a},B)=
\displaystyle\frac{1}{\pi^2}\;T_{\mu\nu}(\hat k,B)\; U(\hat k, k_3,
\displaystyle\frac{y_3}{a})$
\footnote{The 2 vertices are located at space-time points $x$ and $y$. After
the dependence on $x_3-y_3$ has been factored out, a very weak dependence on
$u=y_3/a$ subsists. see also the end of subsection \ref{subsub:graph}.}
in which  $U$ is a universal function that does not depend on the magnetic
field, nor on $\alpha\equiv \displaystyle\frac{e^2}{4\pi(\hbar c)}$,
 that we also encounter when 
no external $B$ is present. $U$ turns out (see eq.~(\ref{eq:Pieff2}) below)
 to be the Fourier transform of the
product of two functions:  the first, $\displaystyle\frac{\sin a l_3}{al_3}$,  is itself  the
Fourier transform of the ``gate function'' corresponding to
the graphene strip along $z$; the second  carries  the remaining
information attached to the confinement of the vertices.

The integration variable $l_3$ of this Fourier transform is the 
component along $B$ of the difference $s-k$ between the momenta  of
the outgoing and incoming photons.
It represents the amount of momentum non-conservation of photons
due to the exchange between them
and (the quantum momentum fluctuations of) electrons.

This factorization can be traced back to $T_{\mu\nu}$  not depending on
$k_3$, for the simple reason that the propagators of electrons inside
graphene should be evaluated at vanishing  momentum along $z$.

An example of how factors combine is the following.
$\Pi_{\mu\nu}$ still includes an integration on the loop momentum $p_3$, which
factors out.
That the interactions of electrons are confined along $B$ triggers quantum
fluctuations of their momentum in this direction. Setting an ultraviolet
cutoff  $\pm\displaystyle\frac{\hbar}{a}$ on the $p_3$ integration (saturating the
Heisenberg uncertainty relation) makes  this integral  proportional to
$\displaystyle\frac{1}{a}$. This factor  completes, inside the integral
$\int dl_3$ defining $U$, the ``geometric'' $\displaystyle\frac{\sin al_3}{al_3}$
evoked above.
Then, the integration $\int dl_3$  gets bounded by the rapid decrease
of $\displaystyle\frac{\sin al_3}{al_3}$ for $|l_3|$ larger than
$\displaystyle\frac{\hbar}{a}$;
this upper bound $|l_3|\leq \displaystyle\frac{\hbar}{a}$ is the same as the one
that we set for quantum fluctuations of the electron momentum along $z$.
Therefore, the energy-momentum
non-conservation between  the outgoing and incoming photons cannot exceed
the uncertainty on the momentum of electrons
due to the confinement of vertices. Exact momentum conservation for the photon
only gets recovered when $a\to\infty$ (limit of ``standard'' QFT).

\subsection{The 1-loop photon propagator in position space}
\label{subsec:gammaprop}

I calculate the 1-loop photon propagator (eq.~(4.1) of \cite{DittrichReuter})
\begin{equation}
\Delta^{\rho\sigma}(x,y)= i\,< 0 \;|\; T A^\rho(x) A^\sigma(y)\;|\; 0>,
\end{equation}
and somewhat lighten the notations, often omitting symbols like T-product,
\ldots, writing for example $G(\hat k)$ instead of $G(\hat k, B)$.

Introducing the coordinates $u=(u_0,u_1,u_2,u_3)$ and $v=(v_0,v_1,v_2,v_3)$
of the two $\gamma\,e^+ e^-$ vertices one gets at 1-loop
\begin{equation}
 \Delta^{\rho\sigma}(x,y)=
 i\int d^4u \int d^4v\; A^\rho(x) \big[(-ie) A^\mu(u) \bar\psi(u)\gamma_\mu
\psi(u)\big]
\big[(-ie)A^\nu(v) \bar\psi(v)\gamma_\nu \psi(v)\big] A^\sigma(y).
\end{equation}
Making the contractions for fermions etc \ldots yields
\begin{equation}
\begin{split}
 \Delta^{\rho\sigma}(x,y) &= ie^2\int d^4u \int d^4v\; Tr
\int \frac{d^4k}{(2\pi)^4}\; e^{ik(u-x)}\Delta^{\rho\mu}(k)
\gamma_\mu \Phi(u,v)\int \frac{d^4p}{(2\pi)^4}\; e^{ip(u-v)}G(p)
\gamma_\nu\cr
& \hskip 3cm \Phi(v,u)\int \frac{d^4r}{(2\pi)^4}\; e^{ir(v-u)}G(r)
\int \frac{d^4s}{(2\pi)^4}\; e^{is(y-v)}\Delta^{\sigma\nu}(s).
\end{split}
\label{eq:start}
\end{equation}
In what follows we shall also often omit the trace symbol ``$Tr$''.

I have inserted in (\ref{eq:start}) the phase $\Phi$ that occurs in a
fermion propagator $G$ in the presence of a constant external magnetic field
\cite{DittrichReuter}
\begin{equation}
\begin{split}
G(x',x'') &= \Phi(x',x'')\int\frac{d^4p}{(2\pi)^4}\;e^{ip(x'-x'')}G(p),\cr
\Phi(x',x'') &= \exp\Big[-ie\int_{x''}^{x'} dx_\mu\;\big(A^\mu(x)+\frac12
F^{\mu\nu}(x'_\nu-x''_\nu)\big)\Big].
\end{split}
\end{equation}
Since the curl of the integrand vanishes, the integral inside the phase is
independent of the path of integration,
which can therefore be chosen as straight $x(t)=x''+t(x'-x''), t\in[0,1]$,
 leading to the familiar
expression
\begin{equation}
\Phi(x',x'')= \exp\Big[ie\int_{x''}^{x'}dx_\mu\;A^\mu(x)\Big].
\label{eq:phase}
\end{equation}
This is the last time that we mention $\Phi$ because it goes away when the
path of integration closes, which is the case for the vacuum polarization.

\subsubsection{``Standard'' $\boldsymbol{(3+1)}$-Quantum Field Theory}
\label{subsub:sQFT}

 One integrates $\int_{-\infty}^{+\infty}
d^4u$ and $\int_{-\infty}^{+\infty} d^4v$  for the four components of $u$
and
$v$. This gives:
\begin{equation}
 \Delta^{\rho\sigma}(x,y)=i\int \frac{d^4k}{(2\pi)^4}\; e^{-ik(x-y)}
\Delta^{\rho\mu}(k)
\Delta^{\nu\sigma}(k)
\underbrace{e^2\int \frac{d^4p}{(2\pi)^4}\; \gamma_\mu G(p) \gamma_\nu
G(p+k)}_{i\Pi_{\mu\nu}(k)}.
\label{eq:stand}
\end{equation}
To obtain the sought for vacuum polarization, the two external photon
propagators
$\Delta^{\rho\mu}(k)$ and $\Delta^{\nu\sigma}(k)$
have to be chopped off, which gives the customary expression
\begin{equation}
 i\Pi_{\mu\nu}(k)= e^2\int \frac{d^4p}{(2\pi)^4}\; \gamma_\mu\, G(p)\,
\gamma_\nu\, G(p+k).
\label{eq:pimunustand}
\end{equation}

\subsubsection{The case of a graphene-like medium: the $\boldsymbol{\gamma\, e^+ e^-}$
vertices  are  confined along $\boldsymbol z$}
\label{subsub:graph}

The coordinates $u_3$ and $v_3$ of the two vertices we do not integrate
anymore
$\int_{-\infty}^{+\infty}$ but only $\int_{-a}^{+a}$. This localizes the
interactions of electrons  with photons inside graphene.
It has been shown in \cite{Machet2016-1} that, in the case of the electron
self-energy at 1-loop. this procedure leads to the same result as reduced
QED$_{3+1}$ on a 2-brane \cite{Gorbar2001} \cite{Pevzner}.

Decomposing in (\ref{eq:start})
 $du = d^3\hat u\, du_3,\; dv=d^3\hat v\, dv_3$, we get by
standard manipulations (see Appendix \ref{section:genform})
\begin{equation}
\begin{split}
\Delta^{\rho\sigma}(x,y)
&=
i\int \frac{dp_3}{2\pi} \int \frac{dk_3}{2\pi}\int \frac{dr_3}{2\pi}
\int \frac{ds_3}{2\pi}\int_{-a}^{+a}du_3\; e^{iu_3(k_3+p_3-r_3)}
\int_{-a}^{+a}dv_3\; e^{iv_3(-p_3+r_3-s_3)}
\cr
& \int \frac{d^3\hat k}{(2\pi)^3}\;
e^{i\hat k(\hat y -\hat x)}
e^{ik_3(-x_3)} e^{is_3(y_3)}\Delta^{\rho\mu}(\hat k,k_3)
\Delta^{\sigma\nu}(\hat k, s_3)\
\underbrace{e^2 \int \frac{d^3\hat p}{(2\pi)^3}\;\gamma_\mu G(\hat p,B)
 \gamma_\nu G(\hat p+\hat k,B)}_{iT_{\mu\nu}(\hat k,B)},
\end{split}
\label{eq:genform}
\end{equation}
in which we introduced the tensor $T_{\mu\nu}(\hat
k,B)$ that is  calculated in section \ref{section:T}.

One of the main difference with standard QFT (subsection \ref{subsub:sQFT})
is that the tensor $T_{\mu\nu}$
does not depend on $k_3$, but only on $\hat k$. The reason is that, as already
mentioned, the propagators of electrons in the loop are evaluated at
vanishing momentum in the direction of $B$, simulating a graphene-like
Hamiltonian.

Notice that, despite the ``classical'' input $p_3=0$ 
 the photon propagator still involves an integration $\int dp_3$ over the
loop momentum $p_3$.

Now,
\begin{equation}
\int_{-a}^{+a} dx\; e^{itx} = 2\;\frac{\sin at}{t},
\end{equation}
such that
\begin{equation}
\begin{split}
& \Delta^{\rho\sigma}(x,y)= 4i\int \frac{dk_3}{2\pi} \int \frac{ds_3}{2\pi}
e^{i(s_3y_3-k_3x_3)} L(a,s_3,k_3)
\int \frac{d^3\hat k}{(2\pi)^3}\; e^{i\hat k(\hat y -\hat x)}
\Delta^{\rho\mu}(\hat k,k_3) \Delta^{\sigma\nu}(\hat k, s_3)\;
iT_{\mu\nu}(\hat k,B),\cr
&\hskip 3cm \text{with}\quad L(a,s_3,k_3)=\int_{-\infty}^{+\infty}
\frac{dp_3}{2\pi}  \frac{dr_3}{2\pi}
\;\frac{\sin a(k_3+p_3-r_3)}{k_3+p_3-r_3}
\; \frac{\sin a(r_3-p_3-s_3)}{r_3-p_3-s_3}.
\end{split}
\end{equation}

Going from the variables $r_3,p_3$ to the variables $p_3, h_3=r_3-p_3$
leads to
\begin{equation}
L(a,s_3,k_3)= \int_{-\infty}^{+\infty} \frac{dp_3}{2\pi}\;
K(a,s_3,k_3),\quad
\text{with}\quad K(a,s_3,k_3)=
\int_{-\infty}^{+\infty} \frac{dh_3}{2\pi}\;\frac{\sin
a(k_3-h_3)}{k_3-h_3}\; \frac{\sin a(h_3-s_3)}{h_3-s_3},
\end{equation}
and the photon propagator at 1-loop writes
\begin{equation}
\begin{split}
\hskip -1cm
\Delta^{\rho\sigma}(a,x,y) &= 4i
\int_{-\infty}^{+\infty} \frac{d^3\hat k}{(2\pi)^3}\,e^{i\hat k(\hat y-\hat
x)}
\int_{-\infty}^{+\infty} \frac{ds_3}{2\pi}
\int_{-\infty}^{+\infty} \frac{dk_3}{2\pi} \;
e^{i(s_3y_3-k_3x_3)}\,
\Delta^{\rho\mu}(\hat k, k_3)\;K(a,s_3,k_3)\;
\Delta^{\nu\sigma}(\hat k, s_3) \;\mu\,
T_{\mu\nu}(\hat k,B),\cr
& \hskip 2cm \text{in which}\quad \mu \equiv \int_{-\infty}^{+\infty}
\frac{dp_3}{2\pi}\quad\text{factors out}.
\end{split}
\label{eq:prop1}
\end{equation}
Last, going to the variable $l_3=s_3-k_3$
(difference of the momentum along $z$ of the incoming and outgoing photon),
one gets
\begin{equation}
K(a,s_3,k_3)\equiv \tilde K(a,l_3)
 = \frac12 \frac{\sin a(s_3-k_3)}{s_3-k_3}
= \frac12 \frac{\sin al_3}{l_3}.
\label{eq:Kexp}
\end{equation}
To define the  vacuum polarization $\Pi_{\mu\nu}$ from
(\ref{eq:prop1}) and (\ref{eq:Kexp}) we proceed like with (\ref{eq:stand})
in
standard QFT by chopping the  two external photon propagators
$\Delta^{\rho\mu}(k)\equiv\Delta^{\rho\mu}(\hat k,k_3)$ and
$\Delta^{\nu\sigma}(k)\equiv\Delta^{\nu\sigma}(\hat k,k_3)$ off
$\Delta^{\rho\sigma}$. The mismatch
between $\Delta^{\nu\sigma}(\hat k,k_3)$ and $\Delta^{\nu\sigma}(\hat
k,s_3\equiv k_3+l_3)$ which occurs in (\ref{eq:prop1}) has to be accounted
for by writing symbolically (see subsection \ref{subsub:Feyn} for the
explicit interpretation)
$ \Delta^{\nu\sigma}(\hat k,k_3+l_3)= \Delta^{\nu\sigma}(\hat k,
k_3)[\Delta^{\nu\sigma}(\hat k, k_3)]^{-1}
\Delta^{\nu\sigma}(\hat k,k_3+l_3)$.
I therefore rewrite the photon propagator (\ref{eq:prop1}) as
\begin{equation}
\begin{split}
\Delta^{\rho\sigma}(a,x,y) &=4i\mu
\int_{-\infty}^{+\infty} \frac{d^4 k}{(2\pi)^4}\,e^{i k(y- x)}
\Delta^{\rho\mu}(k)\; \Delta^{\nu\sigma}(k)\cr
& \left[\int_{-\infty}^{+\infty} \frac{dl_3}{2\pi}  \;
e^{il_3 y_3}\, \tilde K(a,l_3)\;
[\Delta^{\nu\sigma}(\hat k, k_3)]^{-1}\Delta^{\nu\sigma}(\hat k, k_3+l_3)
\right] \; T_{\mu\nu}(\hat k,B).
\label{eq:prop2}
\end{split}
\end{equation}
Cutting off $\Delta^{\rho \mu}$ and  $\Delta^{\nu\sigma}$
leads then to the vacuum polarization $\Pi_{\mu\nu}$:
\begin{equation}
\Pi_{\mu\nu}(\hat k, k_3, \frac{y_3}{a}, B)=
-4\mu\int_{-\infty}^{+\infty} \frac{dl_3}{2\pi}  \;
e^{il_3 y_3}\, \tilde K(a,l_3)\;
[\Delta^{\nu\sigma}(\hat k, k_3)]^{-1}\Delta^{\nu\sigma}(\hat k, k_3+l_3)
 \; T_{\mu\nu}(\hat k,B).
\label{eq:Pieff}
\end{equation}
The factor $\mu$, defined in (\ref{eq:prop1}),
 associated with the electron loop-momentum along $z$,
is potentially ultraviolet  divergent and needs to be regularized.
In relation with the ``confinement'' along $z$ of the $\gamma\, e^+ e^-$
vertices, we shall consider that the electron momentum $p_3$ undergoes
quantum fluctuations
\begin{equation}
\Delta p_3\in[-\frac{\hbar}{a},+\frac{\hbar}{a}],
\end{equation}
which saturate the Heisenberg uncertainty relation $\Delta x\; \Delta p \geq
\hbar$
\footnote{Since many photons and electrons are concerned, the system is
presumably gaussian, in which case one indeed expects the uncertainty
relation to be saturated.}
. The quantum ``uncertainty'' on the momentum of electrons  is
therefore, as expected, inversely proportional to their localization in space 
(at the vertices of their creation or annihilation); it goes to $\infty$
when $a \to 0$ and vice-versa. 

This amounts to taking
\begin{equation}
 p_3^m=\frac{\hbar}{a}
\label{eq:p3mdef}
\end{equation}
as an ultraviolet cutoff for the quantum electron momentum along $z$. Then
\begin{equation}
\mu \approx \frac{1}{2\pi}\;\frac{2\hbar}{a} = \frac{\hbar}{a\pi}.
\label{eq:muval}
\end{equation}
One gets accordingly, using also the explicit expression (\ref{eq:Kexp})
for $\tilde K(a,k_3)$, the following expression for the unrenormalized
$\Pi_{\mu\nu}$ (that we shall call  $\Pi_{\mu\nu}^{bare}$ in section
\ref{section:T})
\begin{equation}
\begin{split}
\Pi^{\mu\nu}(\hat k, k_3, \frac{y_3}{a}, B)
 &= -\frac{1}{\pi^2}\;T^{\mu\nu}(\hat k,B)
\times  U(\hat k,k_3,\frac{y_3}{a}),\cr
\text{with}\quad U(\hat k, k_3, \frac{y_3}{a}) &= \int_{-\infty}^{+\infty}
dl_3  \;
e^{il_3 y_3}\, \frac{\sin al_3}{al_3}\;
[\Delta^{\nu\sigma}(\hat k, k_3)]^{-1}\Delta^{\nu\sigma}(\hat k,
k_3+l_3),\cr
\text{and}\quad T_{\mu\nu}(\hat k,B) &= -ie^2\int_{-\infty}^{+\infty}\frac{d^3\hat
p}{(2\pi)^3}\;Tr[\gamma_\mu G(\hat p,B) \gamma_\nu G(\hat p+\hat k,B)] ,
\end{split}
\label{eq:Pieff2}
\end{equation}
in which 
 $T^{\mu\nu}(\hat k,B)$ can be taken out of the integral because it does
not depend on $k_3$.
This is the announced result, that  exhibits the
transmittance function $U(\hat k,k_3,\frac{y_3}{a})$,
independent of $B$.\newline
* At the limit $a\to \infty$,  the position for  creation
and annihilation of electrons gets an infinite uncertainty but
quantum fluctuations of their momentum in the direction of $B$ shrink to
zero.
Despite the apparent vanishing of $\mu$ at this limit obtained from
(\ref{eq:muval}), the calculation remains meaningful. Indeed,
 the function $\displaystyle\frac{\sin al_3}{al_3}$ goes then to
$\delta(l_3)$, which corresponds to the conservation of the photon momentum
along $z$ (the non-conservation of the photon momentum is thus seen to be
directly related to the quantum fluctuations of the electron momentum).
 This limit also corresponds to ``standard'' QFT, in which  $\hat
K(x)=\delta(x) \Rightarrow L(a,s_3,k_3)= \displaystyle\int_{-\infty}^{+\infty}
\displaystyle\frac{dp_3}{2\pi}\,
\displaystyle\frac{dr_3}{2\pi}\;\delta(k_3+p_3-r_3)\;\delta(r_3-p_3-s_3)
=\int \displaystyle\frac{dp_3}{2\pi}\; \delta(k_3-s_3)$.

* For $a<\infty$, momentum conservation along $z$  is only approximate:
then, the photon can exchange
momentum along $z$ with the quantum fluctuations of the electron momentum.
In general, the $\displaystyle\frac{\sin a l_3}{al_3}$ occurring in $U$
provides for photons, by its fast decrease, the same cutoff $|l_3|
 \equiv |s_3-k_3| \leq \displaystyle\frac{\hbar}{a}=p_3^m$ along $z$ as for
electrons.\newline
* The limit $a\to 0$ would correspond to infinitely thin graphene, infinitely
accurate positioning  of the creation and annihilation of
electrons, but to unbounded quantum fluctuations of their  momentum along
$B$.
Since $\displaystyle\frac{\sin x}{x} \to 1$ when $x\to 0$, no divergence can occur as
$a \to 0$, despite the apparent divergence of $p_3^m$ (\ref{eq:p3mdef}) and $\mu$
(\ref{eq:muval}).

By the choice (\ref{eq:p3mdef}), our model gets therefore suitably
 regularized both in the infrared and in the ultraviolet.

Notice that
the 1-loop photon propagator (\ref{eq:prop1}) still depends on the
difference $\hat y -\hat x$ but no longer depends on $y_3-x_3$ only,
it is now a function of both $y_3$ and $x_3$. Once the dependence on
$y_3-x_3$ has been extracted, there is a left-over dependence on $y_3$. It
is however in practice very weak.

\subsection{The transmittance function $\boldsymbol{U(\hat
k,k_3,\frac{y_3}{a}) = \frac{1-n^2}{a}\; V(n,\theta,\eta,u)}$}

\subsubsection{The Feynman gauge}\label{subsub:Feyn}

We have seen that, when calculating the vacuum polarization
(\ref{eq:Pieff}), the mismatch between  $\Delta^{\nu\sigma}(\hat k,k_3)$
had to be accounted for.
This is most easily done in the Feynman gauge for photons, in which their
propagators write
\begin{equation}
\Delta^{\mu\nu}(k) = -i\;\frac{g^{\mu\nu}}{k^2}.
\end{equation}
The use of a special gauge is certainly abusive, but we take advantage of
the gauge invariance of calculations ``\`a la Schwinger''. Making the same
type of calculations in a general $R_\xi$ gauge would be much more intricate.

Thanks to  the absence of ``$k^\mu k^\nu/k^2$'' terms and as can be easily
checked for each component of $\Delta^{\rho\sigma}$,\break
 $[\Delta^{\nu\sigma}(\hat k, k_3)]^{-1} \Delta^{\nu\sigma}(\hat k,
k_3+l_3)$ can be simply written, then
$\displaystyle\frac{k_0^2-k_1^2-k_2^2 -k_3^2}{k_0^2-k_1^2-k_2^2 -(k_3+l_3)^2}$.
Accordingly, the expression for $U$ resulting from (\ref{eq:Pieff2})
 that we shall use from now onwards is
\begin{equation}
U(\hat k,k_3,\frac{y_3}{a})=\int_{-\infty}^{+\infty} dl_3  \;
e^{il_3 y_3}\, \frac{\sin al_3}{al_3}\;
\frac{k_0^2-k_1^2-k_2^2 -k_3^2}{k_0^2-k_1^2-k_2^2 -(k_3+l_3)^2}.
\label{eq:Pieff4b}
\end{equation}
The analytical properties and pole structure  of the integrand in the complex
$k_3$ plane
play, like for the transmittance in optics (or electronics), an essential
role.  Because they share many similarities, we have given the same name to $U$.

\subsubsection{Going to dimensionless variables : $\boldsymbol{U(\hat
k,k_3,\frac{y_3}{a}) \to V(n,\theta,\eta,u)}$}\label{subsub:dimless}

Let us go to dimensionless variables. We define ($p_3^m$ is given in
(\ref{eq:p3mdef}))
\begin{equation}
\eta=\frac{k_0}{(c)p_3^m}=\frac{ak_0}{(\hbar c)},\quad
u=\frac{y_3}{a}.
\end{equation}
It is also natural to go to the integration variable
$\sigma=al_3=\displaystyle\frac{l_3}{p_3^m}$, and to make appear the refractive index 
\begin{equation}
n= \frac{(c)|\vec k|}{k_0}.
\label{eq:ndef}
\end{equation}
and the angle of incidence $\theta$ according to
\begin{equation}
k_2=0,\quad
k_1= |\vec k| s_\theta = n k_0 s_\theta,\quad
k_3= |\vec k| c_\theta = n k_0 c_\theta,\quad \theta \in\; ]0,\frac{\pi}{2}[.
\end{equation}
This leads to
\begin{equation}
U(\hat k,k_3,\frac{y_3}{a})=\frac{1-n^2}{a} V(n,\theta,\eta,u),\quad
V(n,\theta,\eta,u)  =\int_{-\infty}^{+\infty}
d\sigma\; e^{i\sigma u}\;
\frac{\sin\sigma}{\sigma}\frac{1}{1-n^2-\frac{\sigma}{\eta}
(2n\cos\theta+\frac{\sigma}{\eta})},
\label{eq:UVdef}
\end{equation}
and, therefore, to
\begin{equation}
\Pi^{\mu\nu}(\hat k,k_3,\frac{y_3}{a},B) = -\frac{1}{\pi^2}\;T^{\mu\nu}(\hat
k,B)
\frac{1-n^2}{a}\times  V(n,\theta,\eta,u).
\label{eq:PiV}
\end{equation}
I shall also call $V$ the transmittance function.

As  already deduced in subsection \ref{subsub:graph} from the smooth
behavior of the cardinal sine in the expression (\ref{eq:Pieff2}) of $U$,
the apparent divergence of (\ref{eq:PiV}) at $a\to 0$ is fake;
this can be checked by expanding  $V$ at small $\eta \equiv
ak_0$.
The expansions always start at ${\cal O}(\eta^{\geq 1})$ (see
for example (\ref{eq:expV1})),
 which cancels the $1/a$ in (\ref{eq:PiV}).

Notice that the dependence of $\Pi_{\mu\nu}$ on $k_3$ only occurs inside
the transmittance $V$.

\subsubsection{Analytical expression of the transmittance $\boldsymbol V$}

$V$ as given by (\ref{eq:UVdef})
 is the Fourier transform of the function $x \mapsto
-\displaystyle\frac{\sin x}{x}\,
 \displaystyle\frac{\eta^2}{(x-\sigma_1)(x-\sigma_2)}$, in which
\begin{equation}
\sigma_1 = -\eta \left(n c_\theta - \sqrt{1-n^2 s_\theta^2}\right),\quad
\sigma_2 = -\eta \left(n c_\theta + \sqrt{1-n^2 s_\theta^2}\right)
\label{eq:poles}
\end{equation}
are the poles of the integrand.
The Fourier transform of such a product of a cardinal sine with a
rational function is well known.
The result involves Heavyside functions of the imaginary parts of the
poles $\sigma_1, \sigma_2$, noted $\Theta_i^+$ for $\Theta_i
(\Im(\sigma_i))$ and $\Theta_i^-$ for $\Theta_i (-\Im(\sigma_i))$.
\begin{equation}
\hskip -.5cm V(n, \theta, \eta,u)= \frac{-\pi \eta^2}{\sigma_1 \sigma_2
(\sigma_1 -
\sigma_2)}
\left[ (\sigma_1-\sigma_2)
+ \sigma_2 \big(\Theta_1^- e^{-i \sigma_1 (1-u)}+\Theta_1^+
e^{+i \sigma_1 (1+u)}\big)
-\sigma_1 \big(\Theta_2^- e^{-i \sigma_2
(1-u)}+\Theta_2^+ e^{+i \sigma_2 (1+u)}\big) \right].
\label{eq:transmit}
\end{equation}
$\sigma_1, \sigma_2$
 are seen to control the behavior of $V$, thus  of $n$,
which depends  on the signs of their imaginary parts.

(\ref{eq:transmit}) can also rewrite
\begin{equation}
\begin{split}
& \frac{1-n^2}{\pi}V(n,\theta,\eta,u)=\cr
&\hskip 5mm 1
+\frac{-\big(nc_\theta+\sqrt{1-n^2s_\theta^2}\big)\big(\Theta_1^-e^{-i\sigma_1(1-u)}
+\Theta_1^+e^{i\sigma_1(1+u)}\big)
+\big(nc_\theta-\sqrt{1-n^2s_\theta^2}\big)
\big(\Theta_2^-e^{-i\sigma_2(1-u)}+\Theta_2^+e^{i\sigma_2(1+u)}\big)}
{2\sqrt{1-n^2s_\theta^2}},
\end{split}
\label{eq:V2}
\end{equation}
That the Fourier transform  is well defined
needs in particular that they do not vanish.
This requires either  $n \not\in {\mathbb R}$ or $n \in {\mathbb R}$ and $n s_\theta >1$.

When  $\sigma_1$ and $\sigma_2$ are real, which occurs
for $n\in {\mathbb R}$ and $ns_\theta <1$, the simplest procedure is to
define everywhere in (\ref{eq:transmit}) $\Theta(0)=1/2$.
One gets then
\begin{equation}
\frac{1-n^2}{\pi}\;V(n,\theta, \eta,u)\stackrel{\sigma_1,\sigma_2 \in{\mathbb R}}{=}
1+\frac{\sigma_2\,\cos\sigma_1\;e^{iu\sigma_1}
-\sigma_1\,
\cos\sigma_2\;e^{iu\sigma_2}}{2\eta\sqrt{1-n^2s_\theta^2}}.
\label{eq:Vreal}
\end{equation}

\subsubsection{An important property of $\boldsymbol V$}

From (\ref{eq:V2})  one can deduce
\begin{equation}
\frac{1-n^2}{\pi}\;V(k^2=0)=0.
\label{eq:Vprop}
\end{equation}
Indeed, since we are working in a frame in which $k_2=0$, $k^2=0
\Leftrightarrow k_0^2-k_1^2=k_3^2$. From the definition of $n$ and
$\theta$, $nc_\theta = \displaystyle\frac{k_3}{k_0},
ns_\theta=\displaystyle\frac{k_1}{k_0}$, which
entails $\sigma_1 = 0$ and $\sigma_2 =-2\eta k_3$. Both being real entails
in particular that the arguments of all $\Theta$'s in (\ref{eq:V2}) are
vanishing, such that they all should consistently be taken to $1/2$.
This yields accordingly
$\displaystyle\frac{1-n^2}{\pi}\;V(k^2=0)=
1+\displaystyle\frac{-(2k_3)(\displaystyle\frac12 +\displaystyle\frac12) +0\times(\ldots)}{2k_3}=0$. 
Since $1-n^2= 1-\displaystyle\frac{|\vec k|^2}{k_0^2}=-\displaystyle\frac{k^2}{k_0^2}$
trivially vanishes at $k^2=0$, the important property is that $V$ is not
singular at this limit.

The property (\ref{eq:Vprop}) will prove
essential for the renormalization of $\Pi_{\mu\nu}$ (see section
\ref{section:renorm1}).

\subsubsection{Expansions of $\boldsymbol V$ at $\boldsymbol{\eta \equiv
ak_0 \to 0}$}

$\bullet$\ For $n\in {\mathbb R}$ and $ns_\theta >1$, the expansion of $V$
at $\eta \sim ak_0 \ll 1$ writes

\begin{equation}
\begin{split}
\Re(V) &= -\frac{\pi}{\sqrt{n^2 s_\theta^2 -1}}\,\eta +\frac{\pi}{2}
(1+u^2)\,\eta^2 +{\cal O}(\eta^3),\cr
\Im(V) &= u\,n\,c_\theta\,\frac{\pi}{\sqrt{n^2 s_\theta^2 -1}}\,\eta^2
+{\cal O}(\eta^3).
\end{split}
\label{eq:expV1}
\end{equation}
This is equivalent to
\begin{equation}
\frac{1-n^2}{\pi}\;\frac{V}{a} \stackrel{a\to 0}{\simeq}-\frac{k_0^2 -|\vec
k|^2}{\sqrt{s_\theta^2 |\vec k|^2 -k_0^2}} + {\cal O}(a)
\label{eq:limcomp}
\end{equation}
which does not vanish when $a \to 0$.

$\bullet$\ For $n\in {\mathbb R}$ and
$ns_\theta <1 \Leftrightarrow \sigma_1, \sigma_2 \in{\mathbb R}$,
expanding in powers of $\eta=ak_0 \ll 1$ yields
\begin{equation}
\frac{1-n^2}{\pi}\;\frac{V}{a} \stackrel{a\to 0}{\simeq} \frac12
(1-n^2)(1+u^2) \frac{\eta^2}{a} + {\cal O}(\eta^3)
= \frac12 (1+u^2) (k_0^2 -|\vec k|^2)\; a + \ldots
\label{eq:limreal}
\end{equation}
which vanishes when $a \to 0$.

\subsection{The limit $\boldsymbol{k_0 \to 0}$}

This limit is necessary for studying the scalar potential (see section
\ref{section:scalpot}).

Expanding (\ref{eq:V2}) at $k_0 \to 0$ yields
(we use the notation $\csc\theta =1/\sin\theta$)
\begin{equation}
\begin{split}
& \frac{1-n^2}{\pi}V  \simeq \frac{e^{-a |\vec k| (u+1) (\sin
\theta+i \cos \theta)}}{2}
\left(-1+i \cot\theta +2 e^{a |\vec k| (u+1) (\sin \theta+i \cos \theta)}
-i(-i +\cot\theta)  e^{2 a |\vec k| (u \sin \theta+i \cos \theta)}\right)\cr
&\hskip -1.5cm
 +\frac{\csc \theta e^{-a |\vec k| (u+1) (\sin \theta+i \cos \theta)}
\Big(i (\cot \theta \csc\theta
+a |\vec k| (u+1) (\cot \theta+i))+e^{2 a |\vec k| (u \sin \theta+i \cos
\theta)} (a |\vec k| (u-1) (1+i \cot \theta)-i \cot \theta
\csc \theta)\Big)}{4 |\vec k|^2}k_0^2\cr
& + {\cal O}(k_0^3)
\end{split}
\label{eq:Vk0}
\end{equation}

On the expression above one can in particular confirm that no divergence at
$a \to 0$ occurs for $\frac{1-n^2}{\pi}\frac{V}{a}$:

\begin{equation}
\frac{1-n^2}{\pi}\frac{V}{a} \stackrel{k_0\to 0,a\to 0}{\simeq}
\frac{|\vec k|}{\sin \theta}
+\frac{\cos 2 \theta (\csc \theta)^3}{2 |\vec k|}k_0^2
+ {\cal O}(a),
\label{eq:Vlim2}
\end{equation}
which is the same result as from (\ref{eq:limcomp}) for $n\in {\mathbb
R}$ and $ns_\theta >1$, in which case $\sigma_1$ and $\sigma_2$ are
complex.

For $\sigma_1$ and $\sigma_2$ real $n\in {\mathbb R}$ and $ns_\theta < 1$),
 we have found in (\ref{eq:limreal}) that
$\frac{1-n^2}{\pi}\frac{V}{a}$ vanishes at $a\to 0$. However,
$ns_\theta <1 \Leftrightarrow s_\theta < k_0/|\vec k| \stackrel{k_0\to
0}{\simeq} 0$ such that, in practice, except at $\theta=0$, we can expect a
deviation from Coulomb of the scalar potential when $a \to 0$.

\section{Calculation of the unrenormalized $\boldsymbol{T_{\mu\nu}^{bare}}$}
\label{section:T}

I shall now calculate $T_{\mu\nu}^{bare}$ obtained in (\ref{eq:Pieff2}). To ease the
parallel with \cite{DittrichReuter} we shall switch $k$ to $-k$ and
calculate hereafter
\begin{equation}
T_{\mu\nu}^{bare}(\hat k,B) = -ie^2\int_{-\infty}^{+\infty}\frac{d^3\hat
p}{(2\pi)^3}\;Tr\big[\gamma_\mu G(\hat p,B) \gamma_\nu G(\hat p-\hat
k,B)\big],
\end{equation}
which is similar to eq.~(4.1) of \cite{DittrichReuter}.

The counterterms, which have to be evaluated for $\Pi_{\mu\nu}$, will be
dealt with in section \ref{section:renorm1}.

\subsection{First steps}
\label{subsec:firststeps}

The electron propagator in external $B$ inside the graphene-like medium writes
(see (2.47b) of \cite{DittrichReuter}) in momentum space
\footnote{As already noted in \cite{Machet2016-1}, the correct expression is
that of Tsai (eq.~(6) in \cite{Tsai1974})
\begin{equation}
G(k,B)=i\int_0^\infty ds_1\;e^{\displaystyle -is_1\big(m^2-i\epsilon
+k_\parallel^2
+\frac{\tan z}{z} k_\perp^2 \big)}
\; \frac{e^{\displaystyle i qz\sigma^3}}{\cos z}\Big(m-k\!\!\!/_\parallel
-\frac{e^{\displaystyle -iqz\sigma^3}}{\cos z} k\!\!\!/_\perp \Big),
\quad z=|e|Bs_1,
\label{eq:ferprop}
\end{equation}
in which $q=-1$ and $s_1$ be the Schwinger
parameter associated to the internal electron propagator. It can be
obtained from (\ref{eq:eprop}) by $z \to -z$, which is equivalent to considering,
there, $e=q|e|<0$, such that $z = -|e|Bs$. I shall work with the
conventions of \cite{DittrichReuter} despite their contradiction, which we
checked to have  no far reaching consequence here.}
\begin{equation}
\begin{split}
& \hskip -2cm
G(\hat p, B) =i\int_0^\infty ds\; e^{\displaystyle -is\Big[m^2-i\epsilon+(-p_0^2 +
\xcancel{p_3^2})+\frac{\tan
z}{z}(p_1^2+p_2^2)\Big]}\;\frac{e^{\displaystyle iz\sigma^3}}{\cos z}
\Big(m-(-\gamma_0 p_0 +\xcancel{\gamma_3 p_3})-\frac{e^{\displaystyle -iz\sigma^3}}
{\cos z}(\gamma_1 p_1 + \gamma_2 p_2) \Big),\cr
\text{with}\ z &= eBs.
\end{split}
\label{eq:eprop}
\end{equation}
in which any dependence on $p_3$ is set to $0$.
As already mentioned, as far as the vacuum polarization is concerned
one can forget about the $\Phi$ phases (\ref{eq:phase}).

To calculate $T_{\mu\nu}^{bare}$  we must  redo the calculations of p.~56-72 of
\cite{DittrichReuter}, adapting them to the situation under scrutiny. 
I shall emphasize the steps that differ.

Introducing the two Schwinger parameters $s_1$ for $G(\hat p)$ and $s_2$ for
$G(\hat p-\hat k)$ yields the equivalent of (4.5) of \cite{DittrichReuter}
\begin{equation}
\begin{split}
& \hskip -1cm 
T_{\mu\nu}^{bare}(\hat k, B) =ie^2 \int_0^\infty ds_1\int_0^\infty ds_2\;<
e^{\displaystyle\Big[-is_1\big(m^2+\hat p_\parallel^2 +\frac{\tan z_1}{z_1}p_\perp^2 \big)
-is_2 \big(m^2+(\hat p-\hat k)_\parallel^2 +\frac{\tan z_2}{z_2}(p-k)_\perp^2\big)
\Big]}\;\frac{1}{\cos z_1 \cos z_2}\cr
& Tr\ \Big[\gamma_\mu\Big(\big(m-(\gamma
p)_\parallel\big)\;e^{\displaystyle iz_1\sigma^3}
-\frac{(\gamma \hat p)_\perp}{\cos z_1}\Big)
\gamma_\nu \Big(\big(m-(\gamma(\hat p-\hat
k))_\parallel\big)\;e^{\displaystyle iz_2\sigma^3}
-\frac{(\gamma(p-k))_\perp}{\cos z_2}\Big)
\Big] >,
\end{split}
\label{eq:T1}
\end{equation}
in which one has now $\hat p_\parallel^2 =-p_0^2,\;
p_\perp^2=p_1^2+p_2^2,\; z_1=eBs_1,\; z_2=eBs_2$ and, in  (4.3)
of \cite{DittrichReuter}, $\displaystyle\int \frac{d^4p}{(2\pi)^4}$ must be replaced by
$\displaystyle\int \frac{d^3p}{(2\pi)^3}$, such that the notation $<\quad>$
stands now for
\begin{equation}
<f(\hat p)> = \int \frac{d^3 \hat p}{(2\pi)^3}\;f(\hat p).
\end{equation}

One makes the change of variables $(z_1,z_2) \to (s,v)$ such that
\begin{equation}
\begin{split}
& z_1 = eBs_1 = eBs\;\frac{1-v}{2}=z\;\frac{1-v}{2} = \xi,\cr
& z_2 = eBs_2 = eBs\;\frac{1+v}{2}=z\;\frac{1+v}{2} = \eta,
\end{split}
\end{equation}
that is
\begin{equation}
z=\xi+\eta \Leftrightarrow s=s_1+s_2,\quad
v=\frac{\eta-\xi}{\eta+\xi}=\frac{s_2-s_1}{s_2+s_1},
\end{equation}
and one has
\begin{equation}
\int_0^\infty ds_1 \int_0^\infty ds_2 = \int_0^\infty
s\;ds\int_{-1}^{+1} \frac{dv}{2}.
\end{equation}

$\bullet$\quad A few steps are necessary to demonstrate (4.9), (4.10), (4.11) of
\cite{DittrichReuter}, so as to rewrite the exponential function in (\ref{eq:T1}).

*\ $s_1 p_\parallel^2 +s_2 (p-k)_\parallel^2 =
s\left(p_\parallel-\displaystyle\frac{1+v}{2}k_\parallel\right)^2 +
s\displaystyle\frac{1-v^2}{4}
k_\parallel^2$;

*\ $s_1\displaystyle\frac{\tan \xi}{\xi}p_\perp^2 +s_2 \displaystyle\frac{\tan\eta}{\eta}(p-k)_\perp^2
=\displaystyle\frac{1}{eB}\left[\tan\xi p_\perp^2 + \tan\eta(p-k)_\perp^2\right]
=\displaystyle\frac{\tan\xi+\tan\eta}{eB}\left(p_\perp-\displaystyle\frac{\tan\eta}{\tan\xi+\tan\eta}k_\perp\right)^2
+\displaystyle\frac{k_\perp^2}{eB}\displaystyle\frac{\tan\xi\tan\eta}{\tan\xi+\tan\eta}$;

*\ $\tan\xi\tan\eta= \displaystyle\frac{\cos(\xi-\eta)-\cos(\xi+\eta)}
{\cos(\xi-\eta)+\cos(\xi+\eta)}=\displaystyle\frac{\cos zv -\cos z}{\cos zv
+ \cos z}$;

*\ $\tan\xi + \tan \eta = \tan(\xi+\eta)(1-\tan\xi\tan\eta =\tan
z(1-\tan\xi\tan\eta))$;

*\ $\displaystyle\frac{\tan\xi\tan\eta}{\tan\xi+\tan\eta}=
\displaystyle\frac{cos zv -\cos z}{2\sin z}$;

*\ $\exp\Big[\displaystyle \left(-is_1(m^2+p_\parallel^2)-is_2(m^2+(p-k)_\parallel^2)
-is_1(\tan\xi/\xi)p_\perp^2
-is_2(\tan\eta/\eta)(p-k)_\parallel^2\right)\Big]\newline
=\exp\Big[\displaystyle -is\big(m^2+\frac{1-v^2}{4}k_\parallel^2
+(p_\parallel-\frac{1+v}{2}k_\parallel)^2\big)\Big]
\times
\exp\Big[\displaystyle\big(-i\frac{\tan\xi+\tan\eta}{eB}
\left(p_\perp-\frac{\tan\eta}{\tan\xi+\tan\eta}k_\perp^2
\right)^2 -i \frac{k_\perp^2}{eB}\frac{\cos zv-\cos z}{2\sin
z}\big)\Big]\break
=e^{\displaystyle -is(\varphi_0+\varphi_1)}$,\newline
with $\varphi_0=m^2+\displaystyle\frac{1-v^2}{4}k_\parallel^2
+\displaystyle\frac{\cos zv-\cos z}{2z\sin
z}k_\perp^2,\quad
\varphi_1 = \left(p_\parallel-\displaystyle\frac{1+v}{2}k_\parallel\right)^2
+\displaystyle\frac{\tan\xi+\tan\eta}{z}
\left(p_\perp-\displaystyle\frac{\tan\eta}{\tan\xi+\tan\eta}k_\perp\right)^2$.

In our case, $p_3$ and $p_3-k_3$ have to be set formally to $0$ such that
eqs.~(4.10) and (4.11) of \cite{DittrichReuter} are replaced by
\begin{equation}
\begin{split}
& \varphi_0 = m^2-\frac{1-v^2}{4}\;k_0^2
+ \frac{\cos zv-\cos z}{2z\sin z}\;k_\perp^2,\cr
& \varphi_1 = -\left(p_0-\frac{1+v}{2}k_0\right)^2
+\frac{\tan\xi+\tan\eta}{z}
\left(p_\perp-\frac{\tan\eta}{\tan\xi+\tan\eta}k_\perp\right)^2.
\end{split}
\label{eq:varphi}
\end{equation}
This leads to the equivalent of eq.~(4.12) of \cite{DittrichReuter}
\begin{equation}
\begin{split}
T_{\mu\nu}^{bare}(\hat k, B) &=ie^2\int_0^\infty ds\; s\int_{-1}^{+1}\frac{dv}{2}\;
e^{\displaystyle -is\varphi_0}\;\frac{1}{\cos\xi\cos\eta}\cr
& Tr<e^{\displaystyle -is\varphi_1}\;\left\{
\gamma_\mu\left[(m-\gamma\hat
p_\parallel)e^{\displaystyle i\xi\sigma^3}\;-\frac{\gamma p_\perp}{\cos\xi}\right]
\gamma_\nu
\left[(m-\gamma(\hat p-\hat k)_\parallel) e^{\displaystyle i\eta\sigma^3}\;
-\frac{\gamma(p-k)_\perp}{\cos\eta}\right]\right\}> .
\end{split}
\label{eq:T2}
\end{equation}

$\bullet$\quad One now eliminates $\cos\xi \cos\eta$ in terms of
$<e^{-is\varphi_1}>$ in (\ref{eq:T2})

\begin{equation}
<e^{-is\varphi_1}>=\int\frac{dp_0 dp_1 dp_2}{(2\pi)^3}\;\exp\left[
-is\left(-\left(p_0-\frac{1+v}{2}k_0\right)^2
+\frac{\tan\xi+\tan\eta}{z}
\left(p_\perp-\frac{\tan\eta}{\tan\xi+\tan\eta}k_\perp\right)^2\right)\right].
\label{eq:elim1}
\end{equation}

One can freely shift the integration variables.

$\displaystyle\int_{-\infty}^{+\infty}dx\; e^{\displaystyle -\pm i A x^2} =
e^{\displaystyle \pm
i\pi/4}\sqrt{\displaystyle\frac{\pi}{A}},\ A>0$, therefore, inside
(\ref{eq:elim1}):

*\  $\displaystyle\int\frac{dp_0}{2\pi}$ gives
$\displaystyle\frac{1}{2\pi}\;e^{\displaystyle +i\pi/4}\;
\sqrt{\displaystyle \frac{\pi}{s}}$.\newline
*\  $\displaystyle \int\frac{dp_1 dp_2}{(2\pi)^2}$ gives
$\displaystyle \frac{1}{(2\pi)^2}\Big( e^{\displaystyle
-i\pi/4}\;\sqrt{\displaystyle \frac{\pi
z}{s(\tan\xi+\tan\eta)}}\Big)^2$, \newline
and
$ <e^{\displaystyle -is\varphi_1}>= \displaystyle
\frac{1}{(2\pi)^3}\;e^{\displaystyle -i\pi/4}\;z
\left(\displaystyle \frac{\pi}{s}\right)^{3/2}\displaystyle \frac{1}{\tan\xi+\tan\eta}$.

One then uses $\tan\xi+\tan\eta = \displaystyle \frac{\sin
z}{\cos\xi\cos\eta}$ to get
\begin{equation}
<e^{\displaystyle -is\varphi_1}>=
\displaystyle \frac{1}{(2\pi)^3}\;e^{\displaystyle
-i\pi/4}\;\left(\displaystyle \frac{\pi}{s}\right)^{3/2}
\displaystyle \frac{z}{\sin z}\;\cos\xi\cos\eta.
\label{eq:elim2}
\end{equation}
In $T_{\mu\nu}$, one can therefore replace $\displaystyle \frac{1}{\cos\xi\cos\eta}$ with
$\displaystyle \frac{1}{(2\pi)^3}\;e^{\displaystyle
-i\pi/4}\;\left(\displaystyle \frac{\pi}{s}\right)^{3/2}
\displaystyle \frac{z}{\sin z}\;\displaystyle \frac{1}{<e^{-is\varphi_1}>}$
and get
\begin{equation}
\begin{split}
T_{\mu\nu}^{bare}(\hat k, B) &= i\frac{\alpha}{2\pi}\sqrt{\pi}e^{\displaystyle -i\pi/4}
\int_0^\infty \frac{ds}{\sqrt{s}} \int_{-1}^{+1}\frac{dv}{2}\;
e^{\displaystyle -is\varphi_0}\;\frac{z}{\sin
z}\;\frac{1}{<e^{\displaystyle -is\varphi_1}>}\cr
& \hskip 2cm Tr<e^{\displaystyle -is\varphi_1}\;\left[
\gamma_\mu\Big((m-\gamma\hat
p_\parallel)e^{\displaystyle i\xi\sigma^3}\;-\frac{\gamma p_\perp}{\cos\xi}\Big)
\gamma_\nu
\Big((m-\gamma(\hat p-\hat k)_\parallel) e^{\displaystyle i\eta\sigma^3}\;
-\displaystyle \frac{\gamma(p-k)_\perp}{\cos\eta}\Big)\right]>, \cr
&= \frac{\alpha}{2\pi}\int_0^\infty \frac{ds}{\sqrt{s}}
\int_{-1}^{+1}\frac{dv}{2}\;e^{\displaystyle -is\varphi_0}\;\frac{z}{\sin
z}\;I_{\mu\nu},
\end{split}
\end{equation}
with
\begin{equation}
\begin{split}
&\hskip -1cm I_{\mu\nu}= i\sqrt{\pi}\;e^{\displaystyle
-i\pi/4}\;\displaystyle \frac{1}{<e^{\displaystyle -is\varphi_1}>}
 Tr<e^{\displaystyle -is\varphi_1}\;\left[
\gamma_\mu\Big((m-\gamma\hat
p_\parallel)e^{\displaystyle i\sigma^3\xi}\;-\displaystyle \frac{\gamma p_\perp}{\cos\xi}
\Big)
\gamma_\nu
\Big((m-\gamma(\hat p-\hat k)_\parallel) e^{\displaystyle i\sigma^3 \eta}\;
-\displaystyle \frac{\gamma(p-k)_\perp}{\cos\eta}\Big)\right]> \cr
& \hskip -1cm = i\sqrt{\pi}\;e^{\displaystyle
-i\pi/4}\;\frac{1}{<e^{\displaystyle -is\varphi_1}>}
 Tr<e^{\displaystyle -is\varphi_1}\;\left[
\gamma_\mu\Big((m+\gamma_0 p_0)e^{\displaystyle i\sigma^3\xi}\;
-\frac{\gamma p_\perp}{\cos\xi}\Big)
\gamma_\nu
\Big((m+\gamma_0(\hat p-\hat k)_0) e^{\displaystyle i\sigma^3 \eta}\;
-\frac{\gamma(p-k)_\perp}{\cos\eta}\Big)\right]>.
\end{split}
\label{eq:Imunu}
\end{equation}
One needs therefore
$<e^{\displaystyle -is\varphi_1}p_0>, <e^{\displaystyle
-is\varphi_1}p_0^2>, <e^{\displaystyle -is\varphi_1}p_0
p_{1,2}>, <e^{\displaystyle -is\varphi_1}p_{1,2} p_{1,2}>$.

*\ $<e^{\displaystyle -is\varphi_1} \left(p_0-\displaystyle
\frac{1+v}{2}k_0\right)>=0$ because it is an odd
integral $\displaystyle \int dp_0$.\newline
 Therefore, $<e^{\displaystyle -is\varphi_1} p_0>=
\displaystyle \frac{1+v}{2} k_0
<e^{\displaystyle -is\varphi_1}>$;

*\ $<e^{\displaystyle -is\varphi_1}p_0^2>= <e^{\displaystyle
-is\varphi_1}\Big(p_0-\displaystyle \frac{1+v}{2}k_0\Big)^2>
+k_0(1+v)<e^{\displaystyle -is\varphi_1}p_0> -\Big(\displaystyle
\frac{1+v}{2}\Big)^2 k_0^2
<e^{\displaystyle -is\varphi_1}>$.

$ \displaystyle \int\frac{dp_1dp_2}{(2\pi)^2}\; e^{\displaystyle -is \frac{\tan\xi+\tan\eta}{z}
\Big(p_\perp-\frac{\tan\eta}{\tan\xi+\tan\eta}k_\perp \Big)^2}
\displaystyle \int\frac{dp_0}{2\pi}\;e^{\displaystyle
is\Big(p_0-\frac{1+v}{2}k_0\Big)^2}
\Big(p_0-\frac{1+v}{2}k_0\Big)^2 \newline
= \displaystyle \int\frac{dp_1dp_2}{(2\pi)^2}\; e^{\displaystyle -is \frac{\tan\xi+\tan\eta}{z}
\Big(p_\perp-\frac{\tan\eta}{\tan\xi+\tan\eta}k_\perp \Big)^2}
\Big(-i\displaystyle \frac{1}{2\pi}\displaystyle \frac{d}{ds}e^{\displaystyle
i\pi/4}\sqrt{\displaystyle \frac{\pi}{s}}\Big) =
 \displaystyle \int\frac{dp_0dp_1dp_2}{(2\pi)^3}\;
\displaystyle \frac{i}{2s}e^{\displaystyle -is\varphi_1}$.

So,
$<e^{\displaystyle -is\varphi_1}p_0^2>= \displaystyle
\frac{i}{2s}<e^{\displaystyle -is\varphi_1}>
+k_0^2(\frac{1+v}{2})^2 <e^{-is\varphi_1}>$, which agrees with (4.15)
in \cite{DittrichReuter} with $g_{\mu\nu}^\parallel= g_{00}=-1$.

In a similar way one can get the 3 formul{\ae} (4.15) of
\cite{DittrichReuter}.

Following \cite{DittrichReuter} let us write (we keep the natural $(-)$
signs in $S_3$ and $S_4$) 
\begin{equation}
\begin{split}
I_{\mu\nu} &= i\sqrt{\pi}\;e^{\displaystyle -i\pi/4}\;\displaystyle
\frac{1}{<e^{\displaystyle -is\varphi_1}>}
\sum_{i=1,5} Tr S_i,\cr
& \text{with}\cr
Tr S_1 &= m^2 Tr <e^{\displaystyle -is\varphi_1} \gamma_\mu
e^{\displaystyle i\sigma^3 \xi}  \gamma_\nu
e^{\displaystyle i\sigma^3 \eta}>,\cr
Tr S_2 &=  Tr <e^{\displaystyle -is\varphi_1} \gamma_\mu \gamma_0 p_0
e^{\displaystyle i\sigma^3
\xi}  \gamma_\nu \gamma_0(p_0-k_0) e^{\displaystyle i\sigma^3 \eta}>,\cr
Tr S_3 &=  Tr <e^{\displaystyle -is\varphi_1} \gamma_\mu (-)\gamma_0 p_0
e^{\displaystyle i\sigma^3
\xi}  \gamma_\nu \gamma(\hat p -\hat k)_\perp
> \displaystyle \frac{1}{\cos\eta},\cr
Tr S_4 &=  Tr <e^{\displaystyle -is\varphi_1} \gamma_\mu \gamma \hat p_\perp
 \gamma_\nu (-)\gamma_0(p_0-k_0)
e^{\displaystyle i\sigma^3 \eta}> \displaystyle \frac{1}{\cos\xi},\cr
Tr S_5 &=  Tr <e^{\displaystyle -is\varphi_1} \gamma_\mu \gamma \hat p_\perp
 \gamma_\nu \gamma(\hat p-\hat k)_\perp
>\displaystyle  \frac{1}{\cos\xi\cos\eta}.
\end{split}
\end{equation}
Since $\sigma_3^2=1$, $e^{\displaystyle i\sigma^3 \xi}=\cos\xi + i\sigma^3\sin\xi
= \cos\xi -\displaystyle \frac12 [\gamma_1,\gamma_2] \sin \xi=\cos\xi -\gamma_1 \gamma_2
\sin \xi$.

*\ $Tr S_1= 4m^2 <e^{\displaystyle -is\varphi_1}>\big[
-\cos z\; g_{\mu\nu}+(g_{\mu 1}g_{\nu 1} + g_{\mu_2}g_{\nu 2})(\cos z-\cos
zv) + 2(g_{\mu 1}g_{\nu 2}-g_{\mu 2}g_{\nu 1})\sin zv \big]$.

After integrating $\int dv$, the odd function $\sin zv$ yields a vanishing contribution,
such that we can forget it. Furthermore, $g_{\mu 1}g_{\nu 1} +
g_{\mu_2}g_{\nu 2} = g_{\mu\nu}^\perp$ such that, finally

$Tr S_1= -4m^2 <e^{\displaystyle -is\varphi_1}>\left[g_{\mu\nu}^\parallel \cos z
+ g_{\mu\nu}^\perp\cos zv + odd(v)\right]$, which is the result of
\cite{DittrichReuter}.

*\ $Tr S_2$ needs to be re-calculated because the formula
$\gamma_\parallel^\lambda \gamma_{\parallel \lambda}=-2$ used at the end of
page 66 of \cite{DittrichReuter} is no longer valid since $\gamma_\parallel$
stands now for $\gamma_0$ only.

$Tr S_2 = <p_0(p_0-k_0)e^{\displaystyle -is\varphi_1}> \; Tr \gamma_\mu \gamma_0
e^{\displaystyle i\sigma^3
\xi}\gamma_\nu\gamma_0 e^{\displaystyle i\sigma^3 \eta}$ and one uses
(4.15) of \cite{DittrichReuter} for the $<\ >$.

This gives

$Tr S_2 =
4<e^{\displaystyle -is\varphi_1}>\big(-k_0^2\displaystyle
\frac{1-v^2}{4}+\displaystyle \frac{i}{2s}\big)
\Big[(g_{\mu\nu}+2g_{\mu 0}g_{\nu 0}-g_{\mu 1}g_{\nu 1}-g_{\mu 2}g_{\nu 2})
\cos z
+(g_{\mu 1}g_{\nu 1}+g_{\mu 2}g_{\nu 2})\cos zv  \newline
 -(g_{\mu 1}g_{\nu 2}-g_{\mu 2}g_{\nu 1})\sin zv \Big]$.

Since $g_{\mu\nu}$ is diagonal, $g_{\mu 1}g_{\nu 1}+g_{\mu 2}g_{\nu
2}=g_{\mu\nu}^\perp$ such that, dropping like before
the function odd in $v$, one gets

$Tr S_2 =
4<e^{\displaystyle -is\varphi_1}>\big(-k_0^2\displaystyle
\frac{1-v^2}{4}+\displaystyle \frac{i}{2s}\big)
\Big[(g_{\mu\nu}+2g_{\mu 0}g_{\nu 0}-g_{\mu \nu}^\perp)\cos z
+g_{\mu\nu}^\perp\cos zv  +odd(v) \Big]$

$=4<e^{\displaystyle
-is\varphi_1}>\big(-k_0^2\displaystyle \frac{1-v^2}{4}+\displaystyle
\frac{i}{2s}\big)
\Big[(g_{\mu\nu}^\parallel+2g_{\mu 0}g_{\nu 0})\cos z
+g_{\mu\nu}^\perp\cos zv  +odd(v) \Big]$.

A comparison of $Tr S_2$ with the result $(Tr S_2)^{DR}$ in \cite{DittrichReuter} is
due \footnote{The last term in the expression of $C^{\alpha\beta}$ at the
top of p.66 of \cite{DittrichReuter} should be written
$-\frac{i}{2s}\,g_\parallel^{\alpha\beta}$ instead of
$-\frac{i}{s}\,g_\parallel^{\alpha\beta}$.}.
$(Tr S_2)^{DR}= 4<e^{\displaystyle -is\varphi_1}> \Big[
\displaystyle \frac{1-v^2}{4}k_\parallel^2 (g_{\mu\nu}^\parallel \cos z +
g_{\mu\nu}^\perp \cos zv) -\displaystyle \frac{1-v^2}{2}k_\mu^\parallel
k_\nu^\parallel\cos z +\displaystyle \frac{i}{s}g_{\mu\nu}^\perp \cos zv
\Big] + odd(v)$.

One gets
$\displaystyle \frac{Tr S_2}{4<e^{\displaystyle -is\varphi_1}>}=
\displaystyle \frac{(Tr S_2)^{DR}_{k_3=0}}{4<e^{\displaystyle -is\varphi_1}>}
+k_0^2 \displaystyle \frac{1-v^2}{2}\;\cos z \Big(\displaystyle \frac{k_\mu^\parallel
k_\nu^\parallel}{k_0^2}-g_{\mu 0}g_{\nu 0}\Big)
-\displaystyle \frac{i}{2s}\;g_{\mu\nu}^\perp \cos zv $.\newline
 Since $k_\mu^\parallel$ in our
case can only be $k_0$, the second term vanishes such that

$\displaystyle \frac{Tr S_2}{4<e^{\displaystyle -is\varphi_1}>}=
\displaystyle \frac{(Tr S_2)^{DR}_{k_3=0}}{4<e^{\displaystyle -is\varphi_1}>}
+\displaystyle \frac{i}{2s}\underbrace{\big(\cos z(g_{\mu\nu}^\parallel +2g_{\mu 0}g_{\nu 0})
-g_{\mu\nu}^\perp \cos zv\big)}_{B_{\mu\nu}}$.

*\ $Tr S_3= 4<e^{\displaystyle -is\varphi_1}> \displaystyle
\frac{1}{\cos\eta}\displaystyle \frac{1+v}{2}
\displaystyle \frac{\tan\xi}{\tan\xi+\tan\eta}
\Big[k_0k_1\Big((g_{\mu 0}g_{\nu 1}+g_{\mu 1}g_{\nu 0})\cos\xi + (g_{\mu
0}g_{\nu 2}-g_{\mu 2}g_{\nu 0}) \sin\xi\Big)
+k_0k_2\Big((g_{\mu 0}g_{\nu 2}+g_{\mu 2}g_{\nu 0}) \cos\xi -(g_{\mu
0}g_{\nu
1}-g_{\mu 1}g_{\nu 0})\sin\xi\Big)
\Big]$;

It includes the expressions
$\displaystyle \frac{\cos\xi}{\cos\eta}\displaystyle \frac{\tan\xi}{\tan\xi+\tan\eta}=
\displaystyle \frac{\sin \xi \cos\xi}{\sin z}$,\quad
$\displaystyle \frac{\sin\xi}{\cos\eta}\displaystyle \frac{\tan\xi}{\tan\xi+\tan\eta}=
\displaystyle \frac{\sin^2 \xi}{\sin z}$,
which will be replaced accordingly.

*\ $Tr S_4= 4<e^{\displaystyle -is\varphi_1}> \displaystyle
\frac{1}{\cos\xi}\displaystyle \frac{1-v}{2}
\displaystyle \frac{\tan\eta}{\tan\xi+\tan\eta}
\Big[k_0k_1\Big((g_{\mu 0}g_{\nu 1}+g_{\mu 1}g_{\nu 0})\cos\eta - (g_{\mu
0}g_{\nu 2}-g_{\mu 2}g_{\nu 0}) \sin\eta\Big)
+k_0k_2\Big((g_{\mu 0}g_{\nu 2}+g_{\mu 2}g_{\nu 0}) \cos\eta +(g_{\mu
0}g_{\nu
1}-g_{\mu 1}g_{\nu 0})\sin\eta\Big)
\Big]$;

It includes the expressions
$\displaystyle \frac{\cos\eta}{\cos\xi}\displaystyle \frac{\tan\eta}{\tan\xi+\tan\eta}=
\displaystyle \frac{\sin \eta \cos\eta}{\sin z}$,\quad
$\displaystyle \frac{\sin\eta}{\cos\xi}\displaystyle \frac{\tan\eta}{\tan\xi+\tan\eta}=
\displaystyle \frac{\sin^2 \eta}{\sin z}$,
which will be replaced accordingly.

\smallskip

*\ One gets

$Tr S_3 + Tr S_4 = 4 <e^{\displaystyle -is\varphi_1}>
\Big[k_0k_1 (g_{\mu 0}g_{\nu 1} + g_{\mu 1}g_{\nu 0})
+k_0k_2 (g_{\mu 0}g_{\nu 2} + g_{\mu 2}g_{\nu 0})
\Big] \displaystyle \frac{\sin z \cos zv - v \cos z \sin zv}{2\sin z} +
odd(v)$,

that we compare to the result in \cite{DittrichReuter}
$(Tr S_3 + Tr S_4)^{DR}= 4<e^{\displaystyle -is\varphi_1}>\Big[
 -[k_\mu k_\nu -k_\mu^\perp k_\nu^\perp
-k_\mu^\parallel k_\nu^\parallel]
\displaystyle \frac{\sin z \cos zv - v \cos z \sin zv}{2\sin z} + odd(v)\Big]$.
 When one omits $k_3$ in
the formula of \cite{DittrichReuter}, one gets the same expressions for all
the components, which restricts, then, to $(\mu,\nu)=(0,1),(1,0),(0,2),(2,0)$.

*\ $Tr S_5 =4<e^{\displaystyle -is\varphi_1}>\Big[\displaystyle \frac{\cos zv-\cos z}{2\sin^2 z}
(g_{\mu\nu}k_\perp^2 -2 k^\perp_\mu k^\perp_\nu)
+\displaystyle \frac{iz}{s}\displaystyle \frac{1}{\sin z}\;g_{\mu\nu}^\parallel \Big]
$ which agrees with \cite{DittrichReuter}.

$\bullet$\ One gets finally
\begin{equation}
\begin{split}
T_{\mu\nu}^{bare} &=  \frac{\alpha}{2\pi}\int_0^\infty \frac{ds}{\sqrt{s}}
\int_{-1}^{+1}\frac{dv}{2}\;e^{\displaystyle -is\varphi_0}\;\frac{z}{\sin
z}\;I_{\mu\nu},\cr
I_{\mu\nu}&= i\sqrt{\pi}\;e^{\displaystyle -i\pi/4}\;\frac{1}{<e^{-is\varphi_1}>}
\sum_{i=1,5} Tr S_i\cr
&= 2i\sqrt{\pi}\;e^{-i\pi/4}\;
\Big[I_{\mu\nu}^{DR}\big|_{k_3=0}+\frac{i}{s}
\underbrace{\left((g_{\mu\nu}^\parallel + 2g_{\mu 0}g_{\nu 0})\cos z
-g_{\mu\nu}^\perp \cos zv\right)}_{B_{\mu\nu}} + odd(v)
\Big],\cr
& B_{\mu\nu}\ diagonal,\quad B_{00}=\cos z=B_{33},\ B_{11}=-\cos zv=B_{22},
\end{split}
\label{eq:Imunu1}
\end{equation}
in which $I_{\mu\nu}^{DR}$ is given in (4.25) of \cite{DittrichReuter}:
\begin{equation}
\begin{split}
I_{\mu\nu}^{DR} &=2\sum_{i=1}^5 \frac{Tr\,S_i}{4<e^{\displaystyle -is\varphi_1}>}\cr
&=\Big(-2m^2+\frac{1-v^2}{2}\;k_\parallel^2\Big)\big(\cos z\;
g_{\mu\nu}^\parallel +\cos zv\; g_{\mu\nu}^\perp \big)
+\frac{2i}{s}\Big(\cos zv\; g_{\mu\nu}^\perp +\frac{z}{\sin
z}\;g_{\mu\nu}^\parallel \Big)\cr
&- \big(\cos zv -v \cot z\;\sin zv\big) \big[k_\mu k_\nu
-k_{\perp\mu}k_{\perp\nu}-k_{\parallel\mu}k_{\parallel\nu}\big]
+\frac{\cos zv -\cos z}{\sin^2 z}\big[g_{\mu\nu}k_\perp^2
-2k_{\perp\mu}k_{\perp\nu}\big].
\end{split}
\label{eq:ImunuDR}
\end{equation}
Therefore, the only difference with the expression in standard QED as given
in \cite{DittrichReuter} (evaluated at $k_3=0$) is the 
$B_{\mu\nu}$ term that comes from $Tr\,S_2$ because, in there,
$\gamma_\parallel^\lambda \gamma_{\parallel\lambda}=-2$ has to be
replaced by $\gamma^0 \gamma_0=-\gamma_0^2= -1$.

\subsection{The integrations by parts}

Since the power of the $s$ integration in (\ref{eq:Imunu1})
 is $1/\sqrt{s}$ while it was $1/s$
in \cite{DittrichReuter}, the integrations by parts must be redone.
Their goal is to get rid of the terms
proportional to $m^2$ in $I_{\mu\nu}$ such that it only appears inside
$\varphi_0$. Recall $z=eBs$.

This occurs in $Tr S_1$ and we have to integrate
$\displaystyle \int\frac{ds}{\sqrt{s}}\int_{-1}^{+1}\frac{dv}{2}\;
\underbrace{e^{\displaystyle -is\varphi_0}\;\frac{z}{\sin z}(g_{\mu\nu}^\parallel \cos z
+
g_{\mu\nu}^\perp \cos zv)}_{F(s)}$.

Recall that $\varphi_0$ is given in (\ref{eq:varphi}).

I use $\displaystyle \int \frac{ds}{s^{3/2}}\;F(s) =
\underbrace{-\displaystyle \frac{2}{\sqrt{s}}F(s)\Big|_0^\infty}_{B.T.}
-\displaystyle \int
\frac{-2}{\sqrt{s}}
\frac{d}{ds}F(s)$ and we shall always drop the boundary terms (B.T.).
Since $z=eBs$, they  depend a priori on the external $B$.

\medskip

$\bullet$\ $\displaystyle \frac{d}{ds}e^{\displaystyle -is\varphi_0}=
-ie^{\displaystyle -is\varphi_0}\Big[\varphi_0+\frac{zk_\perp^2}{2}\Big(
\displaystyle \frac{z+\sin z(\cos z-\cos zv)-z\cos z\cos zv-zv\sin z\sin zv}{z^2 \sin^2
z}
\Big) \Big]$;

$\bullet$\ $\displaystyle \frac{d}{ds}(g_{\mu\nu}^\parallel \cos z +
g_{\mu\nu}^\perp \cos zv) = eB(-g_{\mu\nu}^\parallel \sin z -
g_{\mu\nu}^\perp v\sin zv)$;

$\bullet$\ $\displaystyle \frac{d}{ds}\displaystyle \frac{z}{\sin
z}=eB\left(\displaystyle \frac{1}{\sin
z}-\displaystyle \frac{z\cos z}{\sin^2 z} \right)$.

\smallskip

After collecting all terms, simplifying and grouping, one gets
\begin{equation}
\begin{split}
& \int\frac{ds}{s^{3/2}}\int_{-1}^{+1}\frac{dv}{2}\;
e^{\displaystyle -is\varphi_0}\;\frac{z}{\sin z}(g_{\mu\nu}^\parallel \cos z
+ g_{\mu\nu}^\perp \cos zv)  \cr
& \hskip -1cm= \int_0^\infty ds \int_{-1}^{+1}\frac{dv}{2}\;
\frac{2}{\sqrt{s}}\frac{z}{\sin z}\;e^{\displaystyle -is\varphi_0}
\left[
(-i)(g_{\mu\nu}^\parallel \cos z + g_{\mu\nu}^\perp \cos zv)
\Big(m^2 +\frac{1-v^2}{4}k_\parallel^2
+\frac{k_\perp^2}{2\sin^2 z}(1-\cos z\cos zv -v \sin z\sin
zv)\Big)\right.\cr
& \hskip 3cm \left. +g_{\mu\nu}^\parallel \frac{1}{s}\left(\cos
z-\frac{z}{\sin z}\right)
+g_{\mu\nu}^\perp\frac{1}{s}\big(-zv\sin zv + \cos zv(1-z\cot z)\big)
\right] + B.T.
\end{split}
\label{eq:ipart1}
\end{equation}
I now integrate by parts w.r.t. $v$ the last term exactly like is done
p.69 of \cite{DittrichReuter}
\begin{equation}
\hskip -1cm \int_{-1}^{+1}\frac{dv}{2}\;e^{\displaystyle -is\varphi_0}\;\frac{1}{s}
\big(-zv\sin zv + \cos zv(1-z\cot z)\big)
=(-i)\int_{-1}^{+1}\frac{dv}{2}\;e^{\displaystyle -is\varphi_0}\;
\left[
\frac12(v\cos zv -\cot z \sin zv)\Big(vk_\parallel^2+
\frac{\sin zv}{\sin z}k_\perp^2\Big)\right]+B.T.
\end{equation}
which leads to
\begin{equation}
\begin{split}
& \int\frac{ds}{s^{3/2}}\int_{-1}^{+1}\frac{dv}{2}\;
e^{\displaystyle -is\varphi_0}\;\frac{z}{\sin z}(g_{\mu\nu}^\parallel \cos z
+ g_{\mu\nu}^\perp \cos zv)  \cr
& \hskip -1cm= \int_0^\infty ds \int_{-1}^{+1}\frac{dv}{2}\;
\frac{2}{\sqrt{s}}\frac{z}{\sin z}\;e^{\displaystyle -is\varphi_0}
\left[
(-i)(g_{\mu\nu}^\parallel \cos z + g_{\mu\nu}^\perp \cos zv)
\Big(m^2 +\frac{1-v^2}{4}k_\parallel^2
+\frac{k_\perp^2}{2\sin^2 z}(1-\cos z\cos zv -v \sin z\sin
zv)\Big)\right.\cr
& \hskip 3cm \left. +g_{\mu\nu}^\parallel \frac{1}{s}\left(\cos
z-\frac{z}{\sin z}\right)
+g_{\mu\nu}^\perp\frac{-i}{2}
(v\cos zv-\cot z \sin zv)\Big(vk_\parallel^2+
\frac{\sin zv}{\sin z}k_\perp^2\Big)
\right] + B.T.
\end{split}
\label{eq:ipart2}
\end{equation}
(\ref{eq:ipart2}) enables now to eliminate the term proportional to $m^2$ in
$I_{\mu\nu}$.
\begin{equation}
\begin{split}
& \int_{-1}^{+1}\frac{dv}{2}\int ds\; \frac{1}{\sqrt{s}}\frac{z}{\sin
z}\;e^{\displaystyle -is\varphi_0}\;(-2 m^2)(g_{\mu\nu}^\parallel \cos z
+g_{\mu\nu}^\perp
\cos zv)\cr
&= B.T.\  -i\int_{-1}^{+1}\frac{dv}{2}\int \frac{ds}{s^{3/2}}\;
\frac{z}{\sin z}\;e^{\displaystyle -is\varphi_0}(g_{\mu\nu}^\parallel \cos z
+g_{\mu\nu}^\perp \cos zv)\cr
& +\int_{-1}^{+1}\frac{dv}{2}\int ds\;\frac{2}{\sqrt{s}}\frac{z}{\sin
z}\;e^{\displaystyle -is\varphi_0}
\Bigg[
(g_{\mu\nu}^\parallel \cos z +g_{\mu\nu}^\perp \cos zv)
\left(
\frac{1-v^2}{4}k_\parallel^2 +\frac{k_\perp^2}{2\sin^2 z}
(1-\cos z\cos zv -v \sin z\sin zv) \right)\cr
&   +i g_{\mu\nu}^\parallel\frac{1}{s}\left(\cos z-\frac{z}{\sin
z}\right)
 + \frac{g_{\mu\nu}^\perp}{2}(v\cos zv -\cot z\sin
zv)\left(vk_\parallel^2 + \frac{\sin zv}{\sin z}k_\perp^2\right)
\Bigg].
\end{split}
\end{equation}
After collecting all terms, one gets
\begin{equation}
\begin{split}
T_{\mu\nu}^{bare}(\hat k, B) &=
\frac{\alpha}{2\pi}\;2i\sqrt{\pi}\;e^{\displaystyle -i\pi/4}
\int\frac{ds}{\sqrt{s}}\int_{-1}^{+1}\frac{dv}{2}\;e^{\displaystyle -is\varphi_0}\cr
& \Big(
N_0[g_{\mu\nu}\hat k^2 - \hat k_\mu \hat k_\nu]
-N_1[g_{\mu\nu}^\parallel \hat k_\parallel^2-\hat k_\mu^\parallel \hat
k_\nu^\parallel]
+N_2[g_{\mu\nu}^\perp \hat k_\perp^2 -\hat k_\mu^\perp \hat k_\nu^\perp]
+\frac{2i}{s}\frac{z}{\sin z}\; \cos z
(g_{\mu\nu}^\parallel + g_{\mu 0} g_{\nu_0})
\Big)+B.T.,\cr
\varphi_0 &= m^2+\frac{1-v^2}{4}\;\hat k_\parallel^2
+ \frac{\cos zv-\cos z}{2z\sin z}\;k_\perp^2,\quad z=eBs,
\end{split}
\label{eq:Tmunu0}
\end{equation}
in which $N_0,N_1,N_2$ are the same as in (4.28\,c) of \cite
{DittrichReuter}:
\begin{equation}
\begin{split}
N_0 &=\frac{z}{\sin z}\big(\cos zv-v\cot z\;\sin zv\big),\cr
N_1 &= -z\cot z\Big(1-v^2+\frac{v\;\sin zv}{\sin z}\Big) +z\;\frac{\cos
zv}{\sin z}=N_0-(1-v^2)z\;\cot z,\cr
N_2 &= -\frac{z\;\cos zv}{\sin z} + \frac{zv\;\cot z\;\sin zv}{\sin z}
+\frac{2z(\cos zv -\cos z)}{\sin^3 z}
= -N_0 +\frac{2z(\cos zv -\cos z)}{\sin^3 z}.
\end{split}
\label{eq:N0N1N2}
\end{equation}

The last contribution to (\ref{eq:Tmunu0}) involves the tensor
$g_{\mu\nu}^\parallel -g_{\mu 0}g_{\nu 0}$, which is identical to
$\frac{1}{\hat k_\parallel^2}\big[g_{\mu\nu}^\parallel \hat k_\parallel^2
-\hat k_\mu^\parallel \hat k_\nu^\parallel\big]$. It is therefore of the
same type as that proportional to $N_1$, and (\ref{eq:Tmunu0}) rewrites
\begin{equation}
\begin{split}
T_{\mu\nu}^{bare}(\hat k, B) &=
\frac{\alpha}{2\pi}\;2i\sqrt{\pi}\;e^{\displaystyle -i\pi/4}
\int\frac{ds}{\sqrt{s}}\int_{-1}^{+1}\frac{dv}{2}\;e^{\displaystyle
-is\varphi_0}\cr
& \Big(
N_0[g_{\mu\nu}\hat k^2 - \hat k_\mu \hat k_\nu]
-N_1[g_{\mu\nu}^\parallel \hat k_\parallel^2-\hat k_\mu^\parallel \hat
k_\nu^\parallel]
+N_2[g_{\mu\nu}^\perp \hat k_\perp^2 -\hat k_\mu^\perp \hat k_\nu^\perp]
+2i\;\frac{eB}{\hat k_\parallel^2}\;\frac{\cos z}{\sin z}
\big[g_{\mu\nu}^\parallel \hat k_\parallel^2
-\hat k_\mu^\parallel \hat k_\nu^\parallel\big]
\Big)+ B.T..
\end{split}
\label{eq:Tmunu1}
\end{equation}

 I remind that the $\hat k$ notation means that
$k_3$ must be set to $0$ everywhere:
$\hat k^2= -k_0^2 + k_\perp^2,\ \hat k_\parallel^2=-k_0^2,\ 
\hat k_\mu^\parallel= -k_0 g_{\mu 0}$. Of course, for the transverse part,
$\hat k_\mu^\perp = k_\mu^\perp$.

\subsection{Changes of variables. The unrenormalized
$\boldsymbol{T_{\mu\nu}^{bare}}$} \label{subsec:changevar}

I go to $s=-it = se^{-i\pi/2}$. Therefore $\sqrt{s} =
\sqrt{t}\;e^{-i\pi/4}$ and (\ref{eq:Tmunu1}) becomes
\begin{equation}
\begin{split}
T_{\mu\nu}^{bare}(\hat k, B) &= \frac{\alpha}{2\pi}\;2\sqrt{\pi}
\int_0^{i\infty}\frac{dt}{\sqrt{t}}\int_{-1}^{+1}\frac{dv}{2}\cr
& \hskip -1cm e^{\displaystyle -t\varphi_0}\Big(
N_0[g_{\mu\nu}\hat k^2 - \hat k_\mu \hat k_\nu]
-N_1[g_{\mu\nu}^\parallel \hat k_\parallel^2-\hat k_\mu^\parallel \hat
k_\nu^\parallel]
+N_2[g_{\mu\nu}^\perp \hat k_\perp^2 -\hat k_\mu^\perp \hat k_\nu^\perp]
- 2\frac{eB}{\hat k_\parallel^2}\;\frac{\cosh eBt}{\sinh eBt}\;
\big[g_{\mu\nu}^\parallel \hat k_\parallel^2
-\hat k_\mu^\parallel \hat k_\nu^\parallel\big]
\Big)+ B.T.,\cr
& \cr
\varphi_0 &= m^2+\frac{1-v^2}{4}\;\hat k_\parallel^2
- \frac{\cosh eBtv-\cosh eBt}{2eBt\sinh eBt}\;k_\perp^2,\quad
N_0 = \frac{eBt}{\sinh eBt}\left(\cosh eBtv -\frac{v\cosh eBt \sinh
eBtv}{\sinh
eBt}\right),\cr
N_1 &= N_0 -(1-v^2) eBt \frac{\cosh eBt}{\sinh eBt},\quad
 N_2 = -N_0 -\frac{2eBt(\cosh eBtv -\cosh eBt)}{\sinh^3 eBt}.
\end{split}
\label{eq:Tmunu2}
\end{equation}
The integration on $t$ is on the imaginary axis, and its rotation back to
the real axis requires that the integrand vanishes on the infinite 1/4
circle.  The convergence is achieved by the exponential $e^{\displaystyle -t(m^2 +
\ldots)}$ as long as $m\not=0$. I shall suppose that this Wick rotation
stays valid even when $m\to 0$ and we shall hereafter define $\Pi_{\mu\nu}$
accordingly.

The last part of $T_{\mu\nu}$ diverges like $\int_0^{(\ )}
\frac{dt}{t^{3/2}}$. This divergent contribution does not depend on $B$,
such that it can be removed by a $B$-independent counterterm. Transversality
is another matter.

I then go to $y=eBt = ieBs$. By this change, the limits $B\to 0$
and $y\to
0$ become similar
\begin{equation}
\begin{split}
T_{\mu\nu}^{bare}(\hat k, B) &= \frac{\alpha}{2\pi}\;\frac{2\sqrt{\pi}}{\sqrt{eB}}
\int_0^{\infty}\frac{dy}{\sqrt{y}}\int_{-1}^{+1}\frac{dv}{2}\cr
& \hskip -1cm
e^{\displaystyle -\frac{y}{eB}\varphi_0}\Bigg(
N_0[g_{\mu\nu}\hat k^2 - \hat k_\mu \hat k_\nu]
-N_1[g_{\mu\nu}^\parallel \hat k_\parallel^2-\hat k_\mu^\parallel \hat
k_\nu^\parallel]
+N_2[g_{\mu\nu}^\perp \hat k_\perp^2 -\hat k_\mu^\perp \hat k_\nu^\perp]
- 2eB\;\frac{\cosh y}{\sinh y}\;
\underbrace{\frac{g_{\mu\nu}^\parallel \hat k_\parallel^2
-\hat k_\mu^\parallel \hat k_\nu^\parallel}{\hat k_\parallel^2}}_{\equiv
g_{\mu 3}g_{\nu 3}}
\Bigg)+ B.T.,\cr
& \cr
\varphi_0 &= m^2+\frac{1-v^2}{4}\;\hat k_\parallel^2
- \underbrace{\frac{\cosh yv-\cosh y}{2y\sinh y}\;k_\perp^2}_{g(y,v)}
= m^2 +\frac{1-v^2}{4}\;\hat k^2-k_\perp^2
\underbrace{\left(\frac{1-v^2}{4} +\frac{\cosh yv-\cosh y}{2y\sinh
y}\right)
}_{h(y,v)\geq 0},\cr
& \hskip -1cm  N_0 = \frac{y}{\sinh y}\left(\cosh yv -\frac{v\cosh y\, \sinh
yv}{\sinh y}\right),\quad
N_1 = N_0 -(1-v^2)\, y\; \frac{\cosh y}{\sinh y},\quad
 N_2 = -N_0 -\frac{2y(\cosh yv -\cosh y)}{\sinh^3 y}.
\end{split}
\label{eq:Tmunu3}
\end{equation}
We already notice in (\ref{eq:Tmunu3}) that the external $B$ breaks the
$(3+1)$-transversality of $T_{\mu\nu}^{bare}$.

\subsection{Transversality}
\label{subsec:transverse}

This issue will be more extensively studied in connection with the
counterterms (see section \ref{section:alter}).

In standard QED in external $B$ \cite{DittrichReuter}, the 3 contributions
to $\Pi_{\mu\nu}$ (see eq.~(4.32) of \cite{DittrichReuter})
 are all transverse since:\newline
 $k^\mu k^\nu(g_{\mu\nu}k^2 -k_\mu k_\nu)=0$,\newline
$k^\mu k^\nu(g_{\mu\nu}^\parallel k_\parallel^2 - k_\mu^\parallel k_\nu^\parallel)
=k_\nu^\parallel k^\nu k_\parallel^2 -(k^\mu k_\mu^\parallel)^2=
(k_\parallel^2)^2 - (k_\parallel^2)^2=0$,\newline
$k^\mu k^\nu(g_{\mu\nu}^\perp k_\perp^2 - k_\mu^\perp k_\nu^\perp)=
(k_\perp^2)^2 -(k_\perp^2)^2=0$.

This is not the case for the graphene-simulating medium under scrutiny here
since:\newline
 $k^\mu k^\nu(g_{\mu\nu}\hat k^2 -\hat k_\mu \hat k_\nu)= k^2 \hat k^2
- (k^\mu \hat k_\mu)^2 = (\hat k^2 + k_3^2) \hat k^2 - (\hat k^2)^2
= k_3^2 \, \hat k^2= k_3^2(-k_0^2 + k_\perp^2)$,\newline
$k^\mu k^\nu(g_{\mu\nu}^\parallel \hat k_\parallel^2 - \hat k_\mu^\parallel
\hat k_\nu^\parallel)= k_\nu^\parallel k^\nu \hat k_\parallel^2 -(k^\mu
\hat k_\mu^\parallel)^2 =k_\parallel^2 \hat k_\parallel^2-(\hat
k_\parallel^2)^2= (-k_0^2 +k_3^2)(-k_0^2) -(k_0^2)^2 =
-k_3^2\,k_0^2$,\newline
$k^\mu k^\nu(g_{\mu\nu}^\perp \hat k_\perp^2 - \hat k_\mu^\perp \hat
k_\nu^\perp)=0$,\newline
$k^\mu k^\nu(\hat g_{\mu\nu}\hat k^2 -\hat k_\mu \hat k_\nu)=0$,\newline
such that  only $(2+1)$-transversality is satisfied:
\begin{equation}
\hat k^\mu \hat k^\nu T_{\mu\nu}^{bare}=0 = \hat k^\mu \hat k^\nu
\Pi_{\mu\nu}^{bare}.
\end{equation}

From the property (see (\ref{eq:PiV}))
 $\Pi_{\mu\nu}(k,B) =-\frac{1}{\pi^2}\frac{1-n^2}{a} V\;T_{\mu\nu}(\hat k, B)$ one gets
$k^\mu \Pi_{\mu\nu}(k,B) =
\big(k^3\Pi_{3\nu}(k,B)+\hat k^\mu \Pi_{\hat\mu \nu}(k,B)\big)
= -\frac{1}{\pi^2}\frac{1-n^2}{a}
\big( k^3 VT_{3\nu}(\hat k,B)+V\underbrace{\hat k^\mu T_{\hat\mu \nu}(\hat
k,B)}_{=0}\big)$ and, therefore
$k^\mu k^\nu\Pi_{\mu\nu}(k,B)= -\frac{1}{\pi^2}\frac{1-n^2}{a}
\big( k_3^2 V T_{33}(\hat k,B)
+k^3 V \underbrace{\hat k^\nu T_{3\hat\nu}(\hat k,B)}_{=0}\big)
= -\frac{1}{\pi^2}\frac{1-n^2}{a}k_3^2 V T_{33}(\hat k,B)$,
which can only vanish if $k_3=0$, or if $ (1-n^2)V=0 \Leftrightarrow k^2=0$
or if $T_{33}=0$. This last
condition is in general not true, unless one makes an additional
subtraction. Since $T_{33}$ depends on $B$, and if one wants counterterms
to be independent of $B$, $(3+1)$-transversality can only be achieved at a given
$B$, for example $B=0$, by defining the renormalized $T_{\mu\nu}(\hat
k, B)$ as $T_{\mu\nu}^{bare}(\hat k,B) -T_{33}^{bare}(\hat k,B=0)\,g_{\mu 3}g_{\nu 3}$.

From this it follows that the scalar potential, which is obtained from
$\Pi_{00}(k_0=0,B)$ is the same as that calculated from the bare
$T_{\mu\nu}$:
\begin{equation}
T_{00}(\hat k, k_0=0, B) =T_{00}^{bare}(\hat k, k_0=0, B)
=\frac{\alpha}{2\pi}\frac{2\sqrt{\pi}}{\sqrt{eB}}(-k_\perp^2)\int_0^\infty\frac{dy}{\sqrt{y}}\int_{-1}^{+1}\frac{dv}{2}\;N_0\;e^{-\varphi_0
y/eB}\big|_{k_0=0}.
\end{equation}

\section{Renormalization conditions and counterterms
}\label{section:renorm1}

The renormalization condition to be fulfilled is (see (4.29) of
\cite{DittrichReuter})
\begin{equation}
\lim_{k^2 \to 0} \lim_{B\to 0} \Pi_{\mu\nu}(k,B)=0,
\label{eq:rencond}
\end{equation}
which should now be applied to the expression (\ref{eq:PiV}) of
$\Pi_{\mu\nu}$.

When $B\to 0$, $z\equiv eBs \to 0,\ N_0 \to 1-v^2 +{\cal O}(y^2),\ N_1 \to
0+{\cal O}(y^2),\ N_2 \to 0 +{\cal O}(y^2)$, $\varphi_0 \to
m^2+\frac{1-v^2}{4}\hat k^2+{\cal O}(y^2)$ and one gets (we prefer to
use, below, the variable $t = y/eB$)
\begin{equation}
\begin{split}
\Pi_{\mu\nu}(k, B=0) &=-\frac{1}{\pi^2}\; T_{\mu\nu}(\hat k,
B=0)\;\frac{1-n^2}{a}\;V(n,\theta,\eta,u) \cr
&= -\Big[\frac{1}{\pi^2}\;\frac{1-n^2}{a}\;V(n,\theta,\eta,u)\Big]\cr
& \times \frac{\alpha}{2\pi}\;2\sqrt{\pi} \Bigg[
(g_{\mu\nu}\hat k^2 -\hat k_\mu \hat k_\nu)
\int_0^{i\infty} \frac{dt}{\sqrt{t}}\int_{-1}^{+1}\frac{dv}{2}\;(1-v^2)
e^{\displaystyle -t\big(m^2+\frac{1-v^2}{4}\hat k^2\big)}\cr
& + \underbrace{\frac{(g_{\mu\nu}^\parallel \hat k_\parallel^2
-\hat k_\mu^\parallel \hat k_\nu^\parallel)}{\hat k_\parallel^2}}_{\equiv
g_{\mu
3}g_{\nu 3}}
\int_0^{i\infty} \frac{dt}{\sqrt{t}}\int_{-1}^{+1}\frac{dv}{2}\;
e^{\displaystyle -t\big(m^2+\frac{1-v^2}{4}\hat k^2\big)}\;\frac{(-2)}{t}\Bigg]+c.t.
\end{split}
\label{eq:Pimunubare}
\end{equation}
in which we have introduced the counterterms ``c.t.'' that we are going to
determine.

$\ast$\ The unrenormalized first part of $\Pi_{\mu\nu}(k,B=0)$ (3rd line
of (\ref{eq:Pimunubare})) is finite
and vanishes at
$k^2=0$ because of the property (\ref{eq:Vprop}) of the transmittance $V$.
Therefore, unlike in standard QED, no counterterm is needed there.

$\ast$\ The second part (4th line of (\ref{eq:Pimunubare}) is easily seen to be divergent since 
$\displaystyle \int_{t_0}^\infty dt\; \frac{e^{-a t}}{t^{3/2}} =
-\frac{2e^{-at}}{\sqrt{t}}\Big|_{t_0}^{\infty} -2\sqrt{\pi a}\;
Erf[\sqrt{at}]\Big|_{t_0}^{\infty}$. Presently, $t_0=0$. $Erf(\infty)=1,
Erf(0)=0$, which makes the 1st contribution
$-\displaystyle \frac{2e^{-at}}{\sqrt{t}}\Big|_{t_0}^{\infty}$ diverge like $1/\sqrt{t}$ at
$t\to 0$.

$\bullet$\ Since, at $B\to 0$ and at $k^2\equiv \hat k^2 +k_3^2 \to 0$, $\varphi_0 \to m^2
-\displaystyle\frac{1-v^2}{4}k_3^2$,
the most naive renormalization that one could propose is the substitution
$e^{\displaystyle -t(m^2+\frac{1-v^2}{4}\hat k^2)} \to
e^{\displaystyle -t(m^2+\frac{1-v^2}{4}\hat k^2)}-e^{\displaystyle -t(m^2-\frac{1-v^2}{4} k_3^2)}$,
that is, to add the following counterterm to (\ref{eq:Pimunubare})\newline
$-\left[\displaystyle \frac{1}{\pi^2}\;\displaystyle
\frac{1-n^2}{a}\;V(n,\theta,\eta,u)\right]
\displaystyle \frac{\alpha}{2\pi}\;2\sqrt{\pi}\;
\underbrace{\frac{(g_{\mu\nu}^\parallel \hat k_\parallel^2
-\hat k_\mu^\parallel \hat k_\nu^\parallel)}{\hat k_\parallel^2}
}_{\equiv g_{\mu 3}g_{\nu 3}}
\displaystyle \int_0^\infty \frac{dt}{\sqrt{t}}\int_{-1}^{+1}\frac{dv}{2}\;
e^{\displaystyle -t(m^2 -\frac{1-v^2}{4}k_3^2)}\;\displaystyle \frac{(+2)}{t}$.

However,  it has two major problems:\newline
*\ it depends on $k_3$, making the situation extremely
cumbersome because the factorization that we demonstrated in
subsection \ref{subsec:gammaprop} of $\Pi_{\mu\nu}$ into $V \times T_{\mu\nu}$
precisely relied on the property that $T_{\mu\nu}$ did not depend on
$k_3$;\newline
*\  the divergence reappears off mass-shell at $k^2 \not=0$.

$\bullet$\ So, we shall instead take for the counterterm  the opposite
of the limit at $B\to 0$ of the last contribution to (\ref{eq:Pimunubare}),
 independently of the limit $k^2 \to 0$\newline
$-\left[\displaystyle \frac{1}{\pi^2}\;\displaystyle \frac{1-n^2}{a}\;V(n,\theta,\eta,u)\right]
\displaystyle \frac{\alpha}{2\pi}\;2\sqrt{\pi}\;
\underbrace{\frac{(g_{\mu\nu}^\parallel \hat k_\parallel^2
-\hat k_\mu^\parallel \hat k_\nu^\parallel)}{\hat k_\parallel^2}}_{\equiv
g_{\mu 3}g_{\nu 3}}
\displaystyle \int_0^\infty \frac{dt}{\sqrt{t}}\int_{-1}^{+1}\frac{dv}{2}\;
e^{\displaystyle -t(m^2 +\frac{1-v^2}{4}\hat k^2)}\;\displaystyle \frac{(+2)}{t}$.
By definition, since it is evaluated at $B=0$,
 this counterterm does not depend on the external $B$.
It ensures finiteness, and renormalization conditions at $k^2=0$ keep
satisfied because of the factor $(1-n^2)V$ that we have shown to vanish at
$k^2=0$.

This amounts to taking the renormalized $\Pi_{\mu\nu}$ to be (after the Wick rotation
evoked above)
\begin{equation}
\begin{split}
\Pi_{\mu\nu}(k,B) &= -\frac{1}{\pi^2}
\;\frac{1-n^2}{a}\;V(n,\theta,\eta,u)\;
\frac{\alpha}{2\pi}
\frac{2\sqrt{\pi}}{\sqrt{eB}}
\int_0^{\infty}\frac{dy}{\sqrt{y}}\int_{-1}^{+1}\frac{dv}{2}\cr
& \hskip -1.5cm e^{\displaystyle -\frac{y}{eB}\,\varphi_0}\Big(
N_0[g_{\mu\nu}\hat k^2 - \hat k_\mu \hat k_\nu]
-N_1[g_{\mu\nu}^\parallel \hat k_\parallel^2-\hat k_\mu^\parallel \hat
k_\nu^\parallel]
+N_2[g_{\mu\nu}^\perp k_\perp^2 -k_\mu^\perp k_\nu^\perp]\Big)
\cr
& - 2eB\;e^{\displaystyle
-\frac{y}{eB}\,\varphi_0}
\underbrace{\frac{g_{\mu\nu}^\parallel \hat k_\parallel^2
-\hat k_\mu^\parallel \hat k_\nu^\parallel}{\hat k_\parallel ^2}}_{\equiv
g_{\mu
3}g_{\nu 3}}
\Bigg[\underbrace{\frac{\cosh y}{\sinh y}\;
- \frac{1}{y}\;e^{\displaystyle -\frac{y}{eB}(m^2+\frac{1-v^2}{4}\hat
k^2-\varphi_0)}}_{N_3}\Bigg],
\end{split}
\label{eq:Piren1}
\end{equation}
with $y=eBt$, $\varphi_0, N_0, N_1, N_2$ given in (\ref{eq:Tmunu3}), and
\begin{equation}
 N_3 =\frac{\cosh y}{\sinh y}\;
- \frac{1}{y}\;e^{\displaystyle -\frac{y}{eB}(m^2+\frac{1-v^2}{4}\hat
k^2-\varphi_0)}=\frac{\cosh y}{\sinh y}- \frac{1}{y}\;e^{\displaystyle
-\frac{y}{eB}k_\perp^2\, h(y,v)},
\label{eq:Piren2}
\end{equation}
with, as stated in (\ref{eq:Tmunu3}),
\begin{equation}
h(y,v)=\frac{1-v^2}{4}+\frac{\cosh yv - \cosh y}{2y \sinh y}.
\end{equation}
$\Pi_{\mu\nu}$ as written in (\ref{eq:Piren1}) satisfies the following properties:\newline
*\ it vanishes at $B=0$ and $k^2=0$, therefore satisfying the
renormalization conditions (\ref{eq:rencond});\newline
*\ it is finite (no divergence $\simeq \displaystyle\frac{1}{\sqrt{y}}$ when $y\to
0$ occurs any more in the last contribution).

It is important to stress the essential role of the transmittance $V$ for
$\Pi_{\mu\nu}$ to fulfill suitable renormalization conditions. The same
conditions cannot
be satisfied for $T_{\mu\nu}$ alone  as one gets rapidly convinced by
explicit calculations. In particular, the counterterms that one is led,
then, to introduce get divergent when $m\to 0$.

\subsection{The limit $\boldsymbol{eB\to 0}$}
\label{subsec:Bnul}

Thanks to the counterterm,  the  contribution to
(\ref{eq:Piren1}) proportional to $N_3$ vanishes at $eB \to 0$
\footnote{This is easily seen for example by going back to the integration
variable $t = y/eB$.}, while 
\begin{equation}
N_0 \stackrel{B\to 0}{\to} 1-v^2,\quad N_1\stackrel{B\to 0}{\to} 0,\quad
N_2\stackrel{B\to 0}{\to}0,\quad \varphi_0 \stackrel{B\to 0}{\to}
m^2+\frac{1-v^2}{4}\;\hat k^2,
\end{equation}
such that
\begin{equation}
\begin{split}
T_{\mu\nu}(\hat k, B=0) &=\frac{\alpha}{2\pi}\;2\sqrt{\pi}
(g_{\mu\nu}\hat k^2 -\hat k_\mu \hat k_\nu)
\int_0^\infty \frac{dt}{\sqrt{t}}\int_{-1}^{+1}\frac{dv}{2}\;(1-v^2)
e^{\displaystyle-t\big(m^2+\frac{1-v^2}{4}\hat k^2\big)}\cr
&= \alpha(g_{\mu\nu}\hat k^2 -\hat k_\mu \hat k_\nu)
\int_{-1}^{+1}\frac{dv}{2}\;\frac{1-v^2}{\sqrt{m^2+\frac{1-v^2}{4}\hat
k^2}},
\end{split}
\end{equation}
that is
\begin{equation}
T_{\mu\nu}(\hat k, B=0) = \alpha(g_{\mu\nu}\hat k^2 -\hat k_\mu \hat k_\nu)
\frac{1}{\sqrt{\hat k^2}} \frac{(\sqrt{\hat k^2}-2m)^2}{\hat
k^2}\;\arcsin\frac{\sqrt{\hat k^2}}{\sqrt{\hat k^2+4m^2}},
\end{equation}
and, according to (\ref{eq:PiV})
\begin{equation}
\Pi_{\mu\nu}(k,B=0)=
-\frac{1}{\pi^2}\;\frac{1-n^2}{a}\;V(n,\theta,\eta,u)\times
\alpha(g_{\mu\nu}\hat k^2 -\hat k_\mu \hat k_\nu)
\frac{1}{\sqrt{\hat k^2}} \frac{(\sqrt{\hat k^2}-2m)^2}{\hat
k^2}\;\arcsin\frac{\sqrt{\hat k^2}}{\sqrt{\hat k^2+4m^2}},
\end{equation}
in which $V$ is given by (\ref{eq:V2}). It vanishes at $k^2=0$ thanks to
the factor $(1-n^2)V$. The non-vanishing components are
$(0,0), (3,3), (1,1), (2,2), (1,2), (2,1)$.

The limit $m\to 0$ yields the non $(3+1)$-transverse
\footnote{The results obtained \cite{CoquandMachet},  calculated directly at $m=0$,
are very close, since they only differ for
$T_{33}$ which got modified by the
counterterms (overlooked in \cite{CoquandMachet}).}
\begin{equation}
T_{\mu\nu}(\hat k, B=0,m=0) =
\alpha\;\frac{\pi}{2}\,(g_{\mu\nu}\hat k^2 -\hat k_\mu \hat k_\nu)
\frac{1}{\sqrt{\hat k^2}}.
\label{eq:TB0m0}
\end{equation}

\subsection{The limit $\boldsymbol{eB\to \infty}$}
\label{subsec:infinitlim}

As usual in Schwinger-type calculations, one takes first the relevant limit
inside the integrand before worrying about the limits of integration.

When $y \equiv eBt \to\infty$:\newline
*\ $N_0\sim \displaystyle \frac{y}{\sinh y}*(\cosh yv - v \sinh yv)
\sim \displaystyle \frac{y(1-v)}{e^{y(1-v)}}$ exponentially vanishes at $y
 \to \infty$;\newline
*\ $N_1 \stackrel{y\to \infty}{\sim} N_0-(1-v^2)y$ has  a polynomial
growth in $y$;\newline
*\ $N_2 \sim -N_0 +$ exponentially damped terms also vanishes at
$y\to\infty$.

Since $N_0, N_2 \to 0$, one is left with the $N_1$ and $N_3$
contributions. They both only concern the subspace $(3,3)$.
This is obvious
since the projector $g_{\mu\nu}^\parallel \hat
k_\parallel^2 -\hat k_\mu^\parallel \hat k_\nu^\parallel$ 
does not vanish  only for $\mu=3=\nu$.

\subsubsection{The $\boldsymbol{N_1}$ part}

It writes
\begin{equation}
T_{\mu\nu}^{N_1}(\hat k, eB \to \infty) =
\frac{\alpha}{2\pi}\;\frac{2\sqrt{\pi}}{\sqrt{eB}}\;
[g_{\mu\nu}^\parallel \hat k_\parallel^2 -\hat k_\mu^\parallel \hat
k_\nu^\parallel]\;
\int_{-1}^{+1}\frac{dv}{2}\int\frac{dy}{\sqrt{y}}\;
e^{\displaystyle -\frac{k_\perp^2}{2eB}}\;
y(1-v^2)\;e^{\displaystyle -\frac{y}{eB}\big(m^2+\frac{1-v^2}{4}\hat
k_\parallel^2\big)}.
\end{equation}

For $v\not= \pm 1$
\footnote{see the remark at the beginning of subsection
\ref{subsec:infinitlim}.},
 $\cosh yv < \cosh y$. When $y \to \infty$, this becomes
$\cosh yv \ll \cosh y$ such that $g(y,v)
\stackrel{y\to\infty}{\simeq}-1/2y$.
We shall therefore consider (remember $\hat k_\parallel^2 =-k_0^2$)
that\newline
$\varphi_0 \stackrel{y\to\infty}{\to} m^2
-\displaystyle \frac{1-v^2}{4}k_0^2+\displaystyle \frac{k_\perp^2}{2y} \Rightarrow
e^{\displaystyle -\frac{y}{eB}\varphi_0} \to e^{\displaystyle -\frac{k_\perp^2}{2eB}}\;
e^{\displaystyle -\frac{y}{eB}\big(m^2-\frac{1-v^2}{4}k_0^2\big)}$,
without worrying about the limits of integration $v=\pm1$.
This gives
\begin{equation}
\begin{split}
T_{\mu\nu}^{N_1}(\hat k, eB \to \infty) &=
\frac{\alpha}{2}\;eB\;[g_{\mu\nu}^\parallel \hat k_\parallel^2 -\hat
k_\mu^\parallel \hat
k_\nu^\parallel]\;
e^{\displaystyle -\frac{k_\perp^2}{2eB}}\int_{-1}^{+1}\frac{dv}{2}\;
\frac{1-v^2}{\left(m^2-k_0^2\frac{1-v^2}{4}\right)^{3/2}}\cr
&=2\alpha\;\frac{eB}{k_0^3}\;
[g_{\mu\nu}^\parallel \hat k_\parallel^2 -\hat
k_\mu^\parallel \hat k_\nu^\parallel]\;
e^{\displaystyle -\frac{k_\perp^2}{2eB}}
\left[\frac{4mk_0}{4m^2-k_0^2}-\ln\frac{2m+k_0}{2m-k_0}\right]\cr
&= -2\alpha \frac{eB}{k_0}\;g_{\mu 3}g_{\nu 3}\;e^{\displaystyle -\frac{k_\perp^2}{2eB}}
\left[\frac{4mk_0}{4m^2-k_0^2}-\ln\frac{2m+k_0}{2m-k_0}\right].
\end{split}
\label{eq:TN1}
\end{equation}

\subsubsection{The $\boldsymbol{N_3}$ part}

It is
\begin{equation}
\begin{split}
T_{\mu\nu}^{N_3}(\hat k,B) &=
\frac{\alpha}{2\pi}
\frac{2\sqrt{\pi}}{\sqrt{eB}}
\int_0^{\infty}\frac{dy}{\sqrt{y}}\int_{-1}^{+1}\frac{dv}{2}
(- 2eB)\;\frac{g_{\mu\nu}^\parallel \hat k_\parallel^2 - \hat
k_\mu^\parallel \hat k_\nu^\parallel}{\hat k_\parallel^2}
\Bigg[\frac{\cosh y}{\sinh y}\;e^{\displaystyle -\frac{y}{eB}\varphi_0} -
\frac{1}{y}\;e^{\displaystyle -\frac{y}{eB}\big(m^2+\frac{1-v^2}{4}\hat
k^2\big)}\Bigg]\cr
&=\frac{\alpha}{2\pi}\;2\sqrt{\pi}
(-2eB)\;\frac{g_{\mu\nu}^\parallel \hat k_\parallel^2 - \hat
k_\mu^\parallel \hat k_\nu^\parallel}{\hat k_\parallel^2}
\int_0^{\infty}\frac{dt}{\sqrt{t}}\int_{-1}^{+1}\frac{dv}{2}
\Bigg[e^{\displaystyle -t\varphi_0} \;\frac{\cosh eBt}{\sinh eBt}
 -e^{\displaystyle -t \big(m^2+\frac{1-v^2}{4}\hat
k^2\big)}\;\frac{1}{eBt}\Bigg].
\end{split}
\end{equation}
Now, since the argument of the $\sinh, \cosh$ is $eBt$,
 one cannot use everywhere the expansion at $eB\to\infty$ without
worrying about the integration variable $t$.  So, we shall
split the $t$ (or $y$) integration into 2 parts, $[0,t_0]$ and
$[t_0,\infty]$.

\begin{figure}[h]
\begin{center}
\includegraphics[width=6 cm, height=4 cm]{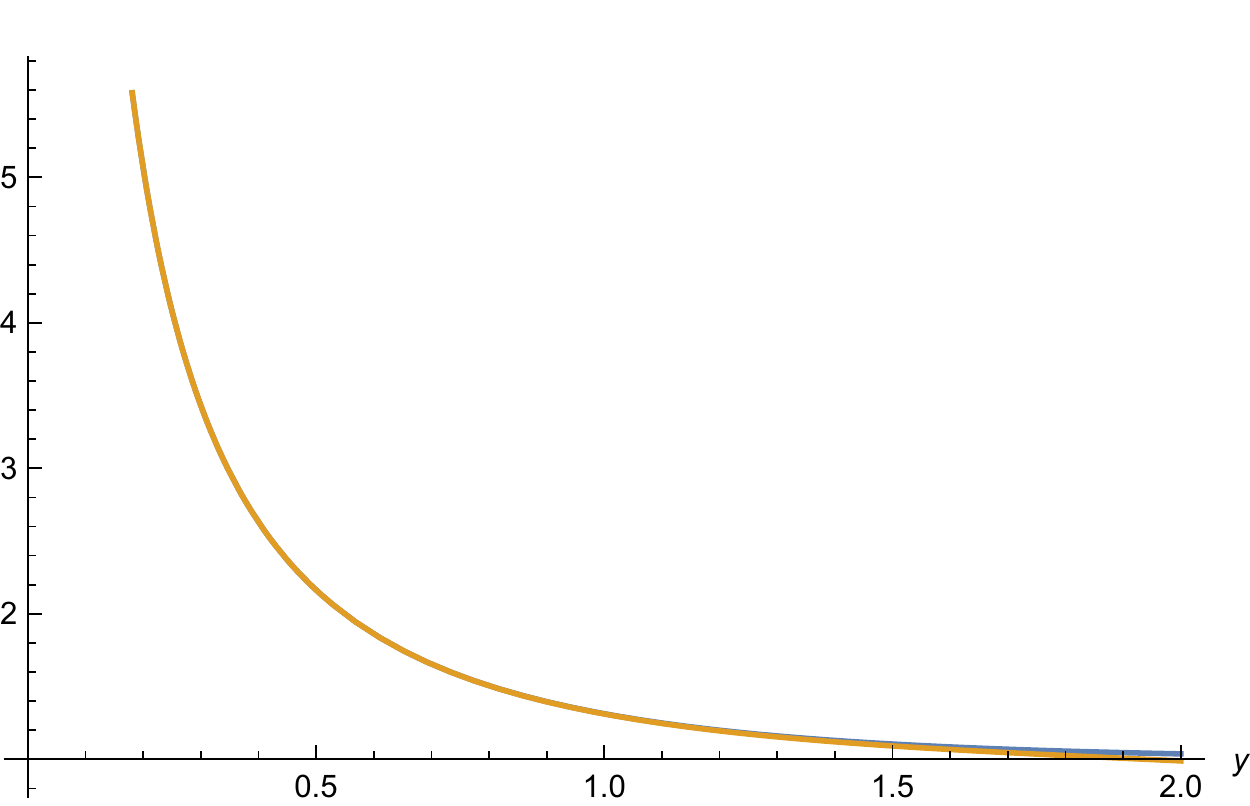}
\caption{$\cosh y/\sinh y$ (blue) and its approximation (yellow), in
practice superposed}
\label{fig:coshsinh}
\end{center}
\end{figure}

$\bullet$\ For $t\in [0,t_0]$ such that $y \equiv eBt_0$ is small, one expands
$\displaystyle\frac{\cosh y}{\sinh y} \simeq \displaystyle\frac{1}{y} +
\displaystyle\frac{y}{3} -\displaystyle\frac{y^3}{45}
+\ldots$, which is a very good approximation up to $y=2$ as shown on Figure
\ref{fig:coshsinh}.
%
%
Since $h(y,v)\stackrel{y\to 0}{\to} 0$ and $h(y,v) \simeq
\frac{(1-v^2)^2}{48} y^2 +\ldots$,
$\varphi_0 \simeq m^2 + \displaystyle\frac{1-v^2}{4} \hat k^2 +
\displaystyle\frac{1}{48}(1-v^2)^2
y^2 + \ldots$, which yields
$e^{\displaystyle -t\varphi_0}\displaystyle \frac{\cosh eBt}{\sinh eBt} \simeq
e^{\displaystyle -t\big(m^2+\frac{1-v^2}{4}\hat k^2\big)}\;\Big(\displaystyle
\frac{1}{y}+\displaystyle \frac{y}{3}
+(1-v^2)^2
\displaystyle \frac{k_\perp^2}{48eB} y^2 -\displaystyle \frac{y^3}{45} +\ldots \Big)$.
One gets the following contribution to $T_{\mu\nu}^{N_3}(\hat k,B)$:
\begin{equation}
\frac{\alpha}{2\pi}\;2\sqrt{\pi}
(-2eB)\;\;\frac{g_{\mu\nu}^\parallel \hat k_\parallel^2 - \hat
k_\mu^\parallel \hat k_\nu^\parallel}{\hat k_\parallel^2}
\int_0^{t_0}\frac{dt}{\sqrt{t}}\int_{-1}^{+1}\frac{dv}{2}\;
e^{\displaystyle -t \big(m^2+\frac{1-v^2}{4}\hat k^2\big)}\Big(\frac{eBt}{3} +(1-v^2)^2
\frac{k_\perp^2}{48eB} (eBt)^2 -\frac{(eBt)^3}{45} +\ldots \Big),
\end{equation}
or, equivalently
\begin{equation}
\frac{\alpha}{2\pi}\;2\sqrt{\pi}
(-2eB)\;\;\frac{g_{\mu\nu}^\parallel \hat k_\parallel^2 - \hat
k_\mu^\parallel \hat k_\nu^\parallel}{\hat k_\parallel^2}
\frac{1}{\sqrt{eB}}
\int_0^{y_0}\frac{dy}{\sqrt{y}}\int_{-1}^{+1}\frac{dv}{2}\;
e^{\displaystyle -\frac{y}{eB} \big(m^2+\frac{1-v^2}{4}\hat k^2\big)}\Big(\frac{y}{3} +(1-v^2)^2
\frac{k_\perp^2}{48eB} y^2 -\frac{y^3}{45} +\ldots \Big).
\end{equation}
Since, in this last expression,  we work at small $y$,
 it is legitimate to expand the exponential. The
leading terms of the integrand are $\frac{y}{3}-\displaystyle \frac{y^3}{45} + \ldots$,
with additional terms damped by $eB$ factors. So, one gets contributions $\propto
\sqrt{eB}, \displaystyle \frac{1}{\sqrt{eB}},\ldots$ $\times$ powers of $y_0$ (which is
$\leq 1$), which, as we shall see, are non-leading.

$\bullet$\ For $t\in [t_0,\infty]$ we consider $\cosh y/\sinh y \approx 1$
and the 2nd contribution to $T_{\mu\nu}$ writes accordingly
\begin{equation}
\frac{\alpha}{2\pi}\;2\sqrt{\pi}
(-2eB)\;\;\frac{g_{\mu\nu}^\parallel \hat k_\parallel^2 - \hat
k_\mu^\parallel \hat k_\nu^\parallel}{\hat k_\parallel^2}
\frac{1}{\sqrt{eB}}
\int_{y_0}^\infty\frac{dy}{\sqrt{y}}\int_{-1}^{+1}\frac{dv}{2}\;
e^{\displaystyle -\frac{y}{eB} \big(m^2+\frac{1-v^2}{4}\hat k^2\big)}\Big(
e^{\displaystyle \frac{y}{eB}k_\perp^2 h(y,v)}-\frac{1}{y}\Big),
\end{equation}
or, equivalently
\begin{equation}
\frac{\alpha}{2\pi}\;2\sqrt{\pi}
(-2eB)\;\;\frac{g_{\mu\nu}^\parallel \hat k_\parallel^2 - \hat
k_\mu^\parallel \hat k_\nu^\parallel}{\hat k_\parallel^2}
\frac{1}{\sqrt{eB}}
\int_{y_0}^\infty\frac{dy}{\sqrt{y}}\int_{-1}^{+1}\frac{dv}{2}\;
e^{\displaystyle -\frac{y}{eB} \big(m^2+\frac{1-v^2}{4}\hat
k_\parallel^2-k_\perp^2g(y,v)\big)}
-e^{\displaystyle -\frac{y}{eB} \big(m^2+\frac{1-v^2}{4}\hat k^2\big)}\frac{1}{y}.
\label{eq:NT4}
\end{equation}

*\  First contribution to  (\ref{eq:NT4}) (main term at  $y>y_0$)

One needs an approximation of $e^{\frac{y}{eB}k_\perp^2
g(y,v)}$ for $y\geq y_0$, which is the most hazardous part. The function
$g(y,v)$ has been defined in (\ref{eq:Piren2}):
\begin{equation}
g(y,v)\equiv \frac{\cosh yv -\cosh y}{2y\sinh y}
\approx \frac{e^{yv}-e^y}{2y e^y}=\frac{1}{2y}\big(e^{-y(1-v)}-1\big).
\end{equation}

\begin{figure}[h]
\begin{center}
\includegraphics[width=6 cm, height=4 cm]{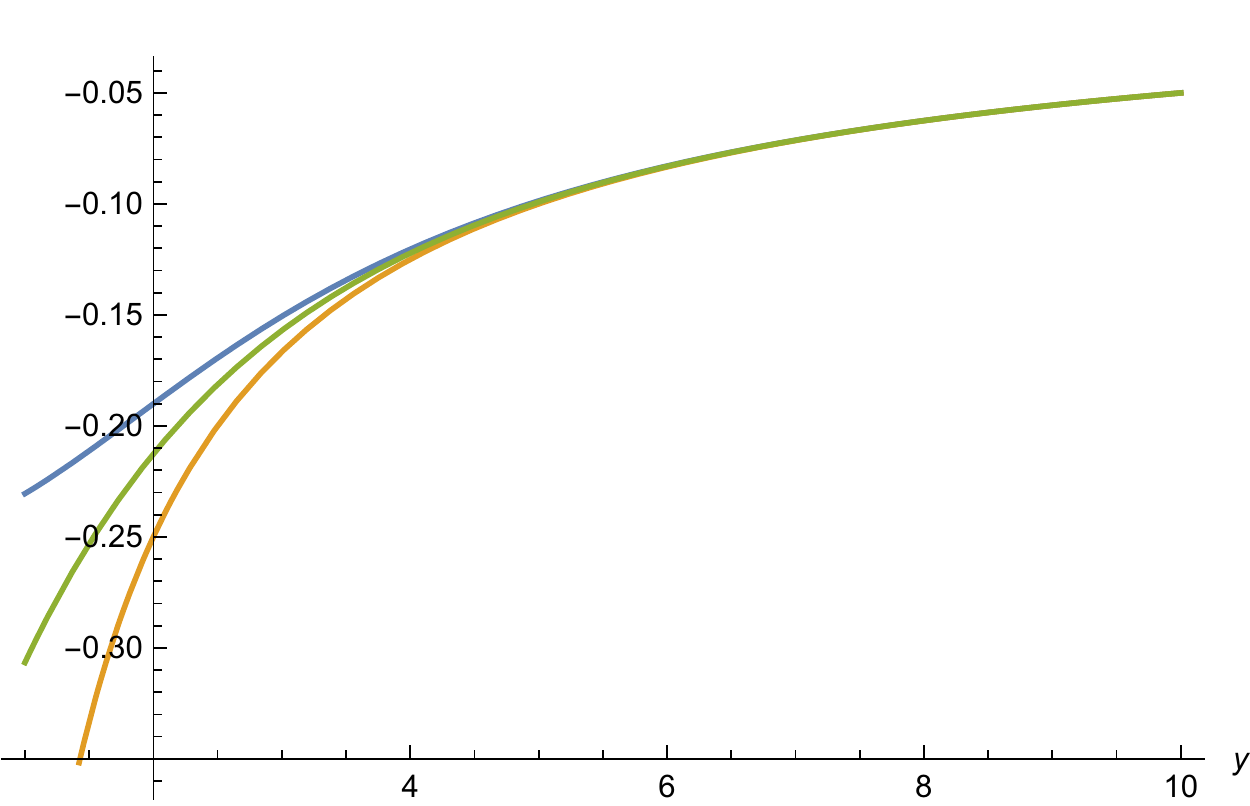}
\includegraphics[width=6 cm, height=4 cm]{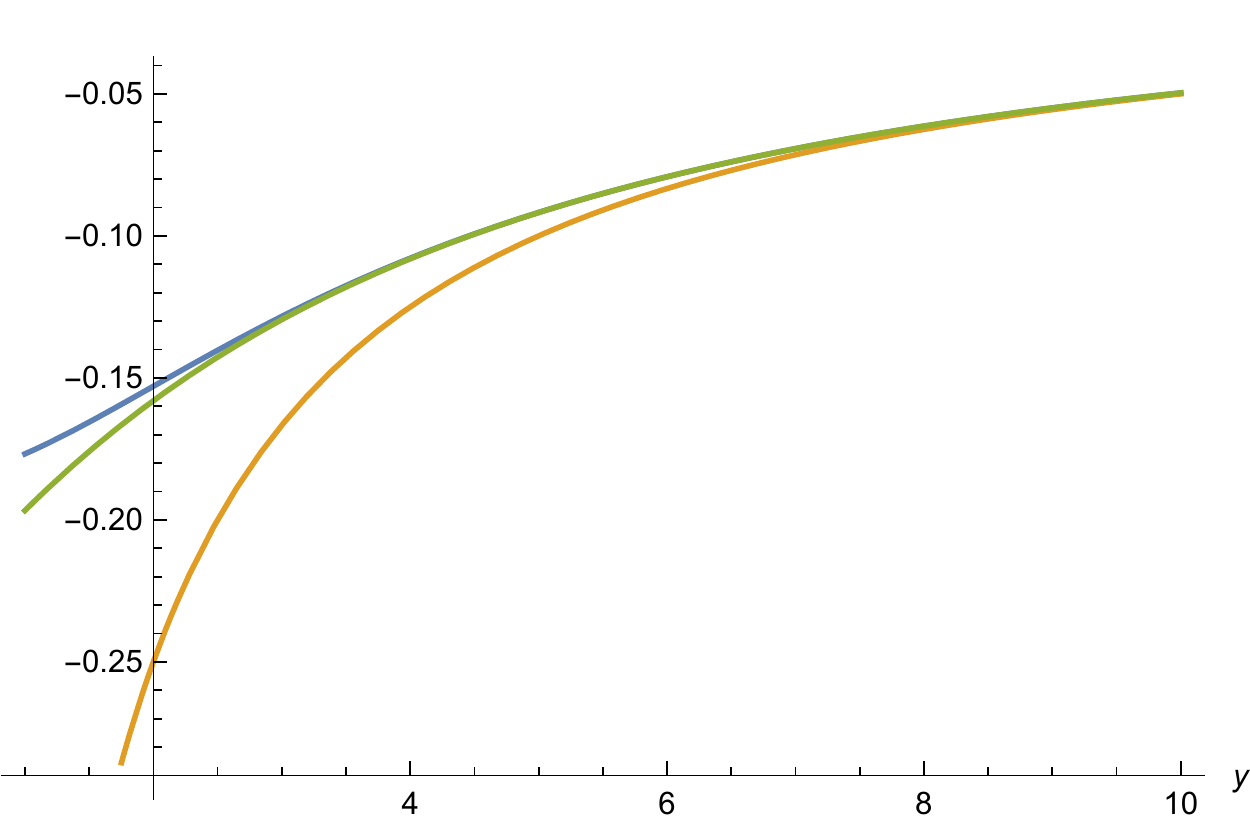}
\caption{$g(y,v)$ (blue), $-1/2y$ (yellow), $(1-e^{-y(1-v)})/2y$ (yellow),
at $v=1/20$ (left) and $v=1/2$ (right)}
\label{fig:gapp}
\end{center}
\end{figure}

I plot in Figure \ref{fig:gapp}  $g(y,v)\;(blue),
 -\displaystyle \frac{1}{2y}\;(yellow),\big(1-e^{-y(1-v)}\big)/2y\;(green)$ at $v=1/20$
(left) and $v=1/2$ (right)
%
%
We notice that $|g|<\displaystyle \frac{1}{2y}$ and $g<0$, therefore
$e^{\displaystyle y\frac{k_\perp^2}{eB}g}\equiv
e^{\displaystyle -y\frac{k_\perp^2}{eB}|g|}>e^{\displaystyle -\frac{k_\perp^2}{2eB}}$.

By replacing $e^{\displaystyle y\frac{k_\perp^2}{eB}g}$ by
$e^{\displaystyle -\frac{k_\perp^2}{2eB}}$
we therefore get a lower bound to the (modulus of the)
 contribution of the main term

\begin{equation}
-T_{\mu\nu}\geq \frac{\alpha}{2\pi}\;2\sqrt{\pi}
(2eB)\;\;\frac{g_{\mu\nu}^\parallel \hat k_\parallel^2 - \hat
k_\mu^\parallel \hat k_\nu^\parallel}{\hat k_\parallel^2}
\frac{1}{\sqrt{eB}}
\int_{y_0}^\infty\frac{dy}{\sqrt{y}}\int_{-1}^{+1}\frac{dv}{2}\;
e^{\displaystyle -\frac{k_\perp^2}{2eB}}
\; e^{\displaystyle -\frac{y}{eB} \big(m^2+\frac{1-v^2}{4}\hat
k_\parallel^2\big)}.
\end{equation}
The exponential under scrutiny is $e^{\displaystyle -\frac{k_\perp^2}{2eB}\frac{\cosh y
-\cosh yv}{\sinh y}}$ and we have used that it is $>
e^{\displaystyle -\frac{k_\perp^2}{2eB}}$. On the other side $\displaystyle \frac{\cosh y
-\cosh yv}{\sinh y} >0$ such that $e^{\displaystyle -\frac{k_\perp^2}{2eB}\frac{\cosh y
-\cosh yv}{\sinh y}}<1$. At large $eB$, the upper and lower bounds are very
close such that our approximation is expected to be quite accurate.

I now use
\begin{equation}
\int_{y_0}^\infty dy\; \frac{e^{\displaystyle -b y}}{\sqrt{y}} =
\sqrt{\frac{\pi}{b}}\;Erf[\sqrt{by}]\Big|_{y_0}^\infty,
\quad b= \frac{1}{eB}\Big(m^2+\frac{1-v^2}{4}(-k_0^2)\Big),
\end{equation}
since $Erf[\infty]=1$, it yields
\begin{equation}
\sqrt{\frac{\pi}{b}}\big(1-Erf[\sqrt{by_0}] \big).
\end{equation}
Furthermore, when $x\to 0$, $Erf[x] \sim \displaystyle \frac{2x}{\sqrt{\pi}}$, which is
the case since $b$ is very small at $B$ large, such that
the result becomes
\begin{equation}
\approx \sqrt{\frac{\pi}{b}}\Big(1 -\frac{2\sqrt{by_0}}{\sqrt{\pi}}\Big)
=\sqrt{\frac{\pi}{b}}-2\sqrt{y_0},
\end{equation}
of which the leading contribution is the 1st one, since $b \propto 1/eB$.

*\ Second contribution to (\ref{eq:NT4}) (counterterm at $y>y_0$)

\begin{equation}
\int_{y_0}^\infty dy\;y \frac{e^{\displaystyle -c y}}{y^{3/2}} =
-\frac{2e^{-cy}}{\sqrt{y}} -2\sqrt{\pi c}\;
Erf[\sqrt{cy}]\Big|_{y_0}^{\infty},\
c=\frac{1}{eB}\big(m^2+\frac{1-v^2}{4}\hat k^2\big),
\end{equation}
which gives
\begin{equation}
\frac{2e^{-cy_0}}{\sqrt{y_0}}-2\sqrt{\pi c}(1-Erf[\sqrt{cy_0}]).
\end{equation}
When $c, cy_0 \to 0$, this can be approximated by
$\displaystyle \frac{2(1-cy_0)}{\sqrt{y_0}} -2\sqrt{\pi
c}\big(1-\displaystyle \frac{2\sqrt{cy_0}}{\sqrt{\pi}}\big)= \displaystyle \frac{2}{\sqrt{y_0}}-2c\sqrt{y_0}
-2\sqrt{\pi c}+4c\sqrt{y_0}=  \displaystyle \frac{2}{\sqrt{y_0}} + 2c\sqrt{y_0}
-2\sqrt{\pi c}$, which are all  sub-leading with respect to the first
contribution.

$\bullet$\ I shall therefore approximate $T_{\mu\nu}^{N_3}$ by its main
term at $y>y_0$ 
\begin{equation}
\begin{split}
& -T_{\mu\nu}^{N_3}(\hat k, B) \geq \frac{\alpha}{2\pi}2\sqrt{\pi}
\;\frac{g_{\mu\nu}^\parallel \hat k_\parallel^2 - \hat
k_\mu^\parallel \hat k_\nu^\parallel}{\hat k_\parallel^2}
\frac{(2eB)}{\sqrt{eB}}e^{\displaystyle -\frac{k_\perp^2}{2eB}}\int_{-1}^{+1}
\frac{dv}{2}\; \frac{\sqrt{\pi eB}}{\sqrt{m^2-k_0^2 \frac{1-v^2}{4}}}\cr
& =\alpha (-2eB)e^{\displaystyle -\frac{k_\perp^2}{2eB}}
\;\frac{g_{\mu\nu}^\parallel \hat k_\parallel^2 - \hat
k_\mu^\parallel \hat k_\nu^\parallel}{\hat k_\parallel^2}
\int_{-1}^{+1}\frac{dv}{2}\; \frac{1}{\sqrt{m^2-k_0^2 \displaystyle \frac{1-v^2}{4}}},
\end{split}
\end{equation}
that is, 
\begin{equation}
-T_{\mu\nu}^{N_3}(\hat k, B) \geq
\alpha\; (2eB)\;e^{\displaystyle -\frac{k_\perp^2}{2eB}}\;
\;\frac{g_{\mu\nu}^\parallel \hat k_\parallel^2 - \hat
k_\mu^\parallel \hat k_\nu^\parallel}{\hat k_\parallel^2}
\frac{1}{k_0}\ln\frac{2m+k_0}{2m-k_0},
\end{equation}
and one must remember that the sign $\geq$ is, in practice, an equality.

\subsubsection{Summing the contributions}
\label{subsub:sumcontribs}

The $N_3$ term cancels the logarithmic contribution of the
$N_1$ part,  and one gets
\footnote{This result is very different from that of \cite{CoquandMachet},
 (except that it is also
finite at $m\to 0$), which has a leading dependence $\propto \sqrt{eB}$ and
satisfies the relation$T_{00}=-T_{33}$.
This difference is due to the counterterm, and also to the fact that
in \cite{CoquandMachet} only the first Landau level was taken into account
 for virtual electrons, while here all of them are  accounted for.
}
\begin{equation}
T_{\mu\nu}(\hat k, B\to\infty)  \simeq
-8 \alpha\; m\; g_{\mu 3}g_{\nu 3} \frac{eB}{4m^2 -k_0^2}.
\end{equation}
The limit $m\to 0$ gives, using (\ref{eq:PiV})
\begin{equation}
\Pi_{\mu\nu}(k, B\to\infty)  \stackrel{m\to 0}{\to} 0.
\label{eq:PiBinf}
\end{equation}
such that radiative corrections to the photon propagator get frozen at
1-loop when $B \to \infty$.

The cancellation of the logarithmic term ensures in particular that the
imaginary part of $\Pi_{\mu\nu}$ vanishes. Its presence would correspond to
the creation of $e^+ e^-$ pairs, in contradiction to the property that an
external magnetic field cannot transfer energy to a charged particle and
cannot trigger such a pair creation (see for example p.83 of \cite{DittrichReuter}).

\section{The scalar potential}\label{section:scalpot}

It does not depend on the choice of counterterms,
which are  $\propto g_{\mu 3}g_{\nu 3}$ (see sections \ref{section:renorm1} and
\ref{section:alter}).

\subsection{A reminder}

See for example \cite{ShabadUsov2008}.

The electromagnetic 4-vector potential produced by the 4-current
$j_\nu(y)$ is
\begin{equation}
A_\mu(x)=-i\int d^4y\; \Delta_{\mu\nu}(x-y)\;j_\nu(y).
\end{equation}
The current of a pointlike charge $q$ placed at $\vec y=0$ is
\begin{equation}
j_\nu(y) = q\,\delta_{\nu 0}\,\delta^3(\vec y).
\end{equation}
Therefore
\begin{equation}
\begin{split}
A_\mu(x) &= -iq\int_{-\infty}^{+\infty}dy_0\;\Delta_{\mu 0}(x_0-y_0,\vec x)\cr
&= -iq\int_{-\infty}^{+\infty}dy_0\;\Delta_{\mu 0}(x_0+y_0,\vec x)\cr
&=-iq\int_{-\infty}^{+\infty}dy_0\;\Delta_{\mu 0}(y_0,\vec x).
\end{split}
\end{equation}
The usual Coulomb potential is easily recovered when one takes the photon
propagator in the Feynman gauge $\Delta_{\mu\nu} \sim
-i\displaystyle\frac{g_{\mu\nu}}{(x-y)^2}$. Only $A_0$ subsists
\begin{equation}
A_0(x) \sim -iq\int dy_0\; \frac{-i}{y_0^2-\vec x^2} \sim \frac{q}{|\vec
x|}.
\end{equation}
In general, in Fourier space
\begin{equation}
\Delta_{\mu\nu}(x)=\frac{1}{(2\pi)^4}\int d^4k\;
e^{ikx}\;\Delta_{\mu\nu}(k),
\end{equation}
such that
\begin{equation}
\begin{split}
A_\mu(x) &=-i\frac{q}{(2\pi)^4}\int dy_0\,d^4k\;e^{i(-k_0y_0+\vec k\vec x)}
\;\Delta_{\mu 0}(k)\cr
&= -i\frac{q}{(2\pi)^4}\int d^4 k\;(2\pi)\delta(k_0)];e^{i(-k_0y_0+\vec k\vec
x)}\;\Delta_{\mu 0}(k)\cr
&= -i\frac{q}{(2\pi)^3}\int d^3k\; e^{i\vec k \vec x}\; \Delta_{\mu
0}(k_0=0,\vec k).
\end{split}
\end{equation}
We see that $k_0$ has to be set to $0$ for  a static charge.

If we are interested in the scalar potential $A_0$
\begin{equation}
\Phi(\vec x) \equiv A_0(\vec x)
=-i\frac{q}{(2\pi)^3}\int d^3k\; e^{i\vec k \vec x}\; \Delta_{0
0}(k_0=0,\vec k).
\end{equation}
So, the geometric series of 1-loop vacuum polarizations that needs to be resummed is that corresponding to
$\Delta_{00}(k_0=0,\vec k)$, which involves $\Pi_{00}((k_0=0,\vec k),B)$
\begin{equation}
\begin{split}
\Phi(\vec k,B) &= -ie[\Delta_{00}(k_0=0,\vec k)] +
 (-ie)[\Delta_{00}(k_0=0,\vec k)(i\Pi_{00}((k_0=0,\vec
k),B)\Delta_{00}(k_0=0,\vec k)]
+ \ldots\cr
&= (-ie)\frac{\Delta_{00}(k_0=0,\vec k)}{1-i\Pi_{00}((k_0=0,\vec k),B)
\Delta_{00}(k_0=0,\vec k)}
\stackrel{Feynman\ gauge}{=}\frac{e}{\vec k^2 +\Pi_{00}((k_0=0,\vec k),B)}.
\end{split}
\label{eq:geom}
\end{equation}

\subsection{Calculating $\boldsymbol{\Pi_{00}((k_0=0,\vec k),B)}$}
\label{subsec:Pi00}

Eq.~(\ref{eq:Piren1}) gives
\begin{equation}
\Pi_{00}((k_0=0,\vec k),B)=-\frac{1}{\pi^2}\frac{1-n^2}{a}
V(n,\theta,\eta,u)\;\frac{\alpha}{2\pi}\frac{2\sqrt{\pi}}{\sqrt{eB}}(-)k_\perp^2
\int_0^\infty\frac{dy}{\sqrt{y}}\int_{-1}^{+1}\frac{dv}{2}\;e^{-(y/eB)\varphi_0}\,N_0
\ \Bigg|_{k_0=0},
\end{equation}
with
\begin{equation}
\begin{split}
\varphi_0\big|_{k_0=0} &= m^2 -\frac{\cosh yv -\cosh y}{2y\sinh
y}\;k_\perp^2,\cr
N_0 &= \frac{y}{\sinh y}\Big( \cosh yv-\frac{v \cosh y\, \sinh yv}{\sinh
y}\Big).
\end{split}
\end{equation}
Using (\ref{eq:Vk0}) yields
\begin{equation}
\begin{split}
\Pi_{00}((k_0=0,\vec k),B) &\stackrel{s_\theta>1/n}{\simeq}
+\frac{\alpha}{\pi^{3/2}}\frac{k_\perp^2}{\sqrt{eB}}
\underbrace{\frac{1-\frac{1-i\cot\theta}{2}e^{-a|\vec k|(1+u)(s_\theta+ic_\theta)}
-\frac{1+i\cot\theta}{2}e^{a|\vec
k|(u-1)(s_\theta-ic_\theta)}}{a}}_{\frac{1-n^2}{\pi}\frac{V}{a}}\cr
& \hskip 5cm
\underbrace{\int_0^\infty\frac{dy}{\sqrt{y}}\int_{-1}^{+1}\frac{dv}{2}\;e^{-(y/eB)\varphi_0}\,N_0
\ \Bigg|_{k_0=0}}_{L(eB, k_\perp^2)},
\end{split}
\end{equation}
which is a convergent integral. It vanishes at $\boldsymbol{B\to
\infty}$ because $N_0 \stackrel{B\to\infty}{\to}0$.

At the limit $a\to 0$ it simplifies to
\begin{equation}
\Pi_{00}((k_0=0,\vec k),B) \stackrel{s_\theta>1/n, a\to 0}{\simeq}
+\frac{\alpha}{\pi^{3/2}}\frac{k_\perp^2}{\sqrt{eB}}
\underbrace{\frac{|\vec k|}{\sin\theta}}_{\frac{1-n^2}{\pi}\frac{V}{a}}
\underbrace{\int_0^\infty\frac{dy}{\sqrt{y}}\int_{-1}^{+1}\frac{dv}{2}\;e^{-(y/eB)\varphi_0}\,N_0
\ \Bigg|_{k_0=0}}_{L(eB, k_\perp^2)},
\end{equation}
with $\varphi_0$ and $N_0$ given in (\ref{eq:Tmunu3}.

In our setup $k_2=0$, $\sin\theta=k_1/|\vec k|$ such that $|\vec
k|/\sin\theta = (k_1^2+k_3^2)/k_1=|\vec k|^2/k_1=|\vec k|^2/k_\perp$.

\subsubsection{The function $\boldsymbol{L(eB,k_\perp^2)}$}

\begin{equation}
L(eB,k_\perp^2)=
\int_0^\infty\frac{dy}{\sqrt{y}}\int_{-1}^{+1}\frac{dv}{2}\;
e^{\displaystyle -\frac{y}{eB}\big(m^2 -\frac{\cosh yv -\cosh y}{2y\sinh
y}\;k_\perp^2\big) }\;\frac{y}{\sinh y}\Big( \cosh yv-\frac{v \cosh y\, \sinh
yv}{\sinh y}\Big).
\end{equation}
In practice, in our setup, $k_\perp^2=k_1^2$.

\begin{figure}[h]
\begin{center}
\includegraphics[width=7cm, height=4cm]{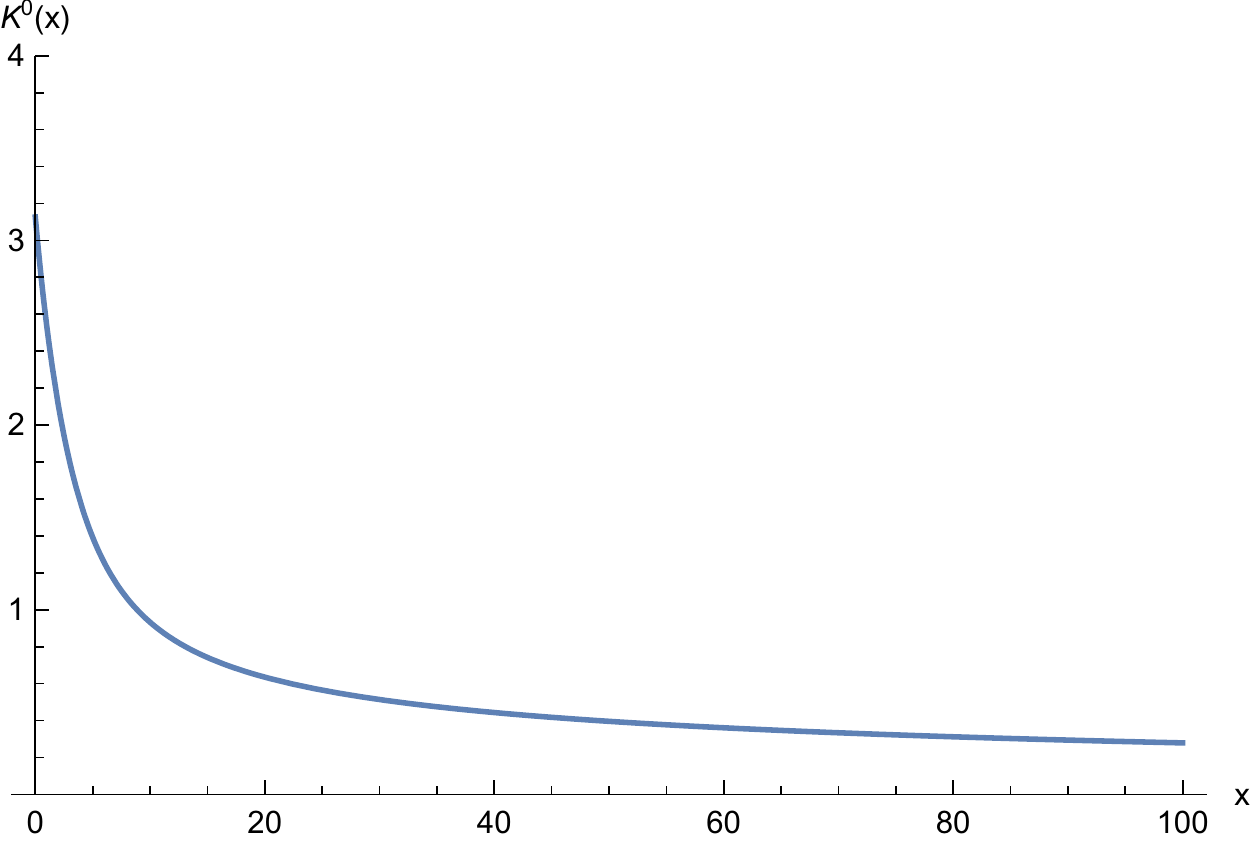}
\hskip 1cm
\includegraphics[width=7cm, height=4cm]{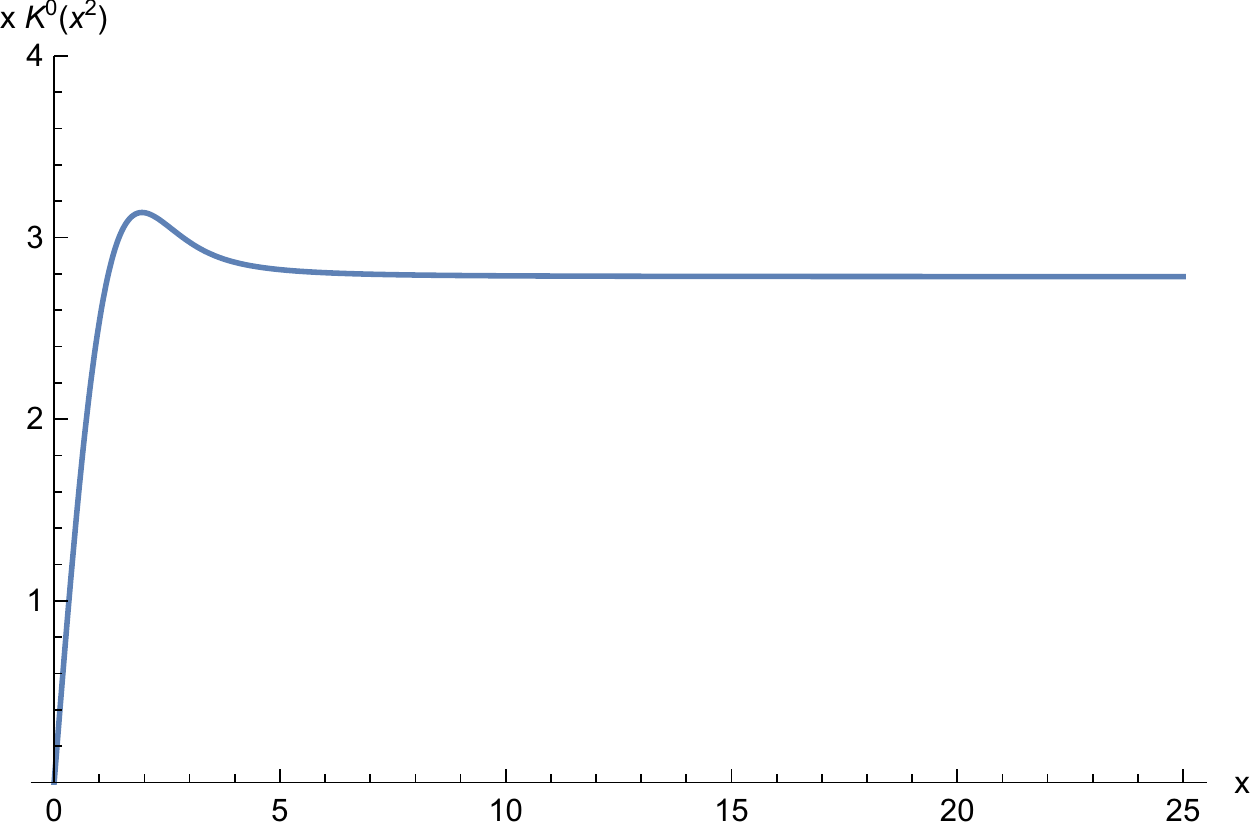}
\caption{$L^0(x)$ (left) and $x L^0(x^2)$ (right).}
\label{fig:K0fig}
\end{center}
\end{figure}

Graphically:\newline
* $f(y,v)= \frac{\cosh yv -\cosh y}{2y\sinh y} \leq 0$, such that, even at
$m=0$ the exponential is convergent;\newline
* $0\leq N_0(y,v) \leq 1$;\newline
therefore the ``singularity $1/\sqrt{y}$ is no problem and  $L$ is a
convergent integral, even at $m=0$. Let $L^0=L|_{m=0}$, such that, at the
limit $a \to 0$
\begin{equation}
\Pi_{00}((k_0=0,\vec k),B=0,m=0) \stackrel{s_\theta>1/n, a\to 0}{\simeq}
+\frac{\alpha}{\pi^{3/2}}\frac{k_\perp^2}{\sqrt{eB}}\frac{|\vec k|}{\sin\theta}
L^0(\frac{k_\perp^2}{eB}),
\label{eq:Pi00}
\end{equation}
with
\begin{equation}
L^0(\frac{k_\perp^2}{eB})=
\int_0^\infty\frac{dy}{\sqrt{y}}\int_{-1}^{+1}\frac{dv}{2}\;
e^{\displaystyle -\frac{k_\perp^2}{eB}\frac{\cosh y -\cosh yv}{2\sinh y}}\;
\frac{y}{\sinh y}\Big(\cosh yv-\frac{v \cosh y\, \sinh yv}{\sinh y}\Big).
\label{eq:K0def}
\end{equation}
In Figure \ref{fig:K0fig}, $L^0(x)$ is plotted for $x\in[0,100]$ on the left,
while on the right is plotted
$x L^0(x^2)$ which will occur when calculating the scalar potential.

\subsubsection{General expression for the scalar potential
$\boldsymbol{\Phi(\vec x, B)}$ at $\boldsymbol{m\to 0}$ and $\boldsymbol{a\to 0}$}

Combining (\ref{eq:geom}) and (\ref{eq:Pi00}) yields, in Fourier space, to
\begin{equation}
\begin{split}
\Phi(\vec x,B) \stackrel{a\to 0,m=0}{=}
 e\int\frac{d^3 \vec k}{(2\pi)^3}\;e^{i\vec k \vec x}\;
\frac{1}{\vec k^2 + \Pi_{00}((k_0=0,\vec k),B,m=0)}
&= e\int\frac{d^3 \vec k}{(2\pi)^3}\;e^{i\vec k \vec x}\;
\frac{1}{\vec k^2 +\frac{\alpha}{\pi^{3/2}}
\frac{k_\perp^2}{\sqrt{eB}}\frac{|\vec k|}{\sin\theta}
L^0(\frac{k_\perp^2}{eB})}\Big|_{k_0=0}\cr
&= e\int\frac{d^3 \vec k}{(2\pi)^3}\;e^{i\vec k \vec x}\;
\frac{1}{(k_3^2+k_\perp^2)\big(1+ +\frac{\alpha}{\pi^{3/2}}
\frac{k_\perp}{\sqrt{eB}}
L^0(\frac{k_\perp^2}{eB})\big)},
\end{split}
\label{eq:scalpot0}
\end{equation}
where, in the last line, we have used the characteristics of our setup,
$\sin \theta=\frac{k_1}{|\vec k|} = \frac{k_\perp}{|\vec k|}$.

\subsection{The scalar potential at $\boldsymbol{eB=0}$}
\label{subsec:potBnul1}

Using  (\ref{eq:TB0m0}) at $k_0=0$
\begin{equation}
T^{00}(\hat k,B=0)\stackrel{m-0}{=}
\alpha\frac{\pi}{2}(-k_\perp^2)\frac{1}{\sqrt{\hat k^2}}
\stackrel{k_0=0}{\to} -\frac{\alpha \pi}{2} k_\perp
\end{equation}
and taking the limit at $a\to 0$ of $\frac{1-n^2}{\pi}\frac{V}{a}$
at $k_0=0$ given in (\ref{eq:Vlim2}) 
yields for $\Pi^{00}$ obtained from (\ref{eq:PiV})
\begin{equation}
\Pi^{00}((0,\vec k),B=0)
\stackrel{a=0,m=0}{=}-\frac{1}{\pi^2}
\underbrace{(-)\frac{\alpha \pi}{2} k_\perp}_{T^{00}}
\underbrace{\frac{\pi |\vec k|}{\sin\theta}}_{\frac{1-n^2}{a} V}
= \frac{\alpha}{2}(k_\perp^2 + k_3^2),
\end{equation}
in which we have used again $\sin\theta=k_1/|\vec k|=k_\perp/|\vec k|$.
One gets accordingly
\begin{equation}
\Phi(\vec x,B=0)\stackrel{a=0,m=0}{=} e\int\frac{d^3 \vec k}{(2\pi)^3}\;e^{i\vec k\vec x}
\frac{1}{k_3^2+k_\perp^2 +\frac{\alpha}{2}(k_\perp^2 + k_3^2)}
=\frac{e}{1+\frac{\alpha}{2}}\int\frac{d^3 \vec k}{(2\pi)^3}\;e^{i\vec
k\vec x}\frac{1}{|\vec k|^2},
\end{equation}
which is the Coulomb potential renormalized by
$\frac{1}{1+\frac{\alpha}{2}}$.

Therefore, even in the absence of external $B$, the Coulomb potential
gets renormalized for a graphene-like medium.

This effect results from a subtle interplay between $T_{00}$, in which the
peculiarities of the graphene hamiltonian play a major role and which does
not vanish at $B=0$, and the ``geometric'' transmittance $V/a$, which, in particular,
does not vanish at $a\to 0$.
The screening effect, small at $\alpha=1/137$, can
become important in a strongly coupled medium $\alpha \sim 1$.

\subsection{The scalar potential at $\boldsymbol{B\to\infty}$}

Because of the factor $1/\sqrt{eB}$  and of the decrease of
$L^0$ at large $B$ (see Figure \ref{fig:K0fig}) that occur in $\Pi^{00}$ (see
(\ref{eq:Pi00})), $\Pi_{00}\stackrel{eB\to\infty}{\to} 0$, such that
\begin{equation}
\Phi(\vec x, B=\infty)= e\int \frac{d^3 k}{(2\pi)^3}\;e^{i\vec k \vec
x}\;\frac{1}{\vec k^2},
\end{equation}
which is the Coulomb potential.

This is due in particular to the property that $\Pi_{00}$ is subleading at
$B\to \infty$ (the leading $\Pi_{33}$ grows instead with $B$, but does not
influence the scalar potential of a static charge).

\subsection{The scalar potential for $\boldsymbol{eB\not = 0}$}
\label{subsec:potB}

I use cylindrical coordinates: $d^3 \vec k = k_\perp dk_\perp d\omega dk_3$ with
$\omega \in [0,2\pi]$. In our setup $k_1 = k_\perp$ up to the sign, which
is accounted for by $\omega \in [0,2\pi]$. So, we can write (\ref{eq:scalpot0}) as
\begin{equation}
\Phi(\vec x,B) = \frac{e}{(2\pi)^3} \int_0^{2\pi} d\omega\int dk_3
\int_0^\infty dk_\perp\;k_\perp
e^{i(k_3 z + k_\perp x_\perp \cos\omega)} \frac{1}{(k_3^2 + k_\perp^2)
\big(1 +\frac{\alpha}{\pi^{3/2}}
\frac{k_\perp}{\sqrt{eB}} L^0(\frac{k_\perp^2}{eB})\big)}.
\label{eq:scalpot2}
\end{equation}

\subsubsection{Potential along $\boldsymbol z$}

I consider (\ref{eq:scalpot2}) at $x_\perp=0$
\begin{equation}
\begin{split}
\Phi(z,B) &= \frac{e}{(2\pi)^3} \int_0^{2\pi} d\omega\int dk_3 \int_0^\infty
dk_\perp\;k_\perp
e^{ik_3 z} \frac{1}{(k_3^2 + k_\perp^2)
\big(1+\frac{\alpha}{\pi^{3/2}}
\frac{k_\perp}{\sqrt{eB}} L^0(\frac{k_\perp^2}{eB})\big)}\cr
&= \frac{e}{4\pi^2}\int dk_3 \int_0^\infty
dk_\perp\;k_\perp
e^{ik_3 z} \frac{1}{(k_3^2 + k_\perp^2)
\big(1+\frac{\alpha}{\pi^{3/2}} \frac{k_\perp}{\sqrt{eB}}
L^0(\frac{k_\perp^2}{eB})\big)}.
\label{eq:scalpotz}
\end{split}
\end{equation}

$\int dk_3\;\frac{e^{ik_3 z}}{k_3^2+c^2}=\frac{\pi}{c}\;e^{-cz}, z>0
\Rightarrow$

\begin{equation}
\Phi(z,B) = \frac{e}{4\pi^2}\int_0^\infty dk_\perp\;k_\perp
\frac{1}{1+\frac{\alpha}{\pi^{3/2}}
\frac{k_\perp}{\sqrt{eB}}L^0(\frac{k_\perp^2}{eB})}\;
\frac{\pi}{c}\;e^{-cz},\quad
c^2 =k_\perp^2.
\end{equation}
It gives
\begin{equation}
\Phi(z,B)= \frac{e}{4\pi}\int_0^\infty dk_\perp
\frac{ e^{-zk_\perp} }
{1+\frac{\alpha}{\pi^{3/2}}
\frac{k_\perp}{\sqrt{eB}}L^0(\frac{k_\perp^2}{eB})}.
\end{equation}
If one neglects all corrections proportional to $\alpha$, one gets
$\Phi(z) \to \frac{e}{4\pi}\int_0^\infty dk_\perp e^{-z k_\perp}=
\frac{e}{4\pi z}$ which is the Coulomb potential.

\begin{figure}[h]
\begin{center}
\includegraphics[width=7cm, height=4cm]{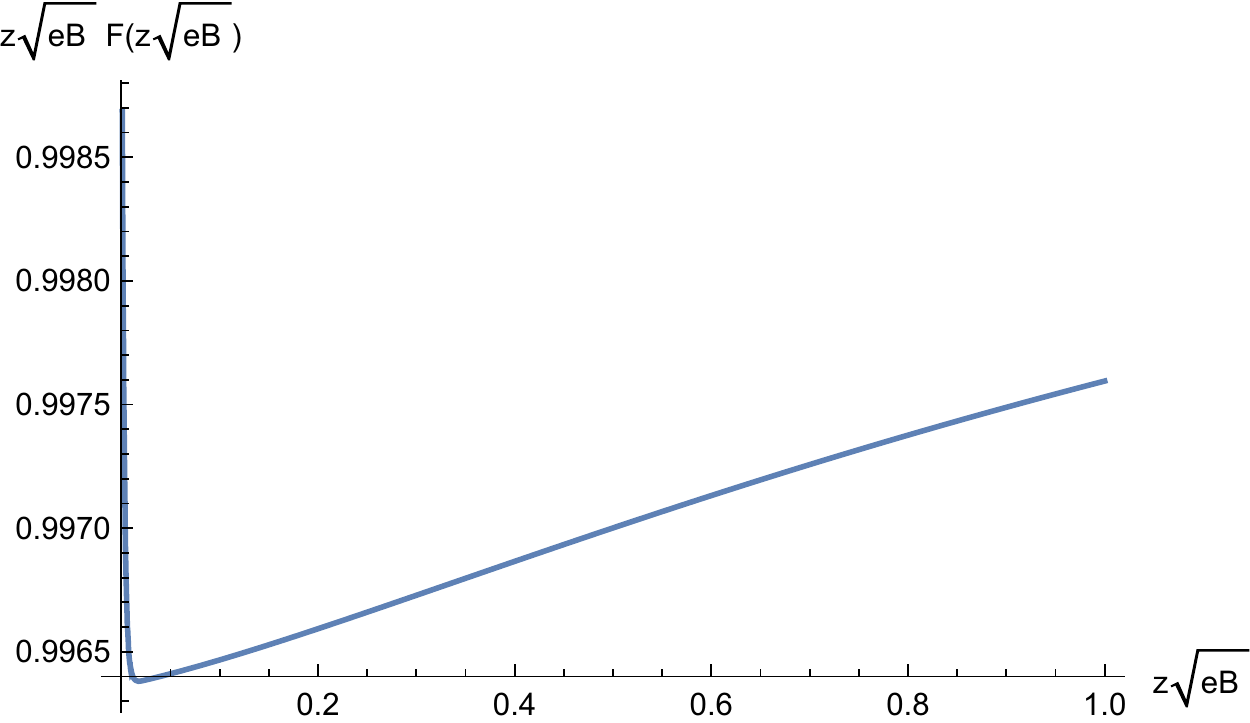}
\hskip 1cm
\includegraphics[width=7cm, height=4cm]{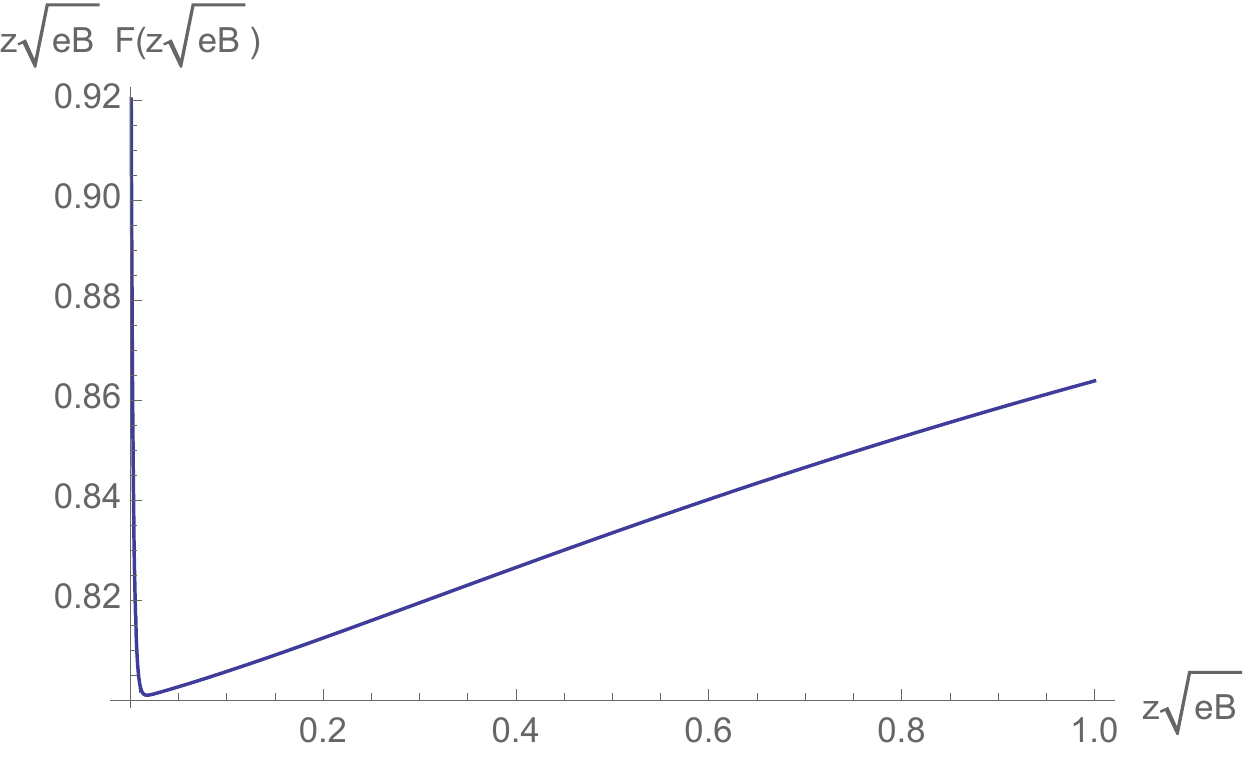}
\caption{the scalar potential along $z$ at the limit $a\to 0$. On the
left $\alpha=1/137$, on the right $\alpha=1/2$.}
\label{fig:scalpotz}
\end{center}
\end{figure}

Going to the integration variable $u=k_\perp/\sqrt{eB}$, $\Phi(z,B)$  rewrites
\begin{equation}
\Phi(z,B)= \frac{e}{4\pi z} z\sqrt{eB}\int_0^\infty du\;
\frac{e^{-z\sqrt{eB} u}}{1+\frac{\alpha}{\pi^{3/2}}\;u\, L^0(u^2)}
=\frac{e}{4\pi z}\; z\sqrt{eB}\;F(z\sqrt{eB}),
\label{eq:zpot}
\end{equation}
in which $z\sqrt{eB}\;F(z\sqrt{eB})$ gives the correction to the Coulomb
potential.
This correction is plotted on Figure \ref{fig:scalpotz}
 for $\alpha=1/137$ (left) and $\alpha=1/2$ (right).

The curves should not be trusted at $z\sqrt{eB} \to 0$. Indeed, 
the limit $eB \to 0$ should be taken before changing to the
integration variable to $y = ieBs$ (see subsection \ref{subsec:changevar}),
 which goes to $0$ with $B$.
 This has been done in subsection
\ref{subsec:Bnul}, with the consequences explicitly studied in subsection
\ref{subsec:potBnul1}. The quasi-straight lines of Figure
\ref{fig:scalpotz} should be
continued till they cross the vertical axes at $\frac{1}{1+1/2\times 137}
\approx .996$ and $\frac{1}{1+1/4} \approx .8$.
  In particular, the singularity of the potential
 at $z=0$ is not canceled, only renormalized.

The scalar potential going to Coulomb at $B\to \infty$, and the curves of
Figure \ref{fig:scalpotz} go asymptotically to $1$.

\subsubsection{Potential at $\boldsymbol{z=0}$ in the transverse plane}
\label{subsub:z0trans}

Setting $z=0$ in (\ref{eq:scalpot0}) yields
\begin{equation}
\begin{split}
\Phi(x_\perp,B) &= e\int\frac{d^3 \vec k}{(2\pi)^3}\;e^{i\vec k_\perp \vec
x_\perp}\;
\frac{1}{\vec k^2 + \Pi_{00}(0,\vec k)}\cr
&=e\int\frac{d^3 \vec k}{(2\pi)^3}\;e^{i\vec k_\perp \vec x_\perp}\;
\frac{1}
{(k_\perp^2 + k_3^2) \big(1 +\frac{\alpha}{\pi^{3/2}}
\frac{k_\perp}{\sqrt{eB}} L^0(\frac{k_\perp^2}{eB})\big)}.
\end{split}
\end{equation}
In cylindrical coordinates as before, $d^3\vec k=
dk_3\; k_\perp dk_\perp\;d\omega$ such that
\begin{equation}
\Phi(x_\perp,B)= \frac{e}{(2\pi)^3}\int dk_3\int_0^\infty dk_\perp\;k_\perp
\int_0^{2\pi}d\omega\; e^{ik_\perp x_\perp \cos\omega}\frac{1}
{(k_\perp^2 + k_3^2)\big(1 +\frac{\alpha}{\pi^{3/2}}
\frac{k_\perp}{\sqrt{eB}} L^0(\frac{k_\perp^2}{eB})\big)}.
\end{equation}
Integrating $\int d\omega$ yields
\begin{equation}
\begin{split}
\Phi(x_\perp,B)
&= \frac{e}{4\pi^2}\int dk_3 \int_0^\infty dk_\perp\;k_\perp
\frac{J_0(k_\perp x_\perp)}
{(k_3^2+k_\perp^2)(1+\frac{\alpha}{\pi^{3/2}}
\frac{k_\perp}{\sqrt{eB}}
L^0(\frac{k_\perp^2}{eB}))}\cr
&= \frac{e}{4\pi^2}\int_0^\infty dk_\perp\;k_\perp
\frac{J_0(k_\perp x_\perp)}{1+\frac{\alpha}{\pi^{3/2}}
\frac{k_\perp}{\sqrt{eB}}
L^0(\frac{k_\perp^2}{eB})}\;\frac{\pi}{k_\perp}\cr
&= \frac{e}{4\pi}\int_0^\infty
dk_\perp\;\frac{J_0(k_\perp x_\perp)}{1+\frac{\alpha}{\pi^{3/2}}
\frac{k_\perp}{\sqrt{eB}}
L^0(\frac{k_\perp^2}{eB})},
\end{split}
\end{equation}
where $J_0$ stands for the Bessel function of 1st kind.
I cast $\Phi(x_\perp,B)$  in the form
\begin{equation}
\begin{split}
\Phi(x_\perp,B)&= \frac{e}{4\pi x_\perp}\; x_\perp \sqrt{eB}
\int_0^\infty\frac{dk_\perp}{\sqrt{eB}}
\frac{J_0(\frac{k_\perp}{\sqrt{eB}}\; x_\perp \sqrt{eB})}
{1+\frac{\alpha}{\pi^{3/2}}\frac{k_\perp}{\sqrt{eB}}\;L^0(\frac{k_\perp^2}{eB})}\cr
&= \frac{e}{4\pi x_\perp}\; x_\perp \sqrt{eB}\ G(x_\perp \sqrt{eB}),\quad
G(x_\perp \sqrt{eB})=\int_0^\infty du\;
\frac{J_0(x_\perp \sqrt{eB}\,u)}
{1+\frac{\alpha}{\pi^{3/2}}\,u\;L^0(u^2)},
\end{split}
\label{eq:transpot}
\end{equation}
in which we have gone to the variable $u = k_\perp/\sqrt{eB}$. With respect
to the scalar potential along the $z$ axis and formula (\ref{eq:zpot}), the
decreasing $\exp[-z \sqrt{eB}\,u]$ has been replaced with the oscillating
and decreasing $J_0(x_\perp\sqrt{eB}\, u)$.

If one neglects the corrections proportional to $\alpha$ one gets
$\Phi(x_\perp) \approx \frac{e}{4\pi x_\perp}\int_0^\infty dk_\perp x_\perp
J_0(k_\perp x_\perp) = \frac{e}{4\pi x_\perp} \times 1$, which is
the Coulomb potential.

Since getting curves for the potential turns out to be very difficult, let us only
understand why the deviations from Coulomb are in general very small.
The corrections to $1$ in the denominator of (\ref{eq:transpot}) are
$\frac{\alpha}{\pi^{3/2}} u L^0(u^2) \approx \frac{\alpha}{5.57}uL^0(u^2)$.
We have seen on Figure \ref{fig:K0fig} that $u K^0(u^2) \leq 3$
which makes this correction $\leq .54\,\alpha$.
One accordingly expects sizable corrections to the Coulomb potential
only in strongly coupled systems. Like before, at $z\sqrt{eB}=0$,
$\Phi(x_\perp) = \frac{Coulomb}{1+\alpha/2}$.

\section{Alternative choices of counterterms}
\label{section:alter}

\subsection{Boundary terms and counterterms}

Counterterms are devised to fulfill suitable renormalization conditions (in
our case the on mass-shell conditions (\ref{eq:rencond})), and in
particular cancel unwanted infinities. In standard QED$_{3+1}$ in external
$B$, this is enough to ensure the $(3+1)$-transversality of the vacuum polarization
$k_\mu k_\nu \Pi^{\mu\nu}=0$ (see for example \cite{Schwinger1951}),
closely connected to gauge invariance and to the conservation of the
electromagnetic current.
However, as  shown in \cite{DittrichReuter} (see p.70 for example),
 this is obtained by including inside the counterterms the
boundary terms of partial integrations. Since
boundary terms obviously depend on the external $B$ (and have no reason to
be transverse), the property that the sum [boundary terms + counterterms]
do not depend on $B$ actually means that the raw counterterms do depend on
it. This is non-standard (see for example \cite{Collins}), but one presumably
cannot state whether this is legitimate or not; along the path followed by Schwinger and
\cite{DittrichReuter}, one is induced to consider
that introducing $B$-dependent counterterms can be necessary. I 
therefore propose below to improve the situation concerning the
transversality of $\Pi_{\mu\nu}$ along this line.

The counterterms should eventually be adapted:\newline
*\ to fulfill of course the renormalization conditions
(\ref{eq:rencond});\newline
*\ to cure the divergence of the  so-called
$B_{\mu\nu}$ of subsection \ref{subsec:firststeps} coming from
classically imposing $p_3=0$ and $p_3-k_3=0$ for internal electron
propagators to match a graphene-like Hamiltonian;\newline
*\ to eventually achieve full $3+1$-transversality
$k_\mu k_\nu \Pi^{\mu\nu}=0$ instead of  restricted $2+1$-transversality
 $\hat k_\mu \hat k_\nu T_{\mu\nu}=0=\hat k_\mu \hat k_\nu \Pi^{\mu\nu}$.

In addition, the production of $e^+ e^-$ pairs should not occur in the sole
presence of a constant external $B$, which sets
constraints on the imaginary part of $\Pi_{\mu\nu}$.

As we have seen in subsection \ref{subsec:transverse}
\begin{equation}
\hat k_\mu \hat k_\nu T^{\mu\nu}(\hat k, B)=0
\Rightarrow
k_\mu k_\nu\Pi^{\mu\nu}= \frac{1}{\pi^2}\frac{1-n^2}{a} k_3^2\;
V(n,\theta,\eta,u)\; T_{33}(\hat k,B),
\end{equation}
such that the non-transversality of $\Pi^{\mu\nu}$ is solely connected to
$T^{33}$. This is why we shall only consider modifying the counterterms in
relation with $T^{33}$.

I shall investigate the two following subtractions, the first being
independent on $B$, the second depending on $B$:\newline
*\ $T_{\mu\nu}^{ren}(\hat k B) = T_{\mu\nu}^{bare}(\hat k, B)-g_{\mu
3}g_{\nu 3}\,T_{33}^{bare}|_{B=0}$
;\newline
*\ $T_{\mu\nu}^{ren}(\hat k B) = T_{\mu\nu}^{bare}(\hat k, B)-g_{\mu
3}g_{\nu 3}\,T_{33}^{bare}$.

In both cases only the indices $\mu=3, \nu=3$ are concerned, such that
$T_{00} =T_{00}^{bare}$, and therefore $\Pi_{00}$, stay unchanged, together
with the scalar potential. The study of their limits at $k_0=0$ and $m=0$
is as done in section \ref{section:scalpot}.

One can only rely here on transversality  to select the counterterms.
However, modifying $\Pi_{33}$ has consequences on other physical
quantities, like  the refractive index (see for example the beginning of
 \cite{CoquandMachet}). It may happen that reasonable results for the refractive
index (and/or agreement with experiments) can only be achieved
at the price of giving up $3+1$-transversality, leaving only the restricted
$2+1$-transversality. Then, deeper investigations should be done to
understand what ``gauge invariance'' truly means for such a medium as
graphene. I leave this for further works.

\subsection{$\boldsymbol{B}$-independent counterterm.
$\boldsymbol{\Pi_{\mu\nu}}$ made $\boldsymbol{(3+1)}$-transverse only
at $\boldsymbol{B=0}$,
non-vanishing at $\boldsymbol{B=0}$ and at $\boldsymbol{B=\infty}$}
\label{subsec:alt1}

\begin{equation}
T_{33}^{bare}(\hat k, B)
=\frac{\alpha}{2\pi}\,2\sqrt{\pi}\int_0^\infty\frac{dt}{\sqrt{t}}
\int_{-1}^{+1} \frac{dv}{2}\;e^{-t\varphi_0}\;\Big[
(N_0-N_1)\hat k_\parallel^2 +N_0k_\perp^2
\underbrace{-2eB\;\frac{\cosh eBt}{\sinh eBt}}_{\to divergence\ at\ t=0}\Big]
\end{equation}
always depends on $B$, and
\begin{equation}
T_{33}^{bare}(\hat k, B=0)=
\frac{\alpha}{2\pi}\,2\sqrt{\pi}\int_0^\infty\frac{dt}{\sqrt{t}}\int_{-1}^{+1}
\frac{dv}{2}\;
e^{-t\big(m^2+\frac{1-v^2}{4}\hat k^2\big)}\;
\Big[(1-v^2)\hat k^2 -\frac{2}{t}\Big].
\end{equation}
$T_{33}^{bare}$ is divergent at $t\to 0$.

One considers ($\varphi_0, N_0, N_1, N_2$ are given in (\ref{eq:Tmunu3}))
\begin{equation}
\begin{split}
T_{\mu\nu}(\hat k,B) &=T_{\mu\nu}^{bare}(\hat k, B)
-g_{\mu 3}g_{\nu 3}\,T_{33}^{bare}(\hat k, B=0)
 = \frac{\alpha}{2\pi}\;2\sqrt{\pi}
\int_0^{\infty}\frac{dt}{\sqrt{t}}\int_{-1}^{+1}\frac{dv}{2}\cr
&  e^{\displaystyle -t\varphi_0}\Bigg[
N_0[g_{\mu\nu}\hat k^2 - \hat k_\mu \hat k_\nu]
-N_1[\underbrace{g_{\mu\nu}^\parallel \hat k_\parallel^2-\hat k_\mu^\parallel \hat
k_\nu^\parallel}_{\hat k_\parallel^2 g_{\mu 3}g_{\nu 3}}]
+N_2[g_{\mu\nu}^\perp \hat k_\perp^2 -\hat k_\mu^\perp \hat k_\nu^\perp]
- 2\frac{eB\cosh eBt}{\sinh eBt}\;
\underbrace{\frac{g_{\mu\nu}^\parallel \hat k_\parallel^2
-\hat k_\mu^\parallel \hat k_\nu^\parallel}{\hat k_\parallel^2}}_{\equiv g_{\mu
3}g_{\nu 3}}
\Bigg]\cr
& -\underbrace{\frac{g_{\mu\nu}^\parallel \hat
k_\parallel^2 - \hat
k_\mu^\parallel \hat k_\nu^\parallel}{\hat k_\parallel^2}}_{\equiv g_{\mu 3}g_{\nu
3}}\;
e^{\displaystyle -t\big(m^2+\frac{1-v^2}{4}\hat k^2\big)}\;
\Big[\underbrace{(1-v^2)\hat k^2}_{new/section\, \ref{section:renorm1}} 
\underbrace{-\frac{2}{t}}_{cancels\; divergence\;at\;t=0}\Big].
\end{split}
\end{equation}

So doing, the corresponding $\Pi_{\mu\nu}$:\newline
*\ vanishes at $k^2=0$ thanks to $(1-n^2)V$, in particular at $B=0$:
the renormalization condition are therefore satisfied;\newline
*\ does not vanish in general at $B=0$;\newline
*\ is  finite thanks to the term $\propto 2/t$ in the counterterms;\newline
*\ is transverse at $B=0$, $k^\mu k^\nu \Pi_{\mu\nu}|_{B=0}=0$, 
but it is so only at $B=0$.

Unlike in subsection \ref{subsec:infinitlim} it does not vanish at $B\to
\infty$.

\subsubsection{At $\boldsymbol{B=0}$}\label{subsub:1caseB0}

Only $T_{33}$ vanishes at $B=0$ because, then, $N_0 \to 1-v^2,N_1\to 0,N_2 \to
0, \varphi_0 \to m^2+\frac{1-v^2}{4}\;\hat k^2$;
one has
\begin{equation}
\begin{split}
T_{\mu\nu}(\hat k, B=0) &=
T_{\mu\nu}^{bare}(\hat k, B=0)-g_{\mu 3}g_{\nu 3}\,T_{33}^{bare}|_{B=0}
 = \frac{\alpha}{2\pi}\;2\sqrt{\pi}
\int_0^{\infty}\frac{dt}{\sqrt{t}}\int_{-1}^{+1}\frac{dv}{2}\cr
& e^{\displaystyle -t\big(m^2+\frac{1-v^2}{4}\hat k^2\big)}
\;(1-v^2)
\Big[\underbrace{
(g_{\mu\nu}\hat k^2 - \hat k_\mu \hat k_\nu)
 - g_{\mu 3}g_{\nu 3}\; \hat k^2}_{\hat g_{\mu\nu}\hat k^2 -\hat k_\mu \hat
k_\nu} \Big],
\end{split}
\end{equation}
which is transverse because $k^\mu k^\nu(\hat g_{\mu\nu}\hat k^2 -\hat k_\mu \hat k_\nu)
\equiv \hat k^\mu \hat k^\nu(\hat g_{\mu\nu}\hat k^2 -\hat k_\mu \hat k_\nu)=0$.
 One gets
\begin{equation}
\begin{split}
T_{\mu\nu}(\hat k, B=0) &= \alpha(\hat g_{\mu\nu}\hat k^2 -\hat k_\mu
\hat k_\nu)
\int_{-1}^{+1}\frac{dv}{2}\;\frac{1-v^2}{\sqrt{m^2+\frac{1-v^2}{4}\hat
k^2}}\cr
&=\alpha(\hat g_{\mu\nu}\hat k^2 -\hat k_\mu \hat k_\nu)
\frac{2}{\sqrt{\hat k^2}}\;\frac12\Big(
\frac{2m}{\sqrt{\hat k^2}}+\big(1-\frac{4m^2}{\hat k^2}\big)
\cot^{-1}\frac{2m}{\sqrt{\hat k^2}}\Big).
\end{split}
\end{equation}
The limit $m\to 0$ is the transverse
\begin{equation}
T_{\mu\nu}(\hat k, B=0,m=0) 
=\frac{\pi\;\alpha}{2}\;\frac{\hat g_{\mu\nu}\hat k^2 -\hat k_\mu \hat k_\nu}
{\sqrt{\hat k^2}}.
\end{equation}

\subsubsection{At $\boldsymbol{B=\infty}$}

$N_0, N_2 \stackrel{B\to\infty}{\to} 0, N_1 \stackrel{B\to\infty}{\to} -y(1-v^2)$.
\begin{equation}
T_{\mu\nu}(\hat k, B=\infty)= T_{\mu\nu}^{bare}(\hat k, B=\infty)
-g_{\mu 3}g_{\nu 3}\,T_{33}^{bare}|_{B=0},
\end{equation}
such that
\begin{equation}
\begin{split}
T_{\mu\nu}(\hat k,B\to\infty) &=
T_{\mu\nu}^{bare}(\hat k, B\to\infty)-g_{\mu 3}g_{\nu
3}\,T_{33}^{bare}|_{B=0}
 = \frac{\alpha}{2\pi}\;2\sqrt{\pi}
\int_0^{\infty}\frac{dt}{\sqrt{t}}\int_{-1}^{+1}\frac{dv}{2}\cr
& e^{\displaystyle -t\varphi_0}\Bigg[
+y(1-v^2)[g_{\mu\nu}^\parallel \hat k_\parallel^2-\hat k_\mu^\parallel \hat
k_\nu^\parallel]
- 2g_{\mu 3}g_{\nu 3}\;\frac{eB\cosh eBt}{\sinh eBt}\;
\Bigg]\cr
& \hskip 2cm -g_{\mu 3}g_{\nu 3}\;
e^{\displaystyle -t\big(m^2+\frac{1-v^2}{4}\hat k^2\big)}\;
\Big[(1-v^2)\hat k^2 -\frac{2}{t}\Big],
\end{split}
\end{equation}
which is non-transverse as expected.
It is what we have already calculated in subsection \ref{subsec:infinitlim} 
to which is added\newline
 $S \equiv \frac{\alpha}{2\pi}\;2\sqrt{\pi}
\int_0^{\infty}\frac{dt}{\sqrt{t}}\int_{-1}^{+1}\frac{dv}{2}(-)g_{\mu
3}g_{\nu 3}\;
e^{\displaystyle -t\big(m^2+\frac{1-v^2}{4}\hat k^2\big)}\;
(1-v^2)\hat k^2$.
\begin{equation}
\begin{split}
S &= -\frac{\alpha}{2\pi}\;2\sqrt{\pi}\;\hat k^2\;g_{\mu 3}g_{\nu 3}
\int_0^{\infty}\frac{dt}{\sqrt{t}}\int_{-1}^{+1}\frac{dv}{2}\;(1-v^2)\;e^{\displaystyle
-t\big(m^2+\frac{1-v^2}{4}\hat k^2\big)}\cr
& =-\alpha\;\hat k^2\;g_{\mu 3}g_{\nu 3}
\int_{-1}^{+1}\frac{dv}{2}\;\frac{1-v^2}{\sqrt{m^2+\frac{1-v^2}{4}\hat
k^2}}\cr
&= -\alpha\;\hat k^2\;g_{\mu 3}g_{\nu 3}\;\frac{2}{\sqrt{\hat k^2}}\;\frac12\Big(
\frac{2m}{\sqrt{\hat k^2}}+\big(1-\frac{4m^2}{\hat k^2}\big)
\cot^{-1}\frac{2m}{\sqrt{\hat k^2}}\Big)\cr
&= -\alpha \;g_{\mu 3}g_{\nu 3}\sqrt{\hat k^2}
\Big(
\frac{2m}{\sqrt{\hat k^2}}+\big(1-\frac{4m^2}{\hat k^2}\big)
\cot^{-1}\frac{2m}{\sqrt{\hat k^2}}\Big).
\end{split}
\end{equation}
One gets accordingly, at the limit $m\to 0$, the non-transverse
\begin{equation}
T_{\mu\nu}(\hat k,B\to\infty) \stackrel{m\to 0}{\to}
-\alpha\;\frac{\pi}{2}\; g_{\mu 3}g_{\nu 3}\;
\sqrt{\hat k^2}
\end{equation}
which, unlike in subsection \ref{subsec:infinitlim}, does not vanish at
$m=0$.

\subsection{$\boldsymbol{B}$-dependent counterterm. $\boldsymbol{\Pi^{\mu\nu}}$
made always $\boldsymbol{(3+1)}$-transverse, non-vanishing 
at $\boldsymbol{B=0}$, vanishing at $\boldsymbol{B=\infty}$} \label{subsec:alt2}

To make $T_{\mu\nu}(\hat k,B)$, and therefore also $\Pi_{\mu\nu}(k,B)$
always $(3+1)$- transverse,
one  drastically subtracts $g_{\mu 3}g_{\nu 3}\,T_{33}(\hat k,B)$ from
$T_{\mu\nu}^{bare}(\hat k, B)$ (this also cancels the divergence).  One then gets 
\begin{equation}
\begin{split}
T_{\mu\nu}(\hat k, B)=
T_{\mu\nu}^{bare}(\hat k, B)-g_{\mu 3}g_{\nu 3}\,T_{33}^{bare}(\hat k, B) &= \frac{\alpha}{2\pi}\;2\sqrt{\pi}
\int_0^{\infty}\frac{dt}{\sqrt{t}}\int_{-1}^{+1}\frac{dv}{2}\cr
& \hskip -5cm e^{-t\varphi_0}\;
\Bigg[N_0\Big(\underbrace{(g_{\mu\nu}\hat k^2 -\hat k_\mu \hat
k_\nu)-\hat k^2 g_{\mu 3}g_{\nu 3}}_{\equiv \hat g_{\mu\nu}\hat k^2 -\hat k_\mu
\hat k_\nu} \Big)
-N_1 \Big(\underbrace{(g_{\mu\nu}^\parallel \hat k_\parallel^2 -\hat
k_\mu^\parallel
\hat k_\mu^\parallel)-\hat k_\parallel^2 g_{\mu 3}g_{\nu 3}}_{=0}  \Big)
+N_2\big(g_{\mu\nu}^\perp k_\perp^2 -k_\mu^\perp k_\nu^\perp \big)
\Bigg],
\end{split}
\end{equation}
in which $\varphi_0, N_0, N_2$ are as usual given in (\ref{eq:Tmunu3}).

\subsubsection{At $\boldsymbol{B=0}$}

The result is of course the same transverse result as in subsection \ref{subsub:1caseB0}.

\subsubsection{At $\boldsymbol{B=\infty}$}

$N_0,N_2 \to 0$ such that $T_{\mu\nu}(\hat k, B=\infty)=0$: the 1-loop
vacuum polarization vanishes at $B \to \infty$ such that quantum
corrections to the photon propagator get frozen at this order.

Unlike in subsection \ref{subsec:infinitlim}, the limit $m\to 0$ is not necessary to
achieve the vanishing of $\Pi_{\mu\nu}$ at $B\to \infty$.

\section{Salient features of the calculation, remarks and conclusion}
\label{section:conclusion}

\subsection{Generalities}

The calculation that we have performed has two main characteristics:\newline
*\ it accounts for all Landau levels of the internal electrons;\newline
*\ it simulates a graphene-like medium of very small thickness $2a$, inside
which the interactions between photons and electrons are localized (at
1-loop); this technique, which was shown in the case of the electron
self-energy, to reproduce the results of reduced QED$_{3+1}$ on a 2-brane, has still more
important consequences for the vacuum polarization (in which the external
photon is not constrained to propagate inside the medium) with the
occurrence of a transmittance function.  The latter plays a crucial role,
in particular to implement on mass-shell renormalization conditions.
No singularity occurs when $a \to 0$, and our calculations of the scalar
potential have been mostly done at this simple limit
\footnote{In a first attempt \cite{CoquandMachet} to determine the 1-loop vacuum polarization
for a graphene-like medium in external $B$, the calculations were performed
directly at $m=0$, and only the first Landau
level of the internal electrons was accounted for. All calculations turned
out to be finite. This seducing property unfortunately induced us to
forget about  counterterms.}.

\subsection{Dimensional reduction}

The widely spread belief \cite{GGMS2002} that reduced QED$_{3+1}$ on a 2-brane
provides a fair description of graphene has been comforted in
\cite{Machet2016-1}
concerning the propagation of an electron; however, in view of the present results,
one can hardly believe that it provides a reliable treatment of the photon
propagation at 1-loop because it skips the transmittance and cannot allow for 
suitable renormalization conditions. In
particular spurious divergences at $m=0$, due to inappropriate
counterterms, are likely to arise, in addition to the divergence of
$\frac{1-n^2}{a}T^{\mu\nu}$ at $a \to 0$ which is no longer canceled by
the transmittance $V$.
$T_{\mu\nu}$ is the part of $\Pi_{\mu\nu}$ that has the closest properties
to reduced QED$_{3+1}$ on a 2 brane (in there  no ``effective'' internal
photon propagator gets involved). It is however very far from giving a
suitable description of the vacuum polarization of the graphene-like medium
under consideration.

One of the motivations for this work was also to study the competing roles of two types
of dimensional reduction.
The first is the equivalence, when $B\to \infty$, of QED$_{3+1}$ with
QED$_{1+1}$ with no $B$ (Schwinger model). It was an essential ingredient
for example in \cite{Vysotsky2010} \cite{MachetVysotsky2011},
 where the screening of the Coulomb potential due to a superstrong $B$
in QED$_{3+1}$ was investigated.
The second is the ``confinement'' of electrons  inside the $(x,y)$
plane for a very thin graphene-like medium.
Which of the two spatial subspaces,  the $z$ axis (along $B$) or the
$(x,y)$ plane of the medium, would win and control the underlying physics 
 was not clear a priori.

We have seen that, as far as the vacuum polarization is concerned, 
only $\Pi_{33}$ survives at the limit $B\to \infty$ (like becoming
$(0+1)$-dimensional).  It can even vanish when
$m\to 0$, depending on the choice of counterterms.
 When it does, radiative corrections to the
photon propagator get frozen at 1-loop when $B \to \infty$.

\subsection{Radiative corrections to the Coulomb potential}

The scalar potential is controlled  by $\Pi_{00}$ which is
non-leading at large $B$ (with the same caveat as above in the case
where  $\Pi_{33}$ vanishes). As a consequence,
its modification by the external $B$ is completely different from what
happens in standard QED$_{3+1}$
(see for example \cite{Vysotsky2010} \cite{MachetVysotsky2011}).

The limit of an infinitely thin graphene-like medium exhibits an intrinsic
renormalization of the Coulomb potential by $1/(1+\alpha/2)$ at $B=0$.
Going to stronger $B$ tends instead to restore the genuine Coulomb
potential.
The interpolation between $B=0$ and $B=\infty$ being smooth,
the scalar potential can
substantially deviate from Coulomb only in a strongly coupled  medium and for
weak or vanishing magnetic fields.

\subsection{Conclusion and prospects}

Basic principles of Quantum Field Theory provide a clean approach
to radiative corrections for  a graphene-like medium in external $B$.
We have exhibited once more (see \cite{Machet2016-1} \cite{Machet2016-2}) the primordial
importance of the renormalization conditions and of the counterterms.

Many aspects remain to be investigated. Let us mention:\newline
* how does the scalar potential depend on the thickness $a$ when it is
taken non-vanishing?\newline
* can there be experimental tests of, for example, the renormalization of
the Coulomb potential and of its non-trivial dependence on $B$?\newline
* how are the optical properties of graphene, which in particular depend on
$\Pi_{33}$, modified at 1-loop by the external $B$? \newline
* can this, or other physical properties or constraints, help fixing the
counterterms?\newline
* can $(3+1)$-transversality and gauge invariance be achieved or should one
accommodate with  ``reduced'' $(2+1)$-transversality? Which type of gauge
invariance is then at play, which electromagnetic current is / is not
conserved?\newline
* is it justified to introduce $B$-dependent counterterms? Do other examples act
in favor of it? \newline
* how does dressing the photon propagator modifies the electron self-energy?
can consistent resummations be achieved, while implementing at each order
suitable renormalization conditions? what comes out for the electron mass?
does a  gap always open in graphene like we witnessed at 1-loop with a bare
photon?

All these we postpone to forthcoming works.

\vskip 1cm
{\em \underline{Aknowledgments}: very warm thanks are due to Olivier Coquand
who has been a main contributor to section \ref{section:propagator}, and to Mikail
Vysotsky for contunuous exchanges.}


\newpage
\appendix

\section{Demonstration of eq.~(\ref{eq:genform})}
\label{section:genform}

I start from (\ref{eq:start}), in which, now, the fermion propagator $G$
depends on $B$. The notations are always $v=(v_0,v_1,v_2,v_3)=(\hat
v,v_3), \hat v=(v_0,v_1,v_2)$.
\begin{equation}
\begin{split}
\Delta^{\rho\sigma}(x,y)&=
e^2\int d^3\hat u \int_{-a}^{+a}du_3 \int d^3\hat v \int_{-a}^{+a}dv_3\cr
& \int \frac{d^4k}{(2\pi)^4}\; e^{ik(u-x)}\Delta^{\rho\mu}(k)\,
\gamma_\mu \int \frac{d^4p}{(2\pi)^4}\; e^{ip(u-v)}G(\hat p,B)\,
\gamma_\nu \int \frac{d^4r}{(2\pi)^4}\; e^{ir(v-u)}G(\hat r,B)
\int \frac{d^4s}{(2\pi)^4}\; e^{is(y-v)}\Delta^{\sigma\nu}(s)\cr
& =e^2\int d^3\hat u \int_{-a}^{+a}du_3 \int d^3\hat v \int_{-a}^{+a}dv_3\;
\int \frac{d^3 \hat k}{(2\pi)^3}\frac{dk_3}{2\pi}\; e^{i\hat k(\hat u-\hat
x)}e^{ik_3(u_3-x_3)}\Delta^{\rho\mu}(k)\cr
& \hskip 1.5cm
\gamma_\mu \int \frac{d^3\hat p}{(2\pi)^3}\, \frac{dp_3}{2\pi}\;
e^{i\hat p(\hat u-\hat v)} e^{ip_3(u_3-v_3)}G(\hat p,B)\,
\gamma_\nu \int \frac{d^3\hat r}{(2\pi)^3}\, \frac{dr_3}{2\pi}\; e^{i\hat
r(\hat v-\hat u)}
e^{ir_3(v_3-u_3)}G(\hat r,B)\cr
& \hskip 1.5cm
 \int \frac{d^3\hat s}{(2\pi)^3}\,\frac{ds_3}{2\pi}\; e^{i\hat s(\hat y
-\hat v )}e^{is_3(y_3-v_3)}\Delta^{\sigma\nu}(s)\cr
& =e^2\underbrace{\int d^3\hat u\; e^{i\hat u(\hat k+\hat p-\hat
r)}}_{(2\pi)^3\delta(\hat p + \hat k - \hat r)}
\int d^3\hat v\; e^{i\hat v(-\hat p+\hat r-\hat s)}
\int_{-a}^{+a}du_3 \int_{-a}^{+a}dv_3\;
\int \frac{d^3\hat k}{(2\pi)^3}\,\frac{dk_3}{2\pi}\; e^{i\hat k(-\hat
x)}e^{iq_3(u_3-x_3)}\Delta^{\rho\mu}(k)\cr
&\hskip 1.5cm
\gamma_\mu \int \frac{d^3\hat p}{(2\pi)^3}\, \frac{dp_3}{2\pi}\;
e^{ip_3(u_3-v_3)}G(\hat p,B)\,
\gamma_\nu \int \frac{d^3\hat r}{(2\pi)^3}\, \frac{dr_3}{2\pi}\;
e^{ir_3(v_3-u_3)}G(\hat r,B)\cr
 & \hskip 1.5cm
\int \frac{d^3\hat s}{(2\pi)^3}\, \frac{ds_3}{2\pi}\; e^{i\hat s(\hat
y)}e^{is_3(y_3-v_3)}\Delta^{\sigma\nu}(s)\cr
& =e^2\underbrace{\int d^3\hat v\; e^{i\hat v(\hat k-\hat
s)}}_{(2\pi)^3\delta(\hat k-\hat s)}
\int_{-a}^{+a}du_3 \int_{-a}^{+a}dv_3\;
\int \frac{d^3\hat k}{(2\pi)^3}\,\frac{dk_3}{2\pi}\; e^{i\hat k(-\hat
x)}e^{ik_3(u_3-x_3)}\Delta^{\rho\mu}(k)\cr
&\hskip 1.5cm
\gamma_\mu \int \frac{d^3\hat p}{(2\pi)^3}\, \frac{dp_3}{2\pi}\;
e^{ip_3(u_3-v_3)}G(\hat p,B)\, \gamma_\nu \int  \frac{dr_3}{2\pi}\;
e^{ir_3(v_3-u_3)}G(\hat p+\hat k,B)\cr
& \hskip 1.5cm
\int \frac{d^3\hat s}{(2\pi)^3}\,\frac{ds_3}{2\pi}\; e^{i\hat s(\hat
y)}e^{is_3(y_3-v_3)}\Delta^{\sigma\nu}(s)\cr
& =e^2\int_{-a}^{+a}du_3 \int_{-a}^{+a}dv_3\;
\int \frac{d^3\hat k}{(2\pi)^3}\, \frac{dk_3}{2\pi}\; e^{i\hat k(-\hat
x)}e^{ik_3(u_3-x_3)}\Delta^{\rho\mu}(\hat k,k_3)\cr
&\hskip 1.5cm
\gamma_\mu \int \frac{d^3\hat p}{(2\pi)^3}\, \frac{dp_3}{2\pi}\;
e^{ip_3(u_3-v_3)}G(\hat p,B)\,
\gamma_\nu \int  \frac{dr_3}{2\pi}\;
e^{ir_3(v_3-u_3)}G(\hat p+\hat k,B)
 \int \frac{ds_3}{2\pi}\; e^{i\hat k(\hat y)}e^{is_3(y_3-v_3)}
\Delta^{\sigma\nu}(\hat k, s_3)\cr
& =e^2\int \frac{dr_3}{2\pi} \int \frac{ds_3}{2\pi}\int_{-a}^{+a}du_3\;
e^{iu_3(k_3+p_3-r_3)}
\int_{-a}^{+a}dv_3\; e^{iv_3(-p_3+r_3-s_3)}\cr
& \hskip 1cm
\int \frac{d^3\hat k}{(2\pi)^3}\, \frac{dk_3}{2\pi}\; e^{i\hat k(-\hat
x)}e^{ik_3(-x_3)}\Delta^{\rho\mu}(\hat k,k_3)\,
\gamma_\mu \int \frac{d^3\hat p}{(2\pi)^3}\, \frac{dp_3}{2\pi}\;
G(\hat p,B)\, \gamma_\nu\, G(\hat p+\hat k,B)
 e^{i\hat k(\hat y)}e^{is_3(y_3)}\Delta^{\sigma\nu}(\hat k, s_3)\cr
& =\int \frac{dp_3}{2\pi} \int \frac{dk_3}{2\pi}\int \frac{dr_3}{2\pi}
\int \frac{ds_3}{2\pi}\int_{-a}^{+a}du_3\; e^{iu_3(k_3+p_3-r_3)}
\int_{-a}^{+a}dv_3\; e^{iv_3(-p_3+r_3-s_3)} \cr
&\hskip 1.5cm
\int \frac{d^3\hat k}{(2\pi)^3}\;
e^{i\hat k(\hat y -\hat x)}
e^{ik_3(-x_3)} e^{is_3(y_3)}\Delta^{\rho\mu}(\hat k,k_3)
\Delta^{\sigma\nu}(\hat k, s_3)\
\underbrace{e^2 \int \frac{d^3\hat p}{(2\pi)^3}\;
\gamma_\mu\, G(\hat p,B)\, \gamma_\nu\, G(\hat p+\hat k,B)}_{iT_{\mu\nu}
(\hat k,B)},
\end{split}
\end{equation}
which is eq.~(\ref{eq:genform}).

\newpage


\end{document}